# What is Ethical: AIHED Driving Humans or Human-Driven AIHED? A Conceptual Framework enabling the 'Ethos' of AI-driven Higher education

**Prashant Mahajan**
R. C. Patel Institute of Technology, Shirpur, India, registrar@rcpit.ac.in, ORCID ID: 0000-0002-5761-5757

## Abstract

Artificial Intelligence (AI) is transforming higher education (HE) by enabling personalized learning, automating administrative processes, and enhancing decision-making. However, AI adoption presents significant ethical and institutional challenges, including algorithmic bias, data privacy concerns, and governance inconsistencies. This study introduces the Human-Driven AI in Higher Education (HD-AIHED) Framework, an adaptive and structured model designed to integrate human intelligence (HI) into every phase of the AI lifecycle—adoption, design, deployment, evaluation, and exploration. Unlike conventional AI models that prioritize automation, HD-AIHED emphasizes human-centered governance, ethical compliance, and participatory decision-making to ensure that AI enhances rather than replaces human agency in HE.

The framework aligns AI applications with both institutional and student needs, fostering trust, adaptability, and transparency. Its dual-layered approach integrates AI across both the AI lifecycle and the student lifecycle, ensuring context-sensitive, equitable, and goal-oriented AI implementation. A key contribution of this study is its regionally adaptable approach, recognizing variations in technological infrastructure and policy landscapes. Additionally, the integration of SWOC (Strengths, Weaknesses, Opportunities, and Challenges) analysis during the adoption phase allows HE institutions to evaluate AI readiness, mitigate risks, and refine governance structures, while the exploration phase ensures long-term adaptability, scalability, sustainability, and AI promotion through continuous research, innovation, and interdisciplinary collaboration.

To ensure AI remains a responsible enabler in HE, this study advocates for the establishment of University/HE Institutional AI Ethical Review Boards, alignment with global regulatory frameworks (e.g., UNESCO AI Ethics Guidelines, GDPR, Sustainable Development Goal 4), and the promotion of inclusive and transparent AI adoption policies.

Key insights from the HD-AIHED model highlight its role in bridging AI research gaps and overcoming real-time challenges in global HE institutions. The framework offers tailored strategies for diverse educational contexts, including developed and emerging countries, ensuring that AI implementation is contextually relevant and ethically sound. By emphasizing interdisciplinary collaboration among policymakers, educators, industry leaders, and students, this study envisions AI as an ethical and equitable force for innovation in HE.

Ultimately, the HD-AIHED model serves as a catalyst for AI inclusivity rather than exclusion and a driver of educational equity rather than disparity by embedding the ethos of AIHED into higher education systems. Future research should focus on the empirical validation of HD-AIHED through institutional case studies, AI bias audits, and longitudinal assessments to ensure ethical integrity, transparency, and sustainability in AI deployment.

**Keywords:** Artificial Intelligence; Higher Education; AI in HE; Human Intelligence; AI Lifecycle; Ethics; AI Adoption; Scalability and Sustainability.

# 1. Introduction

Artificial Intelligence (AI) is no longer just an emerging technology; it is fundamentally transforming higher education (HE). AI-driven applications such as adaptive learning platforms, predictive analytics, and automated grading systems are reshaping the way institutions function. These tools offer personalized learning pathways, improve administrative efficiency, and enable data-driven decision-making in student retention and academic performance tracking [1–3]

The evolution of Artificial Intelligence in Education (AIED) has predominantly focused on K-12 education (primary and secondary), driving advancements in adaptive learning, automated assessments, and intelligent tutoring systems [4,5]. However, these innovations serve as a foundation for AIHED, which extends AI applications to complexity of HE [6]. While AIED research is rooted in general education, its principles apply to HE with necessary contextual adaptations [7].

The rapid integration of AI in HE (AIHED) raises critical ethical and institutional concerns, including algorithmic bias, student data privacy risks, and regulatory inconsistencies [5,8]. AI's potential to amplify existing educational inequities has been widely documented, particularly regarding biased datasets that disadvantage marginalized student groups [9]. Furthermore, AI adoption remains fragmented, with developing nations facing infrastructure constraints and digital divide issues, exacerbating educational disparities [10].

HE, long revered as a beacon of enlightenment, stands at a crossroads in this AI-driven transformation. AIHED offers personalized learning, optimized institutional processes, and expanded access for underserved populations [6,11]. Adaptive learning platforms, predictive analytics, and immersive virtual environments redefine student success by enabling individualized learning trajectories [10,12,13]. Empirical research demonstrates that AI-driven learning systems outperform traditional methods, providing real-time feedback, personalized content, and adaptive learning pathways [4,5,14]. AIHED not only enhances learning outcomes but frees educators to focus on mentorship, creativity, and moral guidance [7]. Additionally, AI fosters global collaboration, breaking barriers of geography, age, and accessibility, pushing HE into a realm of limitless learning and discovery [6,15]. As an extension of human intellect, AIHED is not merely transformative—it is transcendent [3].

By 2030, AI is projected to contribute approximately $15.7 trillion to the global economy, with higher education emerging as a key beneficiary [16–18]. While this projection underscores AI’s potential to democratize access to education, empirical studies indicate that current AI implementations in HE often fail to meet student expectations, primarily due to fairness concerns, regulatory gaps, and algorithmic biases [19–21]. Research has shown that AI-driven assessments and decision-making systems sometimes exhibit discriminatory tendencies, disproportionately affecting marginalized student populations and raising ethical concerns [9]. Additionally, the lack of standardized AI governance models has led to inconsistencies in AI adoption across institutions, impacting access and equity in educational opportunities [10].

Empirical evidence suggests that structured AI governance frameworks—integrating algorithmic fairness reviews, human oversight, and participatory governance—can significantly enhance AI’s effectiveness in HE [22]. To ensure responsible and sustainable AI deployment, HE institutions must prioritize ethical compliance, transparency, and stakeholder engagement, mitigating biases and fostering equitable educational outcomes.

This study proposes the Human-Driven AI in Higher Education (HD-AIHED) Framework—a structured and ethical approach that incorporates human oversight into AI adoption across HE institutions. Unlike conventional AI models that prioritize automation, HD-AIHED ensures that AI remains a tool for empowerment rather than exclusion. The framework is context-sensitive, addressing the distinct needs of diverse educational settings, while aligning AI applications with institutional priorities, global ethical standards, and student-centric learning models.

## 1.1 Problem Statement: Reporting And Analysis

The integration of AI into higher education (HE) is not merely a technological advancement but an ethical reckoning that challenges the foundational principles of HE [23,24]. While AI offers transformative opportunities—such as personalized learning, enhanced administrative efficiency, and expanded accessibility—it also introduces a duality that must be critically examined.

The 2024 Global AI Student Survey highlights this dual nature of AI in HE, revealing both its potential for transformation and its pressing challenges. On the positive side, 86% of students worldwide report incorporating AI tools into their studies, showcasing AI's growing role in HE. However, 60% of respondents express concerns about fairness, citing risks of algorithmic bias in AI-driven decision-making [19,22].

Additionally, privacy emerges as a key issue, with students questioning the transparency and security of their data, raising concerns about institutional accountability. Despite AI's promise, 80% of students believe that institutional AI implementations fail to meet expectations, highlighting gaps between AI innovation and its real-world application in HE. These findings underscore AI's double-edged impact—driving innovation while reinforcing the urgent need for ethical, inclusive, and transparent governance frameworks in HE [19].

Similarly, a McKinsey & Company report found that approximately 85% of AI practitioners identified bias in algorithms as a key challenge for implementation in HE, emphasizing the necessity of robust frameworks for transparency, fairness, and responsible AI adoption (Singh, 2024). Addressing these concerns requires multi-stakeholder collaboration, rigorous policy frameworks, and adaptive AI governance models that align AI advancements with educational equity, privacy protection, and institutional integrity in HE [9,20].

In the global context, institutions exemplify both the successes and challenges of AIHED. This dual impact showcases transformative achievements alongside critical shortcomings. For instance, AI-powered grading tools like Gradescope have streamlined grading processes, reducing grading time by 50% at UC Berkeley [25]. However, similar AI-driven systems, such as the UK's algorithmic grading during the COVID-19 pandemic, revealed systemic biases that disproportionately disadvantaged underprivileged students [26].

Additionally, predictive analytics at Georgia State University improved graduation rates by 22% by identifying and supporting at-risk students [27]. Yet, similar tools like Ellucian Analytics disproportionately flagged minority students as "at-risk", exacerbating existing inequities [25,28].

Furthermore, platforms like Coursera have expanded access to high-quality education for over 100 million learners globally, particularly in developing regions [29]. However, their limited accessibility in under-resourced areas has deepened the digital divide [30]. Similarly, Squirrel AI in China has demonstrated success by personalizing learning for millions in underserved areas, improving educational outcomes [31], yet its invasive tracking of student behaviors without consent has raised significant privacy and ethical concerns [32].

The widespread integration of AIHED challenges fundamental social and human principles, reshaping the dynamic between humanity and technology [2,33,34]. As [35] notes, every technological innovation brings both empowerment and unintended consequences. While AIHED has democratized access to knowledge and fostered inclusivity, it also introduces ethical dilemmas such as algorithmic bias, data privacy violations, and the erosion of interpersonal connections, as revealed in UNESCO's report [36,37].

The focus must shift from merely questioning AI's future influence to thoughtfully directing its integration to align with human values. In HE—a domain that fosters intellectual growth and societal advancement—this balance between technological progress and purposeful implementation is essential [38,39].

Leveraging AI's transformative potential while addressing its unintended consequences will help preserve HE as a human-centered domain that nurtures empathy, creativity, and diversity. To achieve this, human oversight must ensure AI aligns with pedagogical goals and the ethical imperatives of HE [6,37].

However, one of the key challenges in AI adoption is ensuring that governance structures incorporate student agency and participatory decision-making. Existing AI ethics discussions in HE tend to focus primarily on educators, policymakers, and institutions, with limited involvement of students in shaping AI policy frameworks [10,22].

This study proposes an inclusive approach where students actively contribute to AI governance, ensuring that educational AI applications remain human-centered and aligned with user needs [40] [19,41].

As [42,43] posits, technology is never neutral, as it inherently reshapes human interactions and values, illuminating certain possibilities while obscuring others. In this vein, [40] raise a fundamental question about AIHED: Is AI merely performing ethical actions, or is it processing ethically? [36] argue that AI tools are not neutral but embedded with the values and biases of their developers.

Thus, without careful human oversight, AI risks reducing HE to algorithmic decision-making, stripping it of the moral and emotional depth essential for meaningful learning [44].

### 1.2 Framing the Study: Aspirations, Purpose, and Objectives

The integration of AI into HE remains in its nascent stages, underscoring the pressing need for comprehensive research into its adoption, application, and broader impact [37]. This study seeks to address the duality of AIHED by proposing a conceptual framework that emphasizes ethical, inclusive, and human-centered AI integration.

The primary objective is to harmonize the role of AI as both a transformative driver in HE and a humanized collaborative tool anchored in core human values. By achieving this balance, the study aims to design a strategic roadmap that enhances learning outcomes, fosters equity, and safeguards the ethical and emotional dimensions vital to meaningful HE [6,8].

Central to this study is embedding accountability, adaptability, and empathy into every phase of the AI lifecycle, from adoption to deployment. By treating AI as an augmentation of human intelligence rather than a replacement, the framework aligns with arguments that transparency and inclusivity must be prioritized to empower rather than exclude [26]. Additionally, AI tools are not neutral; they reflect the values and biases of their creators [36]. By emphasizing human oversight, the framework seeks to mitigate potential harm and ensure that AI complements rather than diminishes the human essence of HE. To preserve creativity, empathy, and interpersonal connections, the study proposes measures that ensure AI applications align with human-centric values.

Furthermore, as [6] highlight, the lack of stakeholder's involvement in AI development risks undermining pedagogical integrity. This framework prioritizes active educator participation in AI implementation to align AI capabilities with institutional goals and pedagogical principles. Similarly, [45] underscore the importance of addressing the social dimensions of AI, including cultural and societal diversity. By incorporating these factors, the proposed framework ensures inclusivity, scalability, and responsiveness to the diverse needs of learners across various contexts.

Progress in AIHED, as envisioned in this study, is measured not only by learning outcomes but also by the ethical processes and intentionality guiding its adoption. By aligning AI's transformative capabilities with the moral foundations of HE, the study aspires to create a future where innovation enhances humanity while preserving HE's core values [8].

To achieve these objectives, the study proposes the Human-Driven AIHED framework, a conceptual model emphasizing equity, transparency, and accountability. This framework bridges the gap between technological advancements and ethical integrity, ensuring that AI serves as a collaborative tool that upholds ethical and institutional values rather than undermining them. In doing so, it redefines AI's role in HE as a driver of innovation that supports humanity, enhancing HE while safeguarding its foundational principles.

Accordingly, the study's objectives are:

- To identify and analyze gaps in existing AIHED integration frameworks, particularly in terms of ethics, inclusivity, and human agency, and propose effective strategies to address them.
- To investigate the dual role of AIHED in shaping regional AI adoption patterns and examine the ethical challenges and opportunities it presents for higher education (HE) institutions worldwide.
- To develop and empirically validate the Human-Driven AIHED framework, ensuring that it supports ethical, inclusive, and human-centered AI integration within HE systems.
- To examine real-time global challenges that hinder AI integration in HE, and assess the operability, adaptability, and effectiveness of the Human-Driven AIHED model.
- To explore the role of interdisciplinary collaboration and participatory co-design in promoting the ethical, responsible, and sustainable deployment of AIHED in HE institutions.
- To formulate actionable policy and implementation recommendations for key stakeholders, including educators and policymakers, to ensure the ethical, effective, and context-sensitive integration of AIHED in HE.

By addressing these objectives, the study provides a structured pathway for ethically integrating AI in HE, ensuring that technological innovation aligns with humanity's highest ideals and educational values. To achieve the research aims and objectives of promoting a Human-Driven AIHED framework, this study focuses on addressing the following research questions:

1. What gaps exist in current AIHED integration frameworks regarding ethics, inclusivity, and human agency, and how can they be effectively addressed?
2. How does the dual role of AIHED influence regional AI adoption patterns, and what ethical challenges and opportunities does it present for HE institutions worldwide
3. How can the Human-Driven AIHED framework be developed and empirically validated to ensure ethical, inclusive, and human-centered AI integration in HE systems?
4. What real-time global challenges hinder the integration of AI in HE, and how can the operability of the Human-Driven AIHED model be evaluated?
5. How can interdisciplinary collaboration and participatory co-design enhance the ethical and responsible deployment of AIHED?
6. What actionable recommendations can be provided to key stakeholders—including educators, and policymakers—to ensure the ethical and effective integration of AI in HE?

## 2.0 Research Methodology

Figure 1 details the structured research methodology used in this study. It outlines the sequential steps, starting with research methods, literature review, and theoretical framework, followed by data extraction, synthesis, and conceptual model development, culminating in model validation to ensure applicability and reliability.

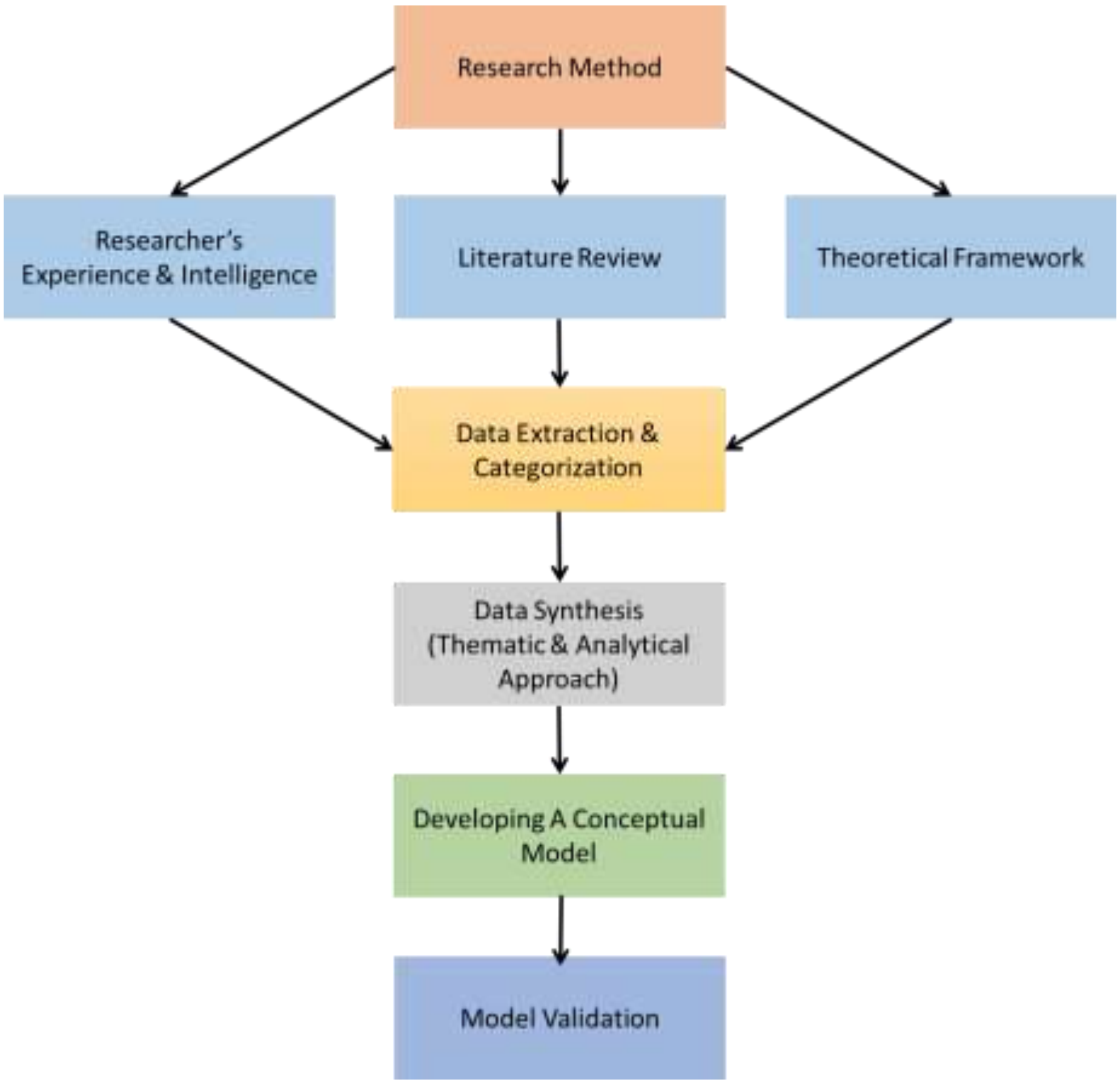

Figure 1: Research Methodology for developing a Conceptual Framework (Own Source)

## 2.1 Nature of Study, Method and Purpose

This study employs a qualitative meta-synthesis approach, integrating findings from multiple qualitative and quantitative studies to identify patterns, contradictions, and gaps in AI adoption within HE [46,47]. The meta-synthesis method is particularly effective for conceptual research [10,48], as it reinterprets existing datasets through theoretical and ethical lenses to develop new governance frameworks for AI in HE [49].

## 2.2 Data Sources and Selection

Secondary data available in English was synthesized from four key sources. Academic literature provided critical insights into the benefits, successes, challenges, and ethical considerations surrounding AI applications in education [8]. Secondary data for this study was sourced from four primary domains:

*Academic Literature* – This included peer-reviewed journals and conference proceedings providing empirical evidence on AI applications, ethical considerations, and institutional challenges in HE [10].

*Global Policy Reports* – Documents from UNESCO, OECD, and IEEE were examined to understand AI's ethical and regulatory landscape [31,50,51].

*Institutional Case Studies* – Real-world applications of AI in HE were analyzed to extract both best practices and challenges in AI implementation [38,39].

*Technology Adoption Theories* – To contextualize AI integration within established educational paradigms, this study reviewed key technology adoption and learning theories. These include the Technology Acceptance Model (TAM) frameworks [52], the Diffusion of Innovation (DoI) Theory [53], the Unified Theory of Acceptance and Use of Technology (UTAUT) [54], and Constructivist Learning Theory (CLT). These models provide insights into user acceptance, innovation diffusion, and pedagogical adaptability in AI-enhanced learning environments.

Additionally, to understand broader socio-educational implications, the study incorporates Trow's Massification Model [55], Bourdieu's Social Capital Theory [56], and Barnett's Higher Education Theory [56]. These theories help frame AI integration within higher education's structural evolution, social dynamics, and institutional governance.

To ensure a comprehensive and high-quality literature review, sources were gathered from reputable academic databases, including: Web of Science, Scopus, ERIC (Education Resources Information Center), EBSCOHost, IEEE Xplore, ScienceDirect, ACM Digital Library, SpringerNature and Google Scholar [10].

The search strategy prioritized studies focusing on ethics, inclusivity, and institutional alignment in AI integration within education or higher education. Research exclusively centered on technical implementations of AI, without pedagogical or ethical considerations, was excluded to maintain the study's education-focused scope [57].

### 2.3 Data Synthesis: Extraction and Categorization

The data synthesis was conducted through two complementary approaches: data-driven descriptive themes and theory-driven analytical themes. The descriptive themes emerged directly from the literature, focusing on observable patterns and recurring insights, providing a practical understanding of AI's current role in HE [40]. Meanwhile, the analytical themes were developed through a conceptual lens, employing established theoretical frameworks to interpret and analyze the data. By integrating these approaches, this study synthesizes practical observations with theoretical insights to develop the HD-AIHED framework, ensuring that AI adoption in higher education remains equitable, scalable, and ethically aligned

#### *2.3.1 Thematic Approach*

Thematic analysis was used to identify and categorize recurring themes in the literature [58,59]. The structured process included:

*Evaluation of Foundational Theories* – AI adoption models were critically examined, revealing gaps in addressing institutional alignment and stakeholder concerns [7,54].

*Highlighting AI's Duality* – AI's potential for efficiency, personalization, and scalability was contrasted against ethical concerns, algorithmic bias, data privacy risks, and inclusivity challenges [26,60–62].

*Comparing Global Frameworks* – UNESCO's AI Ethics Guidelines, OECD's AI Principles, and the European AI Act were analyzed for their applicability in HE, highlighting the need for local adaptation while preserving human integrity [31,50,63].

*Human-Centered Integrated System Approach* – AI integration was examined in the context of human-centered systems, ensuring that technology remains a facilitator of human intelligence rather than a replacement [36,39].

#### *2.3.2 Analytical Approach*

The analytical approach introduced explicit mechanisms to extend primary study content into a theoretical framework. Analytical themes were carefully designed to address gaps identified in the descriptive findings, integrating ethical, operational, and human-centric perspectives to ensure a holistic understanding of AI in HE [31,64]. The process involved:

*Mapping Global Challenges* – Identifying algorithmic bias, digital divide, regulatory fragmentation, and ethical dilemmas in AI adoption [8,26].

*Integrating Ethical Considerations* – Aligning AI adoption with beneficence, non-maleficence, autonomy, justice, and explicability to ensure ethical AI deployment in HE [8,40].

*Stakeholder-Centric AI Adoption* – Ensuring AI integration incorporates educator participation, student agency, and regulatory oversight to balance technological progress with human-centered HE [13,38].

*Decision-Making and Feedback as Human Intelligence* – AI adoption was assessed through decision-making processes, emphasizing the role of human oversight in guiding AI-generated outcomes [6,39].

*AI Life Cycle as Phased Human Intelligence* – The AI integration cycle was examined as a phased approach, incorporating human intelligence at various stages of adoption, design, deployment, analysis, and exploration [13,36]. This approach ensures that AI integration in HE is innovative, ethically sound, and adaptable to the diverse and evolving needs of HE contexts throughout the AI lifecycle [10,49].

*SWOC Analysis for Adoption and Exploration* – A structured SWOC (Strengths, Weaknesses, Opportunities, and Challenges) framework was utilized to evaluate AI adoption strategies and potential areas for future growth [38,40].

*External Intelligence* – AI systems were analyzed in the context of external intelligence, incorporating interdisciplinary perspectives to enhance adaptability and scalability [8,20].

This double powered approach provided a comprehensive framework that aligns AI integration with institutional priorities and ethical imperatives, ensuring that AI adoption remains inclusive, scalable, and accountable [31,50].

### 2.4 Conceptualization of the Framework

The conceptual framework of this study integrates insights from the literature review, theoretical models, and the researcher’s intelligence [39]. It emphasizes the synergy between technological innovation and humanized ethics, promoting accountability, transparency, and scalability. The framework establishes modeling relationships between theories, synthesizing complex theoretical data into thematic and analytical while linking existing research with practical implementation strategies [65].

By prioritizing human oversight, inclusivity, and adaptability, the framework ensures ethically grounded AI applications in HE, aligning innovation with core human values and institutional priorities [31,40].

The framework is designed to distinguish AI's roles across key higher education (HE) functions—teaching, assessment, and administration—ensuring that ethical guidelines are tailored to each domain [11,66]. Additionally, a structured governance roadmap is integrated to define the responsibilities of institutions, policymakers, and AI developers (industry) in promoting responsible AI deployment [10,31,40].

### 2.5 Ethical Considerations and Framework Validation

The proposed framework underwent validation through comparative analysis with global AI ethics frameworks and institutional case studies [31,50]. Validation involved:

*Testing Real-World AI Applications* – Evaluating each phase of the HD-AIHED framework (adoption, design, deployment, analysis, and exploration) using real AI systems to assess feasibility and ethical alignment [8].

*Analyzing Case Studies of Real-World Challenges*– Reviewing institutional experiences with AI to determine the framework’s adaptability across diverse HE settings [38].

*Guidelines for Quantitative Research with Empirical Data for Validation* – Incorporating quantitative research methodologies, including statistical analysis, surveys, and experimental designs, to empirically validate AI implementation outcomes in HE. This ensures that AI models are assessed based on measurable effectiveness, equity, and ethical considerations [3] [37].

These validation efforts ensured that the framework remains applicable, scalable, and aligned with global ethical standards while addressing localized challenges [31,51].

### 2.6 Empirical Validation Strategy

While this study develops a conceptual framework, future empirical testing is required to evaluate the HD-AIHED model's applicability by adopting following strategies.

- Institutional Case Studies: A comparative analysis of HE institutions applying the framework in diverse contexts to assess effectiveness [11,38,67].
- AI Bias Auditing: Implementing a structured evaluation method to identify and mitigate algorithmic biases in AI-driven educational tools [20,68,69].
- Human-AI Interaction Metrics: Developing key performance indicators (KPIs) for inclusivity, transparency, and student engagement to assess AI integration success [10,70]. Future research should focus on mixed-method approaches, incorporating both qualitative and quantitative analyses to assess AI's real-world impact on HE [10,11].

### 2.7 Limitations

While this study provides a strong conceptual foundation, its reliance on secondary data and meta-synthesis introduces limitations regarding empirical validation. Future research should incorporate longitudinal studies, primary data collection, and comparative regional analyses to enhance the framework's practical applicability [10,37]. The study also acknowledges the challenges posed by rapid advancements in AI, requiring continuous updates to the framework to ensure relevance in evolving HE landscapes.

Another limitation is the variability in institutional AI adoption rates, which may influence the applicability of the proposed framework across different HE contexts. Additionally, the integration of external intelligence and human-centered AI approaches requires further empirical validation through experimental and real-world deployment studies [13,20]. Despite these limitations, this study makes a significant contribution to the ethical and human-centric adoption of AI in HE.

## 3. Literature Review and Theoretical Frameworks

The field of Artificial Intelligence in Education (AIED), particularly within Higher Education (HE), has undergone significant growth since 2018, driven by rapid technological advancements and the global shift toward digital learning, particularly accelerated by the COVID-19 pandemic. This expansion has led to innovations in personalized learning, predictive analytics, and administrative automation. However, the ethical implications of AI deployment in education remain a crucial area of academic investigation. Research indicates that China is a leading contributor to AIED scholarship, followed by the United States and Spain, with AIED consistently being a dominant research priority [37].

Aligned with the principles of conceptual framework development [39], this literature review critically assesses the evolution, global trends, and applications of AIED in HE, alongside the ethical challenges it presents. The review interrogates the intersection of human agency and AI, emphasizing how AI's integration in HE is reshaping learning models and institutional governance. By synthesizing findings from empirical studies and theoretical perspectives, the review highlights research gaps and proposes a strategic framework to balance AI's transformative potential with inclusivity, accountability, and equity.

### 3.1 An Overview of AIED

#### *3.1.1 Defining Artificial Intelligence*

Artificial intelligence (AI) has been defined and conceptualized through diverse lenses across academic and industrial domains, reflecting its multifaceted and transformative nature. Broadly, AI is characterized as a system capable of simulating human intelligence to solve problems, analyze complex data, and support decision-making in challenging environments [3]. This perspective underscores AI’s transformative potential, particularly its capacity to process and analyse vast datasets, revolutionizing decision-making processes across industries, including HE.

An alternative definition by [71], frames AI as systems capable of achieving or exceeding human-level performance in various contexts. This view emphasizes AI’s adaptability and self-learning capabilities, which extend beyond static, predefined knowledge. Similarly, [7] describe AI as a cognitive catalyst that solves complex problems through pattern recognition and predictive analytics. This characterization highlights AI’s ability to enhance human cognition, particularly in domains that demand sophisticated analytical and decision-making capabilities.

In line with this, [72] offer a comprehensive understanding of AI, emphasizing its ability to perform intelligent behaviors, including reasoning, learning, and planning. Their foundational work establishes a theoretical basis for understanding AI's applications in various sectors. Additionally, [73] argue that AI encompasses any activity performed by machines that was previously thought to require human intelligence, further broadening the field’s scope.

Despite the nuanced distinctions among these definitions, they converge on the central idea of AI as a powerful enabler that transcends human cognitive limitations. The applications of AI span diverse fields, including HE, where its impact is both profound and transformative. AI’s evolution, marked by rapid advancements in machine learning, natural language processing, and autonomous systems, has transitioned from an evolutionary phase to a revolutionary era, representing a paradigm shift in technology and innovation [74].

### *3.1.2 Emergence of AIHED*

Although the theoretical foundations of AI trace back over seven decades, its integration into HE has gained significant traction only in recent years [75,76]. The COVID-19 pandemic served as a critical inflection point, driving the rapid adoption of AI tools and catalyzing double-digit growth in AI technologies [77]. This unprecedented global crisis accelerated the deployment of AI-driven solutions in teaching, learning, and administrative processes, fundamentally reshaping HE ecosystems [39,78].

The shift toward AI in HE has also been documented by [79], who explore the role of AI in designing personalized learning experiences. They highlight AI’s potential to foster deeper learning by analysing student behaviours and tailoring interventions. Similarly, [5] provide evidence of AI’s role in improving student engagement and retention through adaptive learning systems.

Once considered an enigmatic “black box” technology, AI has evolved into a transparent and integral component of educational ecosystems. Notably, the EDUCAUSE Horizon Report: 2019 Higher Education Edition anticipated the swift proliferation of AI applications in HE, predicting exponential growth and transformative integration, surpassing adoption rates in many other industries [41]. Subsequent trends validated these forecasts, with the AI Index 2024 Report documenting a threefold increase in AIED offerings since 2017, underscoring the sector’s accelerating pace of AI adoption [80].

### *3.1.3 Market Growth and Global Adoption of AIED*

The global AIED market is projected to grow at a compound annual growth rate (CAGR) of 31.2% between 2025 and 2030, signifying substantial investments in AI-driven educational transformation [81]. ‘The Pearson 2024 End of Year AI Report for HE’ corroborates these trends, reporting significant increases in student engagement and academic participation between Spring and Fall 2024 [82]. Additionally, [31] underscores AI’s potential to bridge educational disparities, particularly in underserved regions. However, challenges such as digital access inequalities and AI-induced biases persist, necessitating robust governance frameworks [45].

### 3.2 AIED and AIHED: Convergence and Divergence

The integration of AIED has significantly influenced the development of AIHED, as both domains share foundational principles in personalized learning, student support, and adaptive assessments [6]. While AIED research offers valuable insights into student engagement, automated assessments, and adaptive learning models, AI in HE extends beyond classroom applications to impact institutional governance, faculty recruitment, research integrity, and accreditation processes [26].

AI applications in HE often originate from AIED-driven innovations, particularly in adaptive learning, student support systems, and assessment automation [10,11]. For example, AI-powered platforms such as Khan Academy, Carnegie Learning, and Duolingo have set the foundation for university-level MOOCs like Coursera, edX, and FutureLearn, which provide personalized learning pathways and AI-based skill certification programs [26]. Similarly, AI chatbots such as Georgia State University's Pounce AI, initially developed for K-12 student support, have evolved into predictive analytics systems that enhance university student retention and success [6]. AI-driven grading and plagiarism detection, originally designed for school-level academic integrity enforcement, now play a crucial role in peer review automation, faculty evaluations, and research assessments in HE [8].

Despite these overlaps, AI in HE introduces distinct challenges that necessitate specialized governance frameworks addressing faculty governance, student job placements, research integrity, institutional decision-making and autonomy, external linkages, academic regulatory bodies, industry, accreditations and alumni networks [6,37,83]. While [84] outline the broad potential of AI in education, AI's integration in HE presents specific governance challenges that require structured regulatory frameworks [38].

While early AI research in education [85] emphasized adaptive learning, recent studies in higher education (HE) have shifted focus toward AI's role in academic autonomy, decision-making, stakeholder integration, and research integrity management. AI is increasingly being integrated into institutional policies to streamline decision-making and governance structures [24,38]. Universities are exploring AI-driven strategies to enhance administrative efficiency while maintaining alignment with pedagogical and ethical values [86].

The integration of AI also extends to internal and external stakeholder collaboration, ensuring participatory governance in HEIs. Institutional policies now emphasize structured AI adoption that includes faculty, students, policymakers, and industry partners [22,87]. Additionally, generative AI and learning analytics are being leveraged to create more inclusive, personalized learning environments while addressing ethical concerns [11].

A critical aspect of AI adoption in HE is research integrity management, with scholars highlighting the need for fairness, accountability, and transparency in AI applications [20]. AI-driven tools are being evaluated to ensure that they uphold academic integrity, mitigate bias, and comply with ethical standards [37,88]. As HE institutions navigate this transformation, it is crucial to maintain a balance between AI-driven efficiency and core human-centric values, ensuring responsible and sustainable AI integration.

However, concerns about algorithmic bias, data privacy risks, and AI over-reliance, observed in AI-driven K-12 assessments, continue to persist in HE as well [8]. This is why this study also considers the fundamental approaches of AIED within AIHED to some extent, ensuring that AI adoption in HE remains ethically and strategically aligned with academic goals.

However, given the study's focus on the structured integration of AI in HE, it also underscores the necessity of clear oversight to ensure that universities and HE institutions retain their critical role in knowledge production, intellectual discourse and the preservation of fundamental principles of HE [31].

### 3.3 Theoretical Frameworks Enabling Global Adoption

Artificial Intelligence in Higher Education (AIHED) is influenced by various theoretical models that shape its adoption, governance, and impact. Table 1 presents a global perspective on AIHED theoretical frameworks, highlighting their relevance, applications, and associated challenges in different educational contexts.

**Table 1: Global Perspective on AIHED Theoretical Framework**

| Theory & Founder(s) | Focus Domain & Country | Core Idea & AI Relevance | Impact on AI in Higher Education | Applications in AI-Driven Education | Limitations & Risks of AI in HE |
|---|---|---|---|---|---|
| Trow's Massification Model (1973) [55] | Higher Education (World-wide) | Mass education, scalability, MOOCs, AI tutors | AI democratization, access expansion, scalable models | Coursera, edX, AI tutors | Standardization risk, reduced engagement [6] |
| Bourdieu's Social Capital Theory (1986) [56] | Higher Education (World-wide) | Social capital, networked learning, mentorships | AI-powered networking, peer support, knowledge sharing | LinkedIn Learning, ResearchGate | Digital literacy gaps, academic inequality [83] |
| Barnett's Higher Education Theory (1990) [89] | Higher Education (World-wide) | Knowledge production, student engagement, research collaboration | AI-driven learning paths, lifelong learning, curriculum enhancement | AI-driven advising, research assistants | Overuse of AI, critical thinking reduction [26] |
| Technology Acceptance Model (TAM) (1989) [52] | Higher Education (United States) | User-friendliness, AI adoption, student engagement | Learning efficiency, LMS adoption | AI-enhanced LMS, adaptive learning | Cost, rural-urban divide, privacy concerns [27,80] |
| Unified Theory of Acceptance and Use of Technology (UTAUT) (2003) [54] | Higher Education (European Union) | AI adoption, ethical deployment, performance metrics | AI governance, structured policies, analytics | AI-driven performance tracking, ethical AI | Infrastructure disparity, regulatory delays [27,90] |
| Diffusion of Innovations (DoI) Theory (1962) [91] | Higher Education (China) | AI adoption patterns, innovation scaling, government support | Large-scale AI implementation, government backing | State-led AI platforms, research investment | Urban-rural gap, data privacy concerns [90] |
| Constructivist Learning Theory (CLT) by Jean Piaget and Lev Vygotsky [92,93] | Higher Education (India) | Experiential learning, equity-focused AI, diverse learning needs | AI-personalized education, access expansion | AI literacy programs, mobile AI tools | Infrastructure limits, training gaps, scalability [80] |

The global adoption of AI in higher education (AIHED) is influenced by diverse theoretical frameworks, each shaping AI's role in educational governance and pedagogy [27,80,94]. In the United States, AI integration follows TAM, CLR, and Trow's Massification Model, promoting scalable access but facing challenges related to high costs and privacy concerns. The European Union relies on UTAUT and Barnett's Higher Education Theory, focusing on ethical AI governance and performance metrics, yet struggles with regulatory delays and infrastructural disparities. China's AIHED adoption, guided by DoI and Trow's Model, benefits from strong government support for rapid AI scaling but encounters urban-rural resource gaps and ethical issues. In India, CLR and Bourdieu's Social Capital Theory emphasize equity-focused AI applications and peer networks, though infrastructure limitations and digital inequalities persist. At the global level, AI adoption varies, shaped by governance models like Barnett's Higher Education Theory, but remains hindered by ethical bias, digital disparities, and lack of universal regulation. Thus, these adoption theories explain AI trends in HE but fail to address ethical AI governance and stakeholder-driven decision-making. The proposed framework extends these models by incorporating human oversight, participatory governance, and real-time AI monitoring, ensuring a more ethical and sustainable AI integration strategy.

### 3.4 AIHED Applications and Perspectives: A Global Transformation

Advancements in AIED are revolutionizing HE worldwide, merging technology, pedagogy, and innovation to create inclusive, personalized, and effective learning ecosystems [76]. AI-driven tools are reshaping educational practices by introducing adaptive learning platforms, predictive analytics, and real-time engagement systems that cater to diverse learner needs and institutional priorities [6,67,83,95,96]. However, these innovations must balance technological potential with ethical considerations to ensure equitable and sustainable adoption [83,95]. The integration of AIHED extends beyond functionality—it is a force that redefines human cognition, decision-making, and creativity [22]

#### *3.4.1 Teaching and Learning: A Paradigm Shift*

AIED transforms teaching and learning by introducing personalized learning platforms, predictive analytics, and real-time engagement tools, which collectively create student-centric and adaptive environments.

- *Personalized and Adaptive Learning:*
  AI-powered platforms like Carnegie Learning and Khanmigo—intelligent tutoring systems (ITS)—dynamically assess student performance and personalize content delivery, ensuring targeted support for individual learning needs [7,12]. These AI-driven adaptive learning tools leverage real-time analytics to enhance student engagement, foster individualized instruction, and address learning gaps [17,25]. Empirical studies further indicate that human-centered AI models integrated with ITS platforms improve learning outcomes and accessibility by providing automated, tailored feedback [22,97].
  Global platforms like Coursera and edX extend these capabilities to lifelong learning, aligning educational content with learners' goals to foster career-ready outcomes [11,84]. Platforms like Coursera and edX connect educators and students worldwide, promoting cross-cultural exchanges and aligning with Sustainable Development Goal 4 for equitable access to education [98].
  Global platforms like Coursera and edX extend AI-driven personalized learning to lifelong education, aligning educational content with learners' goals to foster career-ready outcomes [11,84]. These platforms leverage AI-powered analytics and adaptive learning strategies to support individualized skill development and professional growth [25,97]. Furthermore, Coursera and edX serve as global learning hubs, connecting educators and students worldwide, fostering cross-cultural exchanges, and advancing Sustainable Development Goal 4 by promoting equitable access to quality education [98,99].
  Adaptive AI tutors, such as Carnegie Learning's Math Tutor, have demonstrated superior effectiveness over traditional teaching methods by providing real-time feedback and individualized learning paths [79]. Similarly, AI-driven tools like Pounce, a chatbot at Georgia State University, reduce enrollment attrition by offering real-time, personalized support, highlighting AI's potential to enhance student retention and success [11].
  By integrating real-time analytics and predictive modeling, adaptive learning systems continuously refine content delivery, improving accessibility, engagement, and learning outcomes [6,10].

- *Predictive Analytics for Academic Success:*
  Predictive tools in HE empower institutions to anticipate and address challenges by analyzing student behavior and performance data. For example, the University of Guayaquil identifies at-risk students using predictive analytics, enabling timely resource allocation to improve retention [100,101] [37]. In Africa, similar tools address institutional inefficiencies and improve learning outcomes [102].
  Educational Data Mining (EDM) and predictive analytics play a crucial role in identifying at-risk students early, facilitating timely interventions that improve retention and engagement [14]. Additionally, predictive analytics enhance resource management by forecasting enrollment trends and operational needs, ensuring institutional strategies align with student success goals [10]. Predictive analyse and optimize resource allocation, enabling institutions like the University of Melbourne to achieve significant efficiency gains [100]. Squirrel AI has tailored education to millions in China, improving students' outcomes for underserved populations [99].
- *Real-Time Assistance and Engagement:*
  AI-powered chatbots and gamified platforms create interactive, personalized learning experiences that foster engagement and improve outcomes and skills development. Gamification promotes intrinsic motivation and long-term retention, while immersive tools like VR and AR bridge theoretical concepts and practical applications [10,12,37,103,104].

These global studies suggest that AIED has fundamentally transformed traditional teaching and learning practices into dynamic, adaptive environments that emphasize personalization, inclusivity, and engagement. By leveraging adaptive learning platforms, predictive analytics, and immersive technologies, higher education institutions have shifted from standardized methods to tailored, student-centric approaches that address diverse needs and foster better academic outcomes. AI-powered tools not only enable real-time interventions and personalized support but also optimize resource allocation and institutional efficiency. The integration of technologies like gamification and VR/AR bridges theoretical knowledge with practical application, enhancing engagement and preparing students for the challenges of a rapidly evolving global workforce. Ultimately, these advancements mark a paradigm shift in higher education, demonstrating the potential of AIED to create equitable, effective, and impactful learning experiences. As these technologies continue to evolve, they promise to redefine the future of education, ensuring that it is adaptive, inclusive, and aligned with the demands of a globalized and digital society.

### *3.4.2 Administrative Processes*

AI is redefining administrative operations in HE, automating repetitive tasks, enhancing resource management, and enabling strategic innovation. AI automates grading, attendance tracking, and scheduling, reducing workloads and improving efficiency. Platforms like Absorb LMS streamline operations, freeing educators to focus on pedagogy [105].

- *Automated Administrative Tasks:* AI automates grading, attendance tracking, and scheduling, reducing workloads and improving efficiency. Platforms like Absorb LMS streamline operations, freeing educators to focus on pedagogy [103,106,107].
- *Predictive Resource Management:* Predictive analytics forecast enrollment trends and optimize resource allocation, enabling institutions like the University of Melbourne to achieve significant efficiency gains [108,109] This data-driven approach ensures alignment with institutional goals and sustainability.
- *Strategic Innovation:* AI supports curriculum design, faculty development, and interdisciplinary collaboration by aligning strategic priorities with data insights [10,104]. By enabling timely interventions and operational resilience, AI empowers institutions to innovate and thrive in an ever-changing educational landscape.

These global studies revealed that AI is transforming administrative operations in higher education by automating routine tasks, enhancing resource management, and fostering strategic innovation.

Automation tools streamline grading, scheduling, and attendance tracking, reducing workloads and allowing educators to prioritize pedagogy. Predictive analytics further optimize resource allocation and enrollment management, as seen in institutions like the University of Melbourne, ensuring efficiency and sustainability. Moreover, AI supports strategic goals such as curriculum design and faculty development, enabling institutions to align their priorities with data-driven insights. By fostering operational resilience and adaptability, AI empowers higher education institutions to navigate an ever-changing landscape, achieving greater efficiency, innovation, and institutional growth.

### *3.4.3 Inclusivity and Accessibility*

Insights from global research demonstrate that AI plays a pivotal role in fostering inclusivity by addressing barriers faced by marginalized groups and students with disabilities. AI-powered assistive technologies provide meaningful support for learners with cognitive and physical challenges, while multilingual AI tools break down language barriers, extending access to quality education in diverse cultural and linguistic contexts [24,110].

By adopting scalable and innovative AI solutions, institutions are actively reducing educational disparities, ensuring that all learners—regardless of their background or abilities—have equal opportunities to succeed [97]. Empirical studies highlight that AI-powered adaptive learning platforms and personalized learning analytics enhance accessibility, fostering greater inclusivity in digital education [99]. Additionally, human-centered AI models in higher education are transforming digital equity, reinforcing sustainable, accessible, and equitable learning environments on a global scale

Key AI-Driven strategies for inclusive HE includes;

- *Assistive Technologies:* Text-to-speech tools, adaptive content delivery systems, and speech recognition software enable students with cognitive and physical impairments to engage in meaningful learning experiences [111,112].
- *Multilingual Support:* AI-powered translation tools bridge language barriers, extending access to quality education in diverse linguistic and cultural contexts. These tools are particularly impactful in resource-constrained regions, reducing disparities and promoting equity [104,113].

### *3.4.4 Long-Term Benefits*

AIED transcends traditional education boundaries, fostering global collaboration, sustainability, and workforce readiness.

- *Global Collaboration:* Platforms like Coursera and edX connect educators and students worldwide, promoting cross-cultural exchanges and aligning with Sustainable Development Goal 4 for equitable access to education [114,115].
- *Preparing Future-Ready Students:* AI fosters skills such as computational thinking and digital literacy, equipping students for success in an AI-driven economy [104,116].
- *Sustainability and Innovation:* AI-driven educational systems address disparities in resource-constrained regions, ensuring scalable, high-quality learning opportunities [5,10].

Worldwide studies showed that AIED transcends traditional educational boundaries and becoming a catalyst for the future of education by fostering global collaboration, equipping students with future-ready skills, and driving sustainability in education. Platforms like Coursera and edX enable cross-cultural exchanges, aligning with global goals for equitable access to education. Furthermore, AI empowers learners with essential skills such as computational thinking and digital literacy, preparing them for success in an AI-driven economy. By addressing disparities and providing scalable, high-quality learning opportunities in resource-constrained regions, AIED establishes itself as a cornerstone for societal progress, global educational equity, and the sustainable transformation of education systems.

### *3.4.5 Global Dynamics of AIHED Adoption*

HE institutions worldwide are leveraging artificial intelligence to foster innovation, equity, and inclusion. The pace and scope of AI adoption are shaped by regional priorities, economic resources, and policy frameworks. As highlighted in the AI Index Reports [27,80,90], AI is redefining the higher educational landscape on a global scale as illustrated in Table 2. AI adoption in HE varies by region, reflecting unique priorities, as discussed below:

- *Asia:* Countries like China and India leverage AI for scalability and personalized learning, with platforms like Squirrel AI and Byju's addressing urban-rural divides [80,117].
- *Europe:* Initiatives like Finland's Elements of AI democratize AI literacy, while Germany's AI Campus integrates vocational training with AI technologies [84].
- *North America:* Institutions like Stanford lead in deploying predictive analytics and adaptive platforms, emphasizing personalization and workforce readiness [27,80].
- *Africa:* AI-powered tools bridge the digital divide, extending access to rural areas through mobile learning [7].
- *Oceania and the Middle East:* Universities integrate AI with cultural and regional knowledge systems to promote inclusivity and address societal challenges [11,80].

**Table 2: Global Dynamics of AIHED Adoption**

| **Region** | **AIHED Application** | **Area of Focus** | **Literature Support** |
|---|---|---|---|
| Asia | Scalability and Personalization; Adaptive technologies (Squirrel AI, Byju's, Digital Textbook); Multilingual tools (Duolingo, Google Assistant); Lifelong learning platforms (Coursera, edX). | Addresses diverse educational needs; Promotes cross-cultural learning; Reskilling and upskilling opportunities in AI fields. | [23,117–120] |
| Europe | Equity and Digital Competence; National strategies for inclusivity and digital competence; AI literacy programs (Elements of AI); Intelligent tutoring systems (AI Campus). | Ensures equitable learning outcomes; Promotes inclusivity; Advances vocational and applied learning. | [7,10,11,84] |
| North America | Personalization and Workforce Readiness; Adaptive learning platforms (Coursera, ALEKS); Predictive analytics for learning outcomes; Indigenous knowledge integration. | Enhances individualized learning; Prepares workforce for AI-driven economy; Fosters inclusivity with indigenous knowledge. | [27,70,80,121,122] |
| South America | Overcoming Constraints; Adaptive platforms for personalized courses; Predictive analytics to identify at-risk students; International collaborations. | Improves equity and accessibility; Enables proactive student interventions; Strengthens regional collaborations. | [80,123,124] |
| Africa | Bridging the Digital Divide; Mobile learning applications with offline capabilities; Multilingual AI tools for inclusivity; Pan-African AI literacy initiatives. | Expands rural education access; Breaks language barriers; Promotes educator inclusivity. | [24,38,111] |
| Oceania | Cultural Sensitivity and Advanced Analytics; Integration of AI with indigenous knowledge systems; Optimized resource allocation; Lifelong learning platforms. | Optimizes learning through cultural relevance; Promotes equity and personalized education. | [11,39,80] |
| Middle East | Smart Campuses and Problem-Solving; Attendance monitoring, engagement analysis; AI for retention improvement; Research on water scarcity and renewable energy. | Improves student engagement and retention; Addresses region-specific challenges; Supports innovation through research. | [7,17,99,121] |

Above Table 2 illustrates how regional efforts collectively highlight AI's capacity to transform traditional educational paradigms into dynamic, student-centric, and adaptive ecosystems. By addressing unique socio-economic contexts, AI drives a global movement toward more equitable, efficient, and future-ready higher education systems, establishing a foundation for long-term societal progress and innovation.

### 3.5 Global Regulatory Frameworks for Ethical AIHED

Ethical AIED requires robust regulatory frameworks to ensure fairness, transparency, accountability, and inclusivity. Below is an overview of key existing frameworks and guidelines developed by international organizations, national bodies, and regional initiatives, supported with proper references.

#### *3.5.1 UNESCO and OECD's AI Ethics Guidelines*

The UNESCO Recommendation on the Ethics of Artificial Intelligence (2021) emphasizes fairness, inclusivity, accountability, and respect for human rights in AI implementation. It advocates for capacity building, data governance, and monitoring mechanisms, particularly within HE, to ensure ethical AI adoption [24,31,125]. However, a lack of enforceability at the national level has led to inconsistent adoption across countries, limiting the global impact of these principles.

Similarly, the OECD's AI Principles (2019) promote human-centered values, transparency, and robustness in AI systems. Adopted by 42 countries, these principles serve as a global benchmark, influencing regional policies such as the EU's AI framework [50]. Despite their significance, these guidelines lack actionable mechanisms for sector-specific challenges, particularly those unique to higher education, where AI governance must balance ethical compliance with academic autonomy and institutional diversity.

#### *3.5.2 NASSCOM and NITI Aayog's AI Initiatives (India)*

In India, NASSCOM and NITI Aayog have developed initiatives to regulate AI ethically:

- NASSCOM's Responsible AI Hub promotes fairness, accountability, and inclusivity in AI systems, with a focus on higher education [127].
- NITI Aayog's National Strategy for AI (2018) emphasizes AI as a tool for social empowerment, addressing sectors such as higher education, health, and agriculture. It stresses transparency, data privacy, and bias mitigation while advocating for public-private collaborations [128].

While these initiatives drive innovation, they often lack robust regulatory mechanisms to monitor compliance, and their implementation faces challenges in rural and resource-constrained areas.

#### *3.5.3 US Department of Commerce and NIST AI Risk Management Framework*

In the United States, the NIST AI Risk Management Framework (2023) provides risk assessment strategies aligned with ethical principles such as fairness, transparency, and accountability [63]. It encourages participatory processes to ensure diverse stakeholder needs are addressed within higher education, supporting ethical AI deployment [9,20]. Additionally, federal bodies such as the U.S. Department of Higher Education (2021) have issued guidelines to safeguard student data and privacy, reinforcing the importance of institutional compliance and governance [129].

However, the NIST framework lacks specificity in integrating diverse stakeholder inputs and struggles to address the complexities of global data governance and rapidly evolving AI technologies. Empirical studies highlight that AI governance models in HE must integrate adaptive risk management strategies to mitigate bias, enhance algorithmic fairness, and align with national and international AI ethics guidelines [130]. Addressing these limitations requires sector-specific policies tailored to the unique ethical and operational challenges of AI deployment in HE.

### *3.5.4 European AI Ethics and Regulatory Frameworks*

The European Commission's Ethics Guidelines for Trustworthy AI (2019) and the proposed EU Artificial Intelligence Act (2021) focus on regulating high-risk AI systems in higher education. These frameworks emphasize human agency, technical robustness, privacy, transparency, diversity, environmental sustainability, and accountability [131,132].

Despite their rigorous standards, the high compliance costs and tailored focus on well-resourced EU member states pose challenges for smaller institutions and those in lower-income regions.

### *3.5.5 Other Global Ethical Standards for AI*

Various global bodies, such as the Global Partnership on AI (GPAI) and IEEE's Global Initiative on Ethics of Autonomous and Intelligent Systems, have established principles to guide ethical AI deployment.

- GPAI promotes collaboration between governments, academia, and private sectors to ensure inclusivity, transparency, and accountability in AI governance [133].
- IEEE's Ethical Guidelines prioritize accountability, transparency, and human rights protection, offering recommendations for preserving dignity and agency in higher education AI systems [51].

Both frameworks face challenges in enforceability and accessibility, with GPAI's collaborative approach encountering difficulties in aligning diverse stakeholder priorities, and IEEE's technical focus limiting its applicability for non-technical educators and policymakers.

All available global frameworks provide a strong foundation for ethical AIED by emphasizing fairness, transparency, and accountability. However, challenges such as enforceability, inclusivity, and regional disparities limit their practical impact as reflected in Table 3. Effective implementation will require more actionable mechanisms, broader stakeholder engagement, and tailored solutions to meet the diverse needs of global higher education systems.

**Table 3: Global Ethical and Regulatory Frameworks – Focus, Strengths and Limitations**

| Global Ethical and Regulatory Frameworks | Focus and Strengths | Limitations |
|---|---|---|
| UNESCO Recommendation on the Ethics of AI (2021) | Fairness, inclusivity, accountability, human rights; capacity building and monitoring mechanisms. | No enforceability at the national level; inconsistent adoption across countries. |
| OECD AI Principles (2019) | Human-centered values, transparency, robustness; a global benchmark for ethical AI adoption. | Lacks detailed mechanisms for sector-specific challenges, such as those unique to higher education. |
| NASSCOM Responsible AI Hub (India) | Fairness, accountability, inclusivity in AI systems with a focus on higher education. | Limited regulatory mechanisms and accountability; challenges in rural and under-resourced settings. |
| NITI Aayog National Strategy for AI (India, 2018) | Social empowerment through transparency, data privacy, and bias mitigation in key sectors like higher education. | Focuses heavily on public-private collaboration but overlooks diverse stakeholder interests and lacks monitoring compliance. |
| NIST AI Risk Management Framework (USA) | Risk assessment strategies, participatory processes, and safeguarding student data in education technologies. | Lacks specificity for integrating diverse stakeholder inputs; limited adaptability for global data governance. |
| European Commission Ethics Guidelines for Trustworthy AI (2019) | Seven requirements including human agency, technical robustness, transparency, and bias mitigation for high-risk AI systems. | High compliance costs; less accessible for smaller institutions and low-income regions. |
| Global Partnership on AI (GPAI) | Collaboration between governments, academia, and private sectors for ethical AI governance. | Enforceability challenges; delays due to diverse stakeholder priorities. |
| IEEE Global Initiative on Ethics of Autonomous and Intelligent Systems | Accountability, transparency, and human rights protection for preserving dignity and agency in AI systems. | Highly technical focus; limited accessibility for non-technical educators and policymakers. |

### 3.6 Global Ethical Challenges and Barriers

Despite the availability of ethical and regulatory frameworks at the global level, the integration of AI in higher education continues to present complex ethical challenges. These issues have been extensively highlighted in numerous studies [44] and several reviews in global context [6,37,104] [10,36]. These issues arise as institutions adopt advanced technologies, with challenges shaped by regional disparities in technological progress, regulatory frameworks, and socio-economic factors. [36] stress the importance of reflective and ethical considerations in AI adoption, advocating for approaches that balance innovation with equity and inclusivity.

The integration of AI in HE brings significant ethical challenges influenced by regional disparities, regulatory frameworks, and socio-economic contexts. As highlighted by [44] and global reviews [6,10,36,37,104], these challenges underscore the need for reflective approaches that balance technological innovation with equity and inclusivity. Institutions must navigate these complexities to ensure AI adoption aligns with ethical and societal values.

The ethical governance of AI in HE institutions requires structured AI governance frameworks to regulate its use special areas of HE such as in faculty hiring, research integrity, plagiarism detection, and accreditation processes [8].

#### *3.6.1 Basic Ethical Instincts - Privacy, Bias, and Fairness*

AI systems in HE are subject to ethical imperatives of fairness, accountability, transparency, and ethics (FATE), particularly regarding data privacy, algorithmic bias, and equitable implementation [9,97]. Key challenges include privacy breaches, surveillance risks, and biased algorithms influencing admissions, grading, and resource allocation. While frameworks like GDPR enforce data protection, systemic inequities in AI-driven education persist due to biased training datasets and algorithmic opacity [6,26,134,135]. Empirical studies highlight that bias in AI models disproportionately affects marginalized student groups, reinforcing educational inequalities despite regulatory safeguards [20,130].

Addressing these challenges requires enhanced algorithmic transparency, fairness audits, and diverse training data to mitigate discriminatory outcomes in AI-powered learning environments [9]. To ensure accountability and inclusivity, robust governance frameworks must be established, integrating human-centered evaluations [36]. Empirical studies further emphasize that interdisciplinary collaboration, as advocated by UNESCO guidelines and [136], is essential for fostering ethical AI deployment in education. Additionally, contextual and cultural sensitivities should inform policy development to ensure AI technologies align with diverse societal needs [83].

However, a lack of technical expertise within educational institutions can exacerbate ethical risks, including data insecurity and algorithmic biases [60,68,123]. Moreover, the increasing adoption of surveillance technologies, such as facial recognition, raises ethical concerns about privacy and its impact on pedagogical practices [137]. Addressing these challenges requires embedding ethical standards in AI design and ensuring transparency in deployment. Professional development for educators and continuous oversight are equally crucial for fostering equity and fairness.

The following Table 4 provides a summary of literature review on global scale exploring the ethical challenges and barriers associated with AI-driven higher education (HE), along with the key consequences identified in these studies.

**Table 4: Literature Review: Global Ethical Challenges and Barriers**

| Ethical Challenge | Focus of the Study | Impacted Country/Region | Consequences Identified | Literature Support |
|---|---|---|---|---|
| **Bias in AI Systems** | AI models in grading and predictive systems perpetuate biases when training data is unrepresentative. | **Global – Adaptive Learning and Predictive Platforms** | Disadvantages marginalized groups and perpetuates systemic inequities in educational outcomes. | [7,10,20,30,36,90,138] |
| **Privacy and Data Concerns** | AI platforms collect vast student data for learning analytics, raising risks of breaches and misuse. | **Global – Learning Analytics Platforms** | Undermines trust in AI systems, discouraging their adoption in educational contexts. | [25,26,90,139–141] |
| **Scalability and Equity Issues** | AI-enabled tools like MOOCs provide scalable education but are often inaccessible in rural or underfunded areas. | **Developing Countries – Rural Institutions** | Worsens educational inequalities, leaving disadvantaged communities unable to access innovative AI resources. | [7,19,26,30,45,83] |
| **AI for Personalized Learning** | Adaptive platforms like Knewton personalize learning but risk isolating students and reducing peer interaction. | **Global – AI Tools like ALEKS, Knewton** | Enhances individual learning efficiency but overlooks collaborative and social learning dimensions. | [7,11,13,70,90,142,143] |
| **Ethical Challenges in AI Use** | Lack of transparency and ethical frameworks in AI use creates accountability and trust issues. | **Global – AI-driven EdTech platforms** | Raises ethical dilemmas and undermines trust among students and educators. | [8,10,26,121,141,144] |
| **Teacher Readiness for AI** | Many educators lack skills to effectively integrate AI into their teaching and feedback practices. | **Global – Professional Development Programs** | Reduces the potential impact of AI technologies on improving teaching and learning processes. | [12,140,145–147] |
| **Human-AI Collaboration** | AI tools like automated writing assistants and grading systems impact teacher and student autonomy. | **Global – AI-Assisted Academic Writing Tools** | Reduces human engagement and may foster over-reliance on AI for academic tasks. | [17,20,26,27,70] |
| **Impact on Mental Wellbeing** | Overuse of AI in work and learning environments leads to isolation and reduced social connections. | **Global – AI in Higher Education and Workplaces** | Increases loneliness and stress, negatively affecting mental health. | [27,39,122,140,148] |
| **Student Engagement Challenges** | AI fails to address the emotional and motivational needs of learners in virtual learning environments. | **Global – Platforms like Coursera, Kahoot** | Reduces retention and long-term learning outcomes, limiting holistic educational experiences. | [10,17,26,139,142] |
| **Language and Cultural Barriers** | AI systems fail to adapt to diverse linguistic and cultural contexts, marginalizing non-dominant groups. | **India, China – AI in Language Learning** | Limits inclusivity and prevents equal access to AI-driven learning tools for diverse learners. | [13,24,147,149] |
| **Bias in Predictive Analytics** | Predictive systems flagging “at-risk” students disproportionately target specific demographic groups. | **United States – Student Retention Platforms** | Reinforces systemic inequalities, limiting equitable access to educational opportunities. | [25,32,38,103,138] |
| **Ethical Concerns in Mentoring** | AI-supported mentoring systems, such as career advising, risk providing biased or harmful recommendations. | **Global – AI Mentoring Platforms** | Creates ethical dilemmas and reinforces inequities in personalized advice for students. | [8,10,28,109,150] |

### *3.6.2 Sustaining Humanized Ethics*

AI has transformed HE processes, such as grading and administrative workflows, but it also introduces challenges, including job displacement and diminished human engagement. Rather than eliminating jobs, AI often redefines roles. [151] underscores the importance of reskilling to address this transition, while [152] advocate for human-centric AI designs to preserve mentorship and creativity. Ethical frameworks that incorporate societal and cultural nuances are essential, as (Bond et al., 2024) point out, to ensure inclusive workforce adaptation. Proactive measures, such as stakeholder engagement and transparency, foster equitable transitions [6,36].

Another critical concern is over-reliance on AI, which risks diminishing essential human skills like critical thinking and problem-solving. [11] caution against passive intellectual behavior encouraged by AI dominance, and [84] call for frameworks that balance efficiency with active human involvement. [37] highlight the importance of preserving the emotional and cultural dimensions of learning, ensuring AI does not dehumanize education. Embedding ethical norms and meaningful human oversight is essential to maintaining alignment between AI systems and educational values.

For future generations, the benefits of enhanced accessibility and efficiency must be weighed against potential drawbacks such as “intellectual atrophy"; intellectual stagnation and reduced intergenerational knowledge transfer [148,153]. Collaborative frameworks and ethical design, as emphasized by [10] and [142], are key to ensuring AI systems foster creativity and resilience. Equally important is the role of transparency and trust in preserving motivation and engagement for both students and educators [154].

Finally, while AI enhances operational efficiency, excessive automation can erode human connections critical to holistic education. Over-automation risks isolating students, undermining communication and collaboration skills [11]. Embedding emotional resilience and inclusivity in AI-driven systems ensures these tools enhance, rather than replace, the relational aspects of education [33] (Pedro et al., 2019).

## 3.7 Research Gaps and Missing Links:

AIHE has made significant advancements, with frameworks striving to enhance learning experiences, boost operational efficiency, and meet diverse educational needs. However, these developments frequently fail to address critical dimensions essential for holistic and effective AI integration. Drawing on global literature, including journal articles, case studies, and reports, this section identifies the key gaps hindering the successful adoption and implementation of AIED systems. It underscores the necessity for ethical grounding, practical operability, and alignment with human-centric principles to create robust, adaptive, and equitable frameworks for AI in education.

- ***Inclusive and Collaborative AI: Bridging Stakeholder Gaps in Participatory Design***
  AIHED often overlooks ethical challenges such as algorithmic bias, data privacy, and transparency, leading to real-world failures like Amazon’s AI hiring tool perpetuating gender bias [6]. Current frameworks tend to focus narrowly on fairness and privacy while neglecting broader socio-political concerns, such as democratic oversight [135].

Additionally, limited stakeholder involvement—including educators, students, and policymakers—results in AI systems misaligned with diverse needs due to a lack of participatory co-design and feedback mechanisms [37,45,155]. Moreover, AIHED frameworks frequently treat AI as a standalone solution, disregarding the potential of human–AI collaboration. Systems like Georgia Tech's "Jill Watson" AI assistant demonstrate efficiency in handling routine tasks but lack human empathy and contextual adaptability [153]. Research is needed to develop collaborative AI frameworks that integrate AI efficiency with human intuition, creativity, and relational engagement [85]. Addressing these gaps will enable inclusive, participatory, and ethically robust AI systems that better serve diverse educational communities.

- ***Bridging Theory and Practice: Operational Gaps in AI Adoption***
  Many frameworks fail to translate theoretical advancements into actionable solutions. For instance, the UK government's AI grading system during COVID-19 faced public backlash due to inadequate phased testing [7]. Policies often overlook practical guidance for implementation, as noted by [76], leaving institutions without effective workflows to bridge theoretical benefits and practical applications.
- ***Dynamic Feedback: The Foundation for Continuous Improvement***
  A persistent limitation in AIED frameworks is the reliance on static feedback mechanisms, which prevent adaptation to evolving needs. Adaptive learning systems often fail to incorporate real-time feedback from students and educators, resulting in stagnant outcomes [7]. Dynamic feedback loops, as emphasized by [140], are essential for continuous refinement and responsiveness, enabling systems to align with institutional goals and learner requirements.
- ***Adapting to Diversity: Aligning AI with Institutional Goals***
  Current AI frameworks in higher education (HE) often adopt a one-size-fits-all approach, neglecting institutional and cultural diversity [38,96]. Research is needed on adaptive AI models that accommodate varying infrastructure and resources, particularly for institutions with limited AI capacity [66,111]. The use of structured assessments, such as SWOC (Strengths, Weaknesses, Opportunities, Challenges), remains underexplored in guiding AI integration and sustainability across diverse HE settings [10,156]. Additionally, static AI systems struggle in culturally diverse environments, highlighting the need for context-sensitive frameworks that align with regional needs [7,12]. Finally, ensuring AI alignment with institutional missions, inclusivity, and scalability remains a critical challenge, requiring further research on responsive and equitable AI implementation [80].
- ***Automation to Augmentation: Preserving Human Values***
  AIHED frameworks often overemphasize scalability and efficiency, side-lining educational values like human creativity, mentorship, and critical thinking [6,36]. This techno-centric focus risks dehumanizing education, as noted by [8]. Balancing technological advancements with human-centric principles is crucial to preserving equity, intellectual growth, and holistic development.
- ***Missing Links in AI: Resilience and Sustainability for the Future***
  Most frameworks prioritize short-term gains over long-term adaptability, ignoring evolving societal needs and regulatory changes. Over-reliance on AI risks fostering "intellectual atrophy," where human creativity and critical thinking diminish educators [153]. [148] emphasize the need for future-proof strategies, including scenario planning and continuous research, to ensure AIED systems remain adaptable, resilient, and aligned with broader educational values.
- ***Phased Human Intelligence: Filling Gaps in the AI Lifecycle***
  A critical gap in AIED frameworks is the lack of integration between the AI lifecycle (adoption, design, deployment, evaluation and exploration) and phased human intelligence. Human oversight is rarely embedded across these phases, leading to misaligned algorithms and unchecked biases [6,135]. Effective frameworks must involve interdisciplinary teams of educators, AI developers, and policymakers to guide each phase, ensuring alignment with ethical and pedagogical goals [37,140]. Regular audits and dynamic feedback loops can detect inefficiencies and biases, fostering systems that are both transparent and contextually relevant [12,85].

- ***Promotion and Benchmarking AIED: The Need for External Intelligence***
  Most frameworks fail to emphasize multi-stakeholder collaborations, limiting adaptability and scalability. External partnerships with industry, governments, and academia can provide funding, emerging technologies, and diverse perspectives essential for robust AI implementation [6,45]. Furthermore, there is limited documentation of real-world successes and failures, such as the chatbot "Pounce" at Georgia State University [153]. Transparent reporting can foster innovation, build trust, and prevent repeated mistakes.
- ***Need for University / Institutional AI Ethics Committees***
  AIHED lacks standardized ethical governance models, raising concerns about algorithmic bias, academic autonomy, and decision transparency. While institutions like the University of Amsterdam have established AI Ethics Committees [51], many universities lack structured oversight for AI-driven faculty evaluations, admissions, and tenure reviews [26]. The literature reveals critical research gaps in terms of the absence of AI Ethics Review Boards / Committees in universities and HE institutions.
- ***Unaddressed Challenges in AIHED***
  Existing research on AI in Education (AIED) focuses on adaptive learning and student support, while AI in Higher Education (AIHED) requires structured governance, policy frameworks, and institutional oversight [6,26]. AI applications in faculty governance, research integrity, and accreditation remain underexplored [37]. Current studies lack insights into participatory AI governance, academic autonomy, and standardized regulatory frameworks to address bias, transparency, and ethical compliance [20]. This study bridges these gaps by proposing a structured AI governance model that ensures AI adoption aligns with HE's core values and institutional integrity (UNESCO, 2021).
- ***Vision without Action: A Call for Practical and Constructive Solutions***
  Global frameworks, such as those by [31] and [50], often emphasize fairness and accountability but lack localized adaptability and participatory design. Initiatives like NASSCOM and the EU AI Act prioritize compliance but fail to integrate relational engagement and co-design critical for higher education [37]. Practical, actionable strategies are needed to bridge this gap and facilitate safe, ethical AI integration into diverse HE contexts.

Addressing these research gaps is critical to bridging the divide between theoretical aspirations and practical applications of AIHED. By embedding human-centric ethics, dynamic feedback systems, participatory and collaborative frameworks, and long-term strategies, AIED frameworks can align technological advancements with human values, fostering adaptive, equitable, and sustainable systems for HE.

## 4. Global Reality and Experiences: Ethical or Exploitation?

This section critically examines the global realities and experiences of AIHED, assessing whether ethical principles such as fairness, accountability, inclusivity, and human-centric values are genuinely upheld in practice. While AIHED has made significant strides, achieving remarkable progress, its implementation has also revealed a duality of success and setbacks. This analysis underscores the urgent need for a Human-Driven AIHED Framework to ensure the integration of AIHED is meaningful, equitable, and aligned with ethical standards.

### 4.1 Duality in AIHED: Successes and Setbacks

On one hand, AI has empowered higher education institutions to personalize learning, enhance operational efficiency, and democratize access to education. On the other hand, it has introduced significant ethical challenges, including algorithmic bias, privacy violations, and the erosion of interpersonal connections. This duality raises a critical question: Is AIHED merely "doing ethical things right," or is it truly "doing things ethically"? AIHED exemplifies a duality of transformative successes and critical setbacks across global contexts as exemplified below.

- ***Personalized Learning and Accessibility:***
  Platforms like Squirrel AI have tailored education to millions in China, improving outcomes for underserved populations [29,31]. However, invasive tracking of student behaviors without consent has raised concerns about privacy and trust [32]. Similarly, Coursera has democratized access to quality education for over 100 million learners, particularly in developing regions like India and sub-Saharan Africa [19,29]. Yet, infrastructural inequities in under-resourced areas exacerbate the digital divide [30,157].
- ***Efficiency in Grading and Assessments:***
  Tools like Gradescope and Turnitin have streamlined grading and improved writing outcomes at institutions like UC Berkeley and Stanford [25] (Bond et al., 2024). However, the UK's algorithmic grading system during COVID-19 exposed systemic biases, disproportionately disadvantaging students from underprivileged backgrounds [9,26].
- ***Predictive Analytics and Student Success:***
  Predictive analytics at Georgia State University increased graduation rates by 22%, supporting at-risk students through data-driven interventions [27]. Conversely, tools like Ellucian Analytics flagged minority students disproportionately as "at-risk," reinforcing existing inequities [25].
- ***AI and Relational Learning:***
  While tools like Cognitive Tutor have improved STEM learning outcomes by 40% [12], their overuse has diminished meaningful teacher-student interactions, leading to isolation and disengagement [148,153]. Automated systems, like those at Deakin University, have reduced opportunities for faculty engagement, weakening relational aspects of education [13].
- ***Privacy and Surveillance Concerns:***
  Proctoring tools such as Proctorio have raised significant concerns due to invasive surveillance practices that have led to racial bias and trust erosion among marginalized communities [10,141]

These examples highlight the complex landscape of AIHED, illustrating its capacity to transform HE while revealing its inherent risks. They also emphasize that while AIHED often achieves efficiency and scalability, it frequently undermines fundamental educational values, such as equity, trust, and interpersonal connection.

### 4.2 A Call for Action: Emergence of Human-Driven AIHED

The future of HE lies not in the raw power of AI but in the intentionality and ethical rigor with which it is employed. A Human-Driven AIHED Framework is essential to align AI systems with human values, ensuring fairness, inclusivity, and empathy remain at the heart of education. Following are the key considerations for a Human-Driven AIHED Approach. A Human-Driven AIHED Framework is essential to ensure AI adoption prioritizes human values, ethical oversight, and equity in education.

- ***To overcome Ethical and Governance Challenges in AIHED***
  AI-driven decision-making systems can reinforce biases, compromise student privacy, and erode institutional accountability if not designed with ethical safeguards. AI-based grading, predictive analytics, and faculty assessments must be monitored to prevent discriminatory outcomes and ensure transparency, explainability, and fairness [20]. The absence of structured AI governance frameworks in HE leads to disparities in adoption, ethical inconsistencies, and regulatory gaps, making a human-driven approach critical [26].
- ***Ensuring Equity and Inclusivity in AIHED***
  AI must support, not replace, human expertise. AI-driven learning platforms should align with institutional values, cultural diversity, and equitable access to resources [45]. Without human oversight, AI risks exacerbating digital divides, disproportionately affecting underrepresented student populations [7]. Ensuring co-governance with educators, students, and policymakers fosters trust, inclusivity, and balanced AI adoption [19].

- ***The Role of Human Oversight in AI Adoption***
  Embedding human oversight across the AI lifecycle—from design to deployment and evaluation—ensures ethical compliance, institutional accountability, and adaptability [29]. Universities must implement co-design strategies, integrating faculty, administrators, students, and industry experts into AI policymaking to reflect academic and societal needs [157].
- ***Human-Centric AI for Sustainable HE Transformation***
  A Human-Centric framework provides a structured roadmap for responsible AI adoption, balancing technological efficiency with empathy, fairness, and long-term sustainability [148,158]. Ethical AI integration must incorporate dynamic feedback loops, algorithmic fairness audits, and participatory governance models to uphold academic integrity and educational equity.

## 5. Conceptualizing Human-Driven AIHED: Enabling 'Ethos' in AIHED

As [36] emphasize, the adoption of digital technologies in HE should not be seen as an unchecked celebration of progress but rather as an opportunity for critical reflection, ensuring inclusivity, fairness, and sustainability. The HD-AIHED framework provides a structured approach to addressing research gaps and real-time challenges in AI adoption for higher education. By integrating ethical principles and structural components, it ensures AI deployment aligns with institutional values, academic integrity, and equitable governance [159].

Universities and HE institutions face key challenges, including policy gaps, algorithmic biases, governance inconsistencies, and scalability constraints, requiring human oversight, participatory decision-making, and adaptive AI strategies [9,87]. They also face complex challenges regarding special services as discussed in literature review section. The HD-AIHED framework addresses these gaps by embedding ethical AI governance, fairness reviews, transparency mechanisms, and continuous feedback loops to optimize AI integration in diverse educational settings [20].

### 5.1 Structural Components of Proposed Framework and Potential Solutions: Addressing Research Gaps and Global Real-time Challenges:

The figure 2 presents key structural components and potential solutions for addressing global AI challenges in higher education through the Human-Driven framework. It systematically maps real-time challenges, such as algorithmic bias, data privacy concerns, faculty governance gaps, and digital divide, to corresponding research gaps and structural solutions [8,26]. The structural components of the proposed AIHED framework—such as Institutional AI Ethical Review Boards, AI fairness audits, SWOC analysis, participatory governance, and phased AI adoption strategies—are integrated as possible solutions to enhance AI transparency, inclusivity, and adaptability [20,31]. By incorporating internal (faculty, students, IT teams, auditors) and external stakeholders (industry, policymakers, global HE institutions, regulators), the framework ensures a balanced AI adoption process that is ethically responsible, scalable, and human-centered [6,50]. The figure serves as a guiding tool for developing final framework for HE institutions to align AI advancements with sustainable, participatory, and equitable educational policies [7].

**Problems** — **Solutions**

| Global Real-time Challenges | Research Gaps | HD-AIHED Framework (Structural Components and Solutions) |
|---|---|---|
| Algorithmic Bias, Equity & Fairness | Lack of systematic AI bias audits and fairness metrics across institutions. | Institutional AI Ethical Review Boards (students, faculty, third-party auditors), AI fairness reviews, transparency checkpoints, participatory governance models, Institutional Auditors.. |
| Privacy Violations & Data Protection | Gaps in AI privacy frameworks and cross-border data compliance. | Institutional AI Ethical Review Boards, IT Team, Top Management ensuring GDPR, FERPA, UNESCO AI compliance, transparent data governance, Institutional Auditors overseeing compliance. |
| Erosion of Interpersonal Connections & Over-Reliance on Automation | Limited research on AI's emotional intelligence, faculty-AI interactions. | Faculty, Students, Research Team, AI-assisted learning, student engagement models, ethical intervention checkpoints, Institutional Auditors ensuring AI preserves human interactions. |
| Digital Divide & AI Accessibility | Unequal AI adoption due to funding gaps, lack of infrastructure. | SWOC Analysis (AI adoption + AI exploration), IT Team, Government, Industry, NGOs, Global HE Institutions ensuring equitable AI access, Institutional Auditors monitoring implementation. |
| Cultural & Societal Context Adaptation | AI's inability to reflect diverse cultural and linguistic needs. | Academic Experts, Faculty, Research Teams, Government Bodies, AI customized for cultural inclusivity, Institutional Auditors ensuring regional AI adaptability. |
| Scalability & Institutional Adaptability | AI models lack scalability in different educational settings. | Phased Human Intelligence Across AI Lifecycle, SWOC Analysis, Institutional AI Ethics Boards, Global HE Institutions ensuring AI adaptability, Institutional Auditors conducting performance assessments. |
| Ethical Framework Gaps & Institutional AI Governance | Inconsistent AI policies and lack of unified regulations. | Institutional AI Ethical Review Boards, External Regulators (UNESCO, OECD), Government ensuring ethical AI policies, Institutional Auditors conducting bias and compliance audits. |
| Faculty Governance & Institutional Decision-Making | Limited research on AI's role in faculty hiring and evaluations. | Top Management, Faculty, Institutional AI Ethics Review Boards ensuring fair AI-driven governance, Institutional Auditors evaluating governance effectiveness. |
| AI in Student Career & Job Placements | Lack of AI-driven equity assessments in career advising. | Industry, Alumni, Government Agencies, NGOs, Research Teams ensuring fair and inclusive AI-driven career support, Institutional Auditors assessing AI-driven hiring fairness.. |
| Research Integrity & AI Transparency | Ethical risks in AI-supported peer review and research automation. | Research Teams, Institutional AI Auditors, Ethical Review Boards, External Regulators (UNESCO, OECD) ensuring transparency and accountability. |
| External Linkages & AI Collaborations | Limited regulatory oversight in AI-industry partnerships. | Industry, Global HE Institutions, Alumni, Ethical & Regulatory Bodies ensuring responsible AI collaborations, Institutional Auditors monitoring compliance. |
| Continuous AI Improvement & Future-Proofing | AI models lack adaptive learning and real-time self-improvement. | Unified Human Intelligence (Participatory Decision-Making + Continuous Feedback), Research Teams, Institutional Auditors ensuring AI adaptability and long-term effectiveness. |

Figure 2: Structural Components of Framework and Potential Solutions for Addressing Research Gaps and Global Real-time Challenges

### 5.2 Integrated System: Ethical Essence and Essential Structural Components

The proposed framework operates on distinctive system principles, encompassing inputs, processing, and outputs facilitated through seamless integration of AI and unified human intelligence, as depicted in the accompanying Figure 2. The effectiveness of AI in higher education depends on an integrated system where AI, human intelligence, and external intelligence function as a unified structure, fostering continuous interaction between automated processes and human oversight. Unlike isolated AI applications that operate independently, this framework embeds AI within a structured ecosystem that emphasizes ethical integrity, transparency, and accountability [8,26]. By interlinking AI operations with human decision-making and continuous feedback mechanisms, the system mitigates automation risks such as bias, ethical lapses, and the erosion of contextual adaptability [8,36].

This integration ensures that AI does not function as an autonomous entity but rather as an adaptive, human-aligned tool that evolves alongside institutional needs [85]. A structured flow from input to processing and output is maintained through unified human intelligence, ensuring that AI-driven insights remain relevant, ethical, and operationally effective [29,31]. The incorporation of external intelligence further enhances this model by introducing interdisciplinary perspectives, regulatory compliance, and benchmarking against global best practices, making AI implementations more adaptable and responsive to dynamic institutional and societal expectations [19,30].

The necessity of an integrated system lies in its ability to harmonize AI's computational efficiency with human judgment, ensuring that decision-making remains aligned with the core values of education [33,70]. Integrated system approach (refer Figure 3) fosters trust by maintaining transparency in AI-driven decision-making and reinforcing accountability through structured oversight [26,80]. By embedding AI within a holistic and adaptive system, the framework ensures that technology functions as an enabler of education rather than a disruptor, preserving equity, inclusivity, and institutional integrity while optimizing learning and administrative processes [39,157].

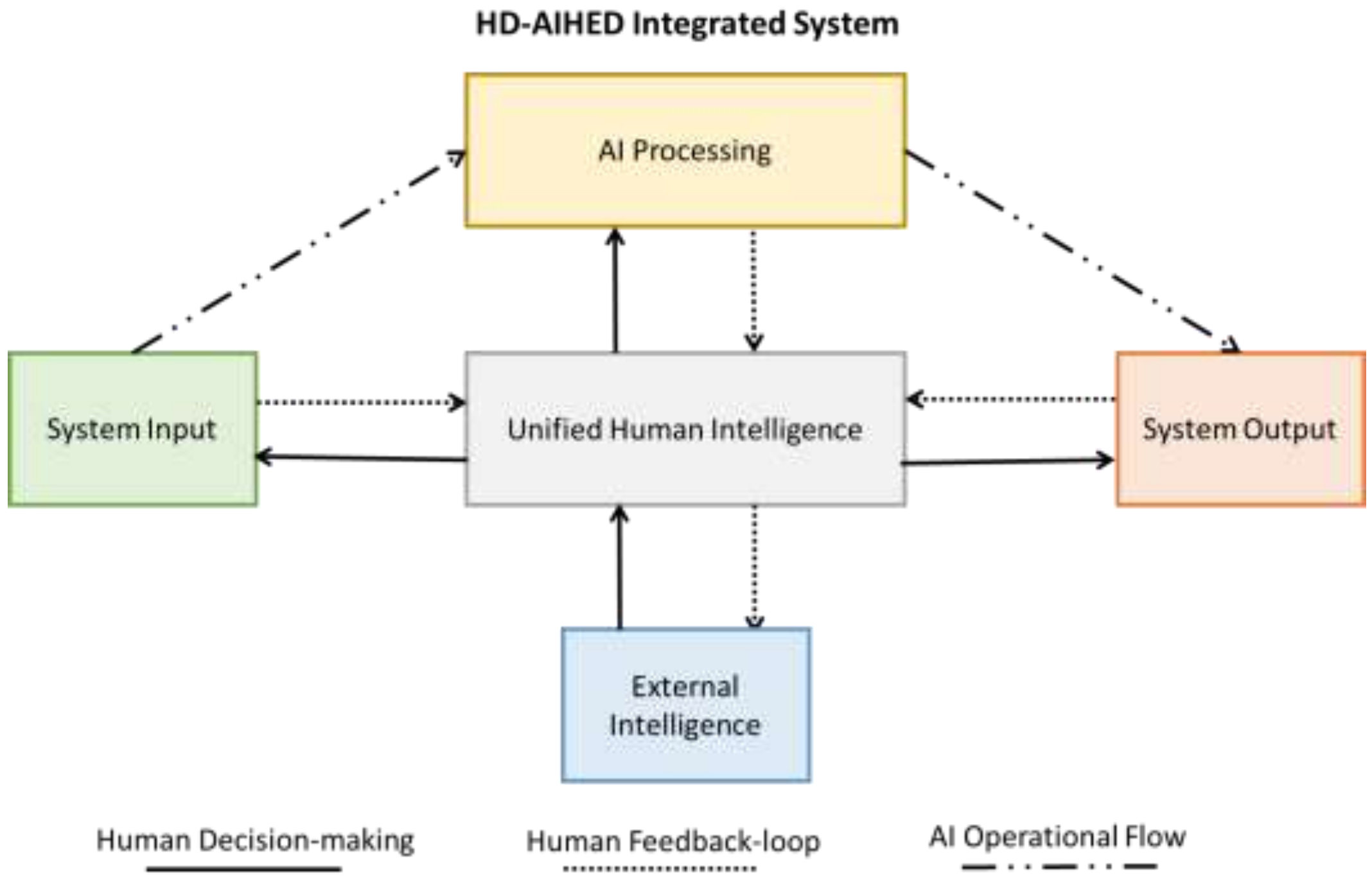

Figure 3: HD-AIHED as an Integrated System

#### *5.2.1 AI System Input*

At the foundation of the system is the system input, which consists of structured institutional data, including student records, administrative policies, resource allocation, and regulatory frameworks. This phase ensures that human oversight is embedded in data selection, curation, and pre-processing to uphold ethical compliance, mitigate biases, and ensure fairness in AI operations [6,26]. By integrating human intelligence at the input stage, study's framework safeguard AI from ingesting flawed, outdated, or contextually irrelevant data, thereby enhancing the accuracy and integrity of AI-driven decisions [11,36]

Institutions leveraging AI for decision-making must ensure that the data feeding AI systems is diverse, current, and free from historical biases that could reinforce systemic inequities [30,135]. Without human intervention in the input stage, AI models risk perpetuating inaccurate patterns that can disproportionately affect marginalized groups, compromise institutional inclusivity, and hinder equitable student outcomes [8]. The study's framework emphasizes the importance of incorporating human intelligence during the data input phase to mitigate potential risks. By involving human oversight, it ensures contextual validation of data, aligns AI operations with institutional policies, and adapts processes to meet ethical and regulatory standards effectively. This integration provides a critical layer of accountability, fostering the responsible and compliant use of AI systems. [11].

Additionally, AI systems in HE rely on data from multiple institutional repositories, such as learning management systems, assessment records, and faculty evaluations. However, disparities in data accessibility, inconsistencies in reporting structures, and variations in institutional policies can create challenges in ensuring uniformity across AI-driven systems [70]. Human oversight in the input stage helps reconcile these discrepancies, fostering a structured and standardized approach to AI-driven decision-making while maintaining the integrity of academic governance [33,80]. By embedding human expertise at this foundational level, institutions not only enhance AI accuracy and adaptability but also reinforce their commitment to ethical, transparent, and inclusive AI adoption in HE [19,31].

### *5.2.2 AI Processing*

The AI Processing phase represents the analytical core of the integrated system, where raw input data is transformed into actionable insights using machine learning algorithms, predictive analytics, and decision-making models. This stage plays a pivotal role in enhancing institutional and pedagogical efficiency by optimizing operations across the student lifecycle, including admissions, personalized learning pathways, early identification of at-risk students, and streamlining administrative processes [11,160]. Key AI applications, such as predictive modeling, automated grading, and resource allocation, facilitate data-driven decisions, improving both educational outcomes and operational effectiveness [6,121].

Despite its transformative potential, AI processing alone cannot guarantee ethical compliance, fairness, or contextual sensitivity. The proposed framework addresses these limitations by embedding Unified Human Intelligence as an integral regulatory mechanism, ensuring that AI-driven outputs align with institutional goals, ethical principles, and pedagogical objectives [25,26]. Human oversight mitigates risks such as automation bias, algorithmic opacity, and the perpetuation of inequities by ensuring that AI systems adhere to fairness, transparency, and accountability standards [8].

Dynamic feedback loops are integrated to enable real-time refinements of AI models, fostering continuous learning and adaptability to meet changing pedagogical demands and regulatory requirements [12,33]. These iterative mechanisms prevent stagnation in AI functionality and ensure the recalibration of algorithms to reflect contextual shifts, thereby addressing the challenges of algorithmic misinterpretation and static decision-making processes [140].

By embedding ethical oversight and dynamic feedback within the AI Processing phase, the framework mitigates risks associated with biased decision-making and algorithmic opacity, ensuring that AI operates as a complementary tool rather than an autonomous determinant of educational outcomes [6,85]. This structured interaction between AI-driven analytics and human intelligence reinforces trust, accountability, and adaptability, positioning AI processing as a cornerstone for the ethical and effective integration

#### *5.2.3 AI System Output*

The "System Output" component of the proposed framework represents the culmination of integrated AI processes, producing actionable insights and tools that enhance decision-making, operational efficiency, and personalized learning experiences in HE. This segment of the ecosystem delivers outputs such as personalized recommendations for students, administrative insights for resource optimization, and accessibility tools that cater to diverse learner needs. These outputs are presented in user-friendly formats, including dashboards, reports, alerts, transcriptions, and interactive interfaces such as chatbots and virtual assistants, enabling seamless interpretation and application of data [6,11].

A key strength of the framework lies in embedding human intelligence throughout the output process. Human oversight is integral to ensuring the accuracy, ethical compliance, and contextual relevance of AI-driven outputs. This involves validating data accuracy, refining algorithms based on feedback, and interpreting insights to align outputs with institutional values and objectives. For instance, personalized learning recommendations are tailored to meet individual student needs while avoiding algorithmic bias, and administrative outputs support equitable resource allocation across departments [26,160].

Furthermore, the framework ensures that outputs lead to actionable outcomes. These include intervention strategies for at-risk students, adjustments to institutional curricula, and data-informed policy updates that drive improvements in teaching, learning, and administration. By fostering inclusivity and addressing the diverse needs of stakeholders, the framework transforms raw AI-generated outputs into meaningful, institutionally aligned actions [33].

The integration of dynamic feedback loops within the output process ensures continuous refinement and adaptability of AI systems. Feedback from educators, students, and administrators helps fine-tune the outputs, ensuring they remain responsive to changing institutional priorities and societal demands. For example, real-time updates to predictive analytics can guide faculty in providing timely interventions, while adaptive dashboards support decision-makers in tracking progress and identifying areas for improvement [11,140].

By emphasizing transparency and inclusivity in the generation and application of outputs, the framework mitigates risks such as bias, misinterpretation, and inefficiency. It reinforces institutional trust and accountability by aligning AI-driven outcomes with ethical standards and stakeholder expectations. This ensures that AI outputs not only optimize operational processes but also advance the broader mission of equity, accessibility, and excellence in higher education [6,45].

The System Output component, therefore, represents a pivotal stage in the framework, translating AI's computational capabilities into actionable, human-centered insights that drive meaningful and sustainable progress in HE.

#### *5.2.4 SWOC Analysis*

In line with [100], the proposed framework adopts a predictive AI preparedness model, reinforced by [130]’s roadmap for institutional competitiveness. Through SWOC analysis, the framework ensures AI adoption aligns with strategic goals while maintaining scalability and long-term adaptability in the AI exploration phase [38,40].

*AI Adoption Phase:* The SWOC analysis framework integrates AI adoption with an institution’s strategic goals, resources, and socio-cultural context [38,40]. During the AI adoption phase, institutional strengths such as faculty expertise, existing infrastructure, and organizational readiness facilitate smooth AI integration [104]. Weaknesses, including gaps in technological infrastructure, limited funding, and lack of staff training, are mitigated through capacity-building initiatives and interdisciplinary collaboration [6,85]. Opportunities arise from expanding global reach, improving student outcomes, and innovating pedagogy through partnerships with NGOs, funding agencies, and industry experts [26,31]. Challenges such as resistance to change, governance complexities, and regulatory compliance are addressed through proactive feedback loops and inclusive decision-making [156].

*AI Exploration Phase:* At the AI exploration stage, institutions benefit from strengths such as accumulated AI expertise, expanding research collaborations, and enhanced digital infrastructure. Weaknesses like regulatory uncertainties and integration complexities are tackled through policy framing, iterative governance, and benchmarking [31,161]. Opportunities include academic expansion, international collaboration, and innovation in AI-driven pedagogy. Challenges focus on ensuring scalability and long-term adaptability while aligning with sustainable development goals. By embedding SWOC analysis at both phases, institutions can strategically plan AI implementation while fostering resilience, scalability, inclusivity, and equity [26,38].

#### *5.2.5 Human Intelligence*

AI governance in HE is designed to be inclusive and participatory, integrating faculty, students, administrators, and external stakeholders into AI decision-making processes. A participatory and collaborative approach ensures AI governance remains adaptive to regional contexts, facilitating cost-effective implementation across diverse educational landscapes [8,30,85].

***Internal Stakeholders:***

Internal stakeholders play a direct role in AI implementation, ensuring AI aligns with educational objectives and ethical considerations [66].

*Management, Educators, Researchers, IT Technicians and Administrators:* University / HE institutional top management—including presidents, vice-chancellors, provosts, and governing board members—plays a strategic role in AI integration, ensuring alignment with institutional vision, regulatory frameworks, and global educational trends. They are responsible for driving the long-term AI strategy, ensuring AI integration aligns with institutional goals and global best practices [162]. They oversee institutional AI governance structures, ensuring compliance with national and international regulatory frameworks such as UNESCO AI Ethics Guidelines, GDPR, and OECD AI Principles [31,103]. Senior leadership allocates resources for AI infrastructure development, ensuring scalability, sustainability, and ethical AI adoption [38]. They fosters cross-institutional collaborations with universities, AI research institutes, and industry partners, enabling access to cutting-edge AI technologies and research funding [11].

Educators (faculty, deans, and academic experts) serve as primary users of AI-driven tools for teaching and assessment, ensuring AI supports rather than dictates teaching methodologies [85]. Faculty also integrate AI into curriculum design while upholding academic integrity and ethical compliance [163]. Administrators oversee AI policy formulation and institutional strategies, ensuring AI aligns with institutional goals, regulatory compliance, and financial considerations [38]. They facilitate AI-driven decision-making and feedback for academic and administrative improvements, including student retention, operational efficiency, and institutional governance [130].

The research team advances AI innovation by conducting empirical studies, bias assessments, and ethical evaluations [20]. They collaborate with faculty, students, and AI governance boards to test AI models, refine policies, and ensure transparent, ethical AI deployment [8]. Research teams also secure AI-related grants and partnerships, keeping institutions at the forefront of ethical and responsible AI adoption [162].

The IT team is responsible for AI system deployment, cyber security, data governance, and infrastructure optimization to align AI tools with institutional policies [26]. They ensure secure AI integration, prevent data breaches, and maintain compliance with GDPR and UNESCO AI Ethics Guidelines. IT specialists facilitate AI-driven automation, interoperability, and student analytics, ensuring seamless AI adoption in learning management and administrative systems [6].

*Students:* Students, as the primary recipients of AI-driven services, provide direct experiential feedback, shaping AI's effectiveness and highlighting potential biases [164,165]. Their inclusion as a partner in AI ethical review boards and as AI auditors is critical to ensuring equity and transparency in AI governance. Actively involving students in AI decision-making and feedback ensures AI-driven solutions remain user-centric, fair, and aligned with diverse educational needs.

*Institutional AI Ethical Review Boards And Auditors:* Institutional AI Ethical Review Boards govern responsible AI use, mitigating automation bias, data privacy concerns, and academic equity issues [23,166]. Research teams refine AI applications for HE, ensuring transparency, adaptability, and fairness [11,167]. Institutional auditors enhance AI oversight by monitoring effectiveness, identifying biases, and reinforcing ethical compliance in AI-driven decision-making [168,169]. The Institutional AI Ethical Review Board comprises advisory panels, compliance bodies, faculty experts, legal professionals, and student representatives. It approves AI adoption strategies, ensures ethical compliance, and aligns AI policies with global regulatory frameworks, including UNESCO AI Ethics, GDPR, and OECD AI Principles [31,103]. The board ensures institutions uphold global ethical standards while safeguarding student data privacy [8,26].

Institutional AI Ethical Review Boards ensure responsible AI use, addressing automation bias, data privacy concerns, and academic equity [23,166]. Research teams refine AI applications for transparency, adaptability, and fairness [11,167]. Institutional auditors strengthen AI oversight by evaluating effectiveness, identifying biases, and ensuring ethical compliance [168,169].

Comprising advisory panels, compliance bodies, faculty experts, legal professionals, and student representatives, the Institutional AI Ethical Review Board approves AI adoption strategies, enforces ethical compliance, and aligns AI policies with global regulatory frameworks such as UNESCO AI Ethics, GDPR, and OECD AI Principles [31,103]. It also safeguards student data privacy while ensuring institutions uphold global ethical standards [8,26].

Institutional auditors monitor AI integrity, reinforcing transparency, fairness, and regulatory compliance [168]. Regular AI bias audits assess fairness, supported by clear policies, oversight mechanisms, and external reviews. Ethical AI auditing and bias mitigation strategies ensure AI-driven assessments remain accountable and impartial, with compliance officers conducting ongoing evaluations. To mitigate data misuse and systemic inequities, the framework integrates validation mechanisms and dynamic oversight, ensuring AI models continuously evolve to meet emerging challenges [6,121].

***External Stakeholders:***

The External Intelligence component is crucial for aligning AI integration with global trends, technological advancements, and societal expectations. By engaging industry experts, academic professionals, global institutions, ethical and regulatory bodies, government agencies, and society, the framework strengthens institutional capacity, enhances AI governance, and fosters trust among stakeholders [130,170]. The framework facilitates industry and academic collaborations to provide technological expertise, allowing AI tools to be tailored to institutional needs while enhancing operational efficiency and accessibility [171,172].

*Ethical AI Regulatory Bodies / Boards:* International organizations and ethical regulatory bodies establish AI governance benchmarks, ensuring responsible AI deployment in higher education. Organizations such as UNESCO, OECD, and the European Union set AI adoption principles emphasizing transparency, fairness, and accountability [130,173]. These regulatory bodies provide ethical guidelines that align AI governance policies with global human rights frameworks, ensuring compliance with data privacy laws such as GDPR and AI ethics principles [26,31]. By defining ethical AI standards, these institutions support universities in implementing AI systems that are fair, unbiased, and accountable.

*Alumni:* Alumni play a key role in AI integration and governance by leveraging their industry expertise, professional networks, and funding potential. Their contributions foster AI-driven mentorship programs, research funding, and industry partnerships. By serving as advisory members on institutional AI governance boards, they ensure AI applications align with real-world industry needs. Their engagement strengthens global institutional networks, enhancing higher education's adaptability to AI advancements [170].

*Government Agencies:* Government agencies and funding organizations help shape AI policies and provide financial assistance for AI integration. Investments in AI infrastructure and technology empower institutions, particularly in under-resourced regions, to implement AI effectively and ensure equitable access to AI-driven higher education [174].

*Global Academic Institutions:* International organizations establish AI governance benchmarks, ensuring ethical AI deployment in higher education. UNESCO, OECD, and similar bodies set AI adoption principles that emphasize transparency, fairness, and accountability [130,173]. Their policies help align AI governance frameworks with global ethical and regulatory standards, ensuring responsible AI integration [26].

*Industry Partnerships:* Public-private partnerships advance AI solutions that address digital inequalities and drive AI-powered innovation. Collaboration with industry leaders fosters technological advancements, ensuring AI solutions remain relevant, adaptable, and impactful within diverse educational settings [11,170].

*Societal Engagement:* Engaging society through transparent communication and public discourse fosters institutional credibility and trust in AI governance. Knowledge-sharing initiatives and benchmarking efforts further refine AI strategies, strengthening global collaboration and continuous improvements in AI adoption [6,121]. Encouraging interdisciplinary cooperation ensures AI policies remain inclusive, equitable, and aligned with societal expectations [26,30].

The synergy of human intelligence (internal and external stakeholders) ensures AI in HE remains ethical, transparent, and aligned with institutional values. By combining internal and external intelligence, AI governance becomes collaborative, adaptive, and inclusive. Unified human intelligence bridges technological advancements with ethical governance, ensuring AI serves as a tool for enhancement rather than automation-driven disruption. Sustainable AI governance relies on continuous stakeholder engagement, regulatory alignment, and participatory oversight, reinforcing trust, fairness, and accountability in AI adoption

***Collaborative Human Intelligence: Synergy Between Internal and External Stakeholders:***

The integration of internal and external intelligence ensures a country's technological readiness and policy maturity. This framework enables institutions in developed countries to emphasize advanced AI research, regulatory compliance, and industry collaboration, while institutions in underdeveloped regions focus on scalable, cost-effective AI solutions that align with local infrastructure and resource availability [26,45]. The interplay between internal and external stakeholders ensures AI governance remains both institutionally responsive and globally aligned. Internal stakeholders, including faculty, administrators, and students, oversee AI's implementation and impact within educational institutions. Meanwhile, external stakeholders—such as policymakers, industry leaders, and regulatory bodies—provide oversight, technological advancements, and ethical benchmarks. This collaborative structure prevents ethical lapses while promoting innovation and institutional adaptability.

#### *5.2.6 Unified Human Intelligence: Integrating Decision-making and Feedback Mechanism for AI Governance*

The Unified Human Intelligence component of the proposed framework establishes a structured approach to AI governance in HE, ensuring that AI integration remains ethical, accountable, and strategically aligned with institutional objectives. This system functions through a continuous interaction between decision-making and feedback loops, creating an adaptive mechanism that refines AI-driven processes while maintaining institutional and ethical integrity (refer Figure 4). Decision-making directs institutions toward their strategic goals, while feedback mechanisms assess the impact of those decisions, ensuring AI remains under human oversight [130]. As illustrated in Figure 3, these interdependent pillars prevent AI from operating in isolation, reinforcing human intelligence to build self-awareness, trust, collaboration, accountability, and ethical compliance. This synergy safeguards against automation bias, ethical lapses, and misalignment with educational values, ensuring AI remains a transparent and human-centered tool [88,163,175].

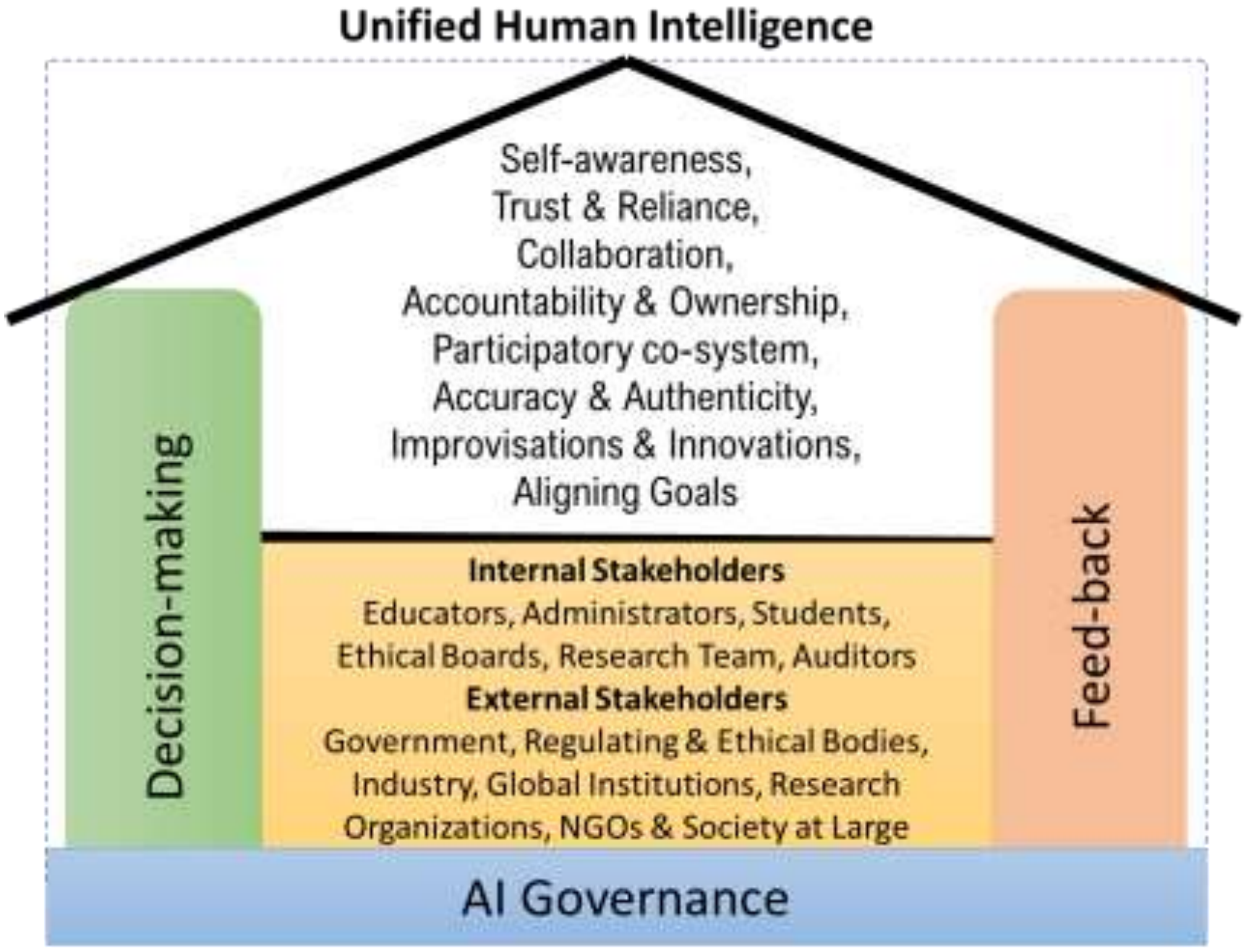

Figure 4: Unified Human Intelligence: Collaborative Decision-making and Feedback Mechanism for AI Governance

- ***Decision-Making as the Driver of Ethical AI Implementation***

  Empirical studies highlight the need for structured AI governance, addressing gaps in AI curriculum implementation and ethics compliance across universities / HEIs [22,98]. The European Commission’s Ethics Guidelines for Trustworthy AI (2019) reinforce HD-AIHED’s emphasis on human agency, bias mitigation, and regulatory compliance, ensuring responsible and sustainable AI adoption in HE [37,38].

  To address these challenges, this study’s framework, the HD-AIHED model, adopts a multi-tiered decision-making approach—collaborative, directive, and participatory—to ensure ethical AI governance in HE. By fostering fairness, transparency, and accountability, this model aligns with global AI ethics standards, promoting the responsible deployment and long-term sustainability of AI in HEIs [9,20].

  Collaborative decision-making involves internal and external stakeholders, including educators, students, industry experts, regulatory bodies, and policymakers, ensuring AI systems address diverse educational needs while mitigating algorithmic biases [176]. Directive decision-making assigns clear governance responsibilities to institutional leadership, reinforcing structured execution, transparency, and accountability [166]. Participatory decision-making fosters direct engagement between faculty, administrators, and students in the co-development of AI tools, ensuring AI serves as an enhancement to pedagogy and administration rather than a directive force [167,177]. By integrating structured decision-making, the framework ensures AI remains aligned with human values, reinforcing educational integrity while fostering innovation [169].

- ***Feedback Loops for AI Optimization and Refinement***

The framework embeds dynamic feedback loops to enhance AI's adaptability, accuracy, and alignment with institutional goals. These mechanisms facilitate real-time refinements, enabling institutions to detect algorithmic inconsistencies, recalibrate AI strategies, and mitigate emerging biases [6]. By ensuring AI governance remains responsive to regulatory changes, technological advancements, and shifting demographics, these feedback processes reinforce long-term AI sustainability [176]. This self-correcting AI ecosystem fosters responsible evolution, maintaining alignment with institutional, ethical, and regulatory principles [88,166,167]. Furthermore, [176] explore decision-making models in AI, reinforcing the necessity of adaptive feedback mechanisms to support ethical governance and strategic oversight.

- ***Synergy Between Decision-Making and Feedback***

  [99] emphasize that upholding institutional credibility and ethical integrity in HE requires a harmonized approach to AI decision-making and feedback, ensuring alignment with regulatory and ethical standards. This harmony ensures that institutional AI strategies remain dynamic, transparent, and adaptable. While decision-making establishes strategic direction and governance frameworks, feedback mechanisms continuously refine AI-driven outputs, fostering a cycle of learning, optimization, and trust-building [6]. This interdependent process enhances AI's accuracy, institutional credibility, and stakeholder confidence, positioning AI as an adaptive, ethical, and human-centered educational tool [176]. As a result, the framework enables a strategically governed, human-aligned transformation that upholds the core values of higher education [169,175].

### *5.2.7 Phased Human Intelligence Corresponding to AI Lifecycle*

- *Synchronizing Phased Human Intelligence Across the AI Lifecycle*
  Ethical AI implementation in higher education requires systematic and structured synchronization and integration of human intelligence across every stage of the AI lifecycle to ensure accountability, transparency, and value alignment [161]. A phased approach to AI governance is essential for ethical compliance and sustainable impact in HE. [22] advocate for a structured AI-driven model with continuous governance, while [110] highlight AI's role in advancing Sustainable Development Goals (SDGs) through phased implementation. Expanding on this, [6] emphasize the necessity of continuous monitoring and adaptive feedback loops to refine AI deployment and mitigate risks. Aligning with these insights, the HD-AIHED model incorporates a self-correcting governance framework, ensuring real-time adjustments while upholding academic integrity. Building on these principles, the proposed framework introduces phased Human Intelligence, strategically mapped across the AI lifecycle—adoption, design, deployment, evaluation, and exploration—to embed structured oversight, institutional adaptability, and ethical safeguards at every stage as illustrated in Figure 5 [31].
  The AI lifecycle, as shown in the Figure 4, consists of five interconnected phases: adoption, design, deployment, analysis, and exploration. Corresponding to these, the framework integrates Phased Human Intelligence components, ensuring strategic governance, ethical compliance, and institutional trust at each stage. As depicted in the Figure 4, this framework integrates decision-making and feedback loops, fostering ethical, transparent, and accountable AI implementation in higher education.

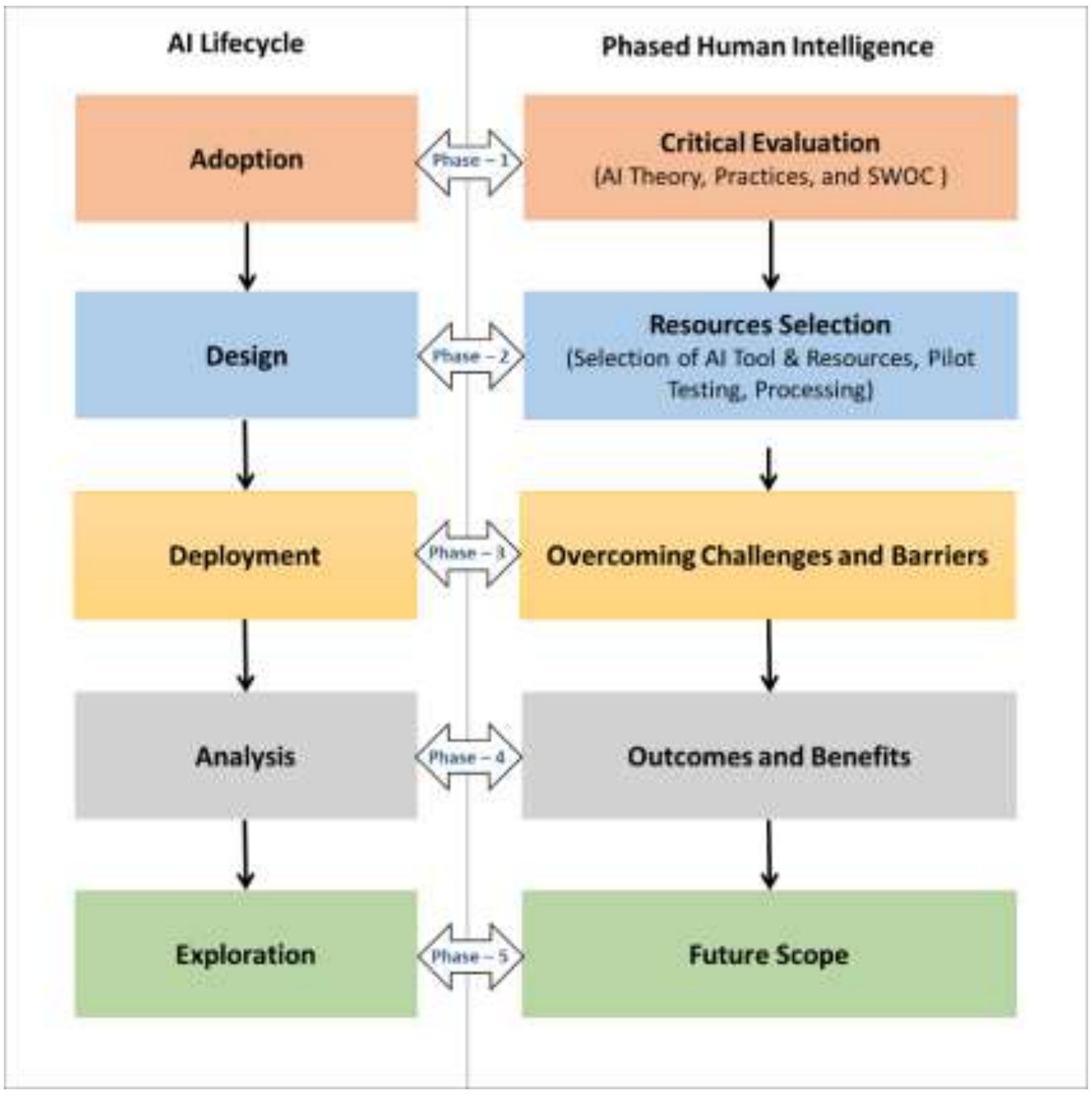

Figure 5: Phased Human Intelligence corresponding to AI Lifecycle

- *Phased Unified Human Intelligence for Addressing AI Lifecycle Challenges*
  Table 7 presents a structured mapping of Phased Human Intelligence within the AI lifecycle, ensuring ethical, inclusive, and transparent AI governance from feasibility assessment to long-term sustainability. The validation criteria establish governance benchmarks, reinforcing AI's accountability, fairness, and effectiveness in higher education.
  The framework structures AI governance into five key phases, integrating human and technological aspects to align AI implementation with institutional goals. As illustrated in Figure 4, each AI lifecycle stage is paired with a specific governance mechanism, enabling continuous evaluation, decision-making, and feedback integration. This adaptive model mitigates biases and ethical risks while fostering trust, scalability, and institutional resilience in AI-driven education.
  In the adoption phase, Critical Evaluation focuses on ethical decision-making and capacity building to assess AI readiness. During design, Resource Selection emphasizes bias minimization, participatory design, and pilot testing to validate AI's role. The deployment phase involves Overcoming Challenges, ensuring ethical compliance, privacy protection, and institutional trust. In analysis, Evaluation of Outcomes measures data-driven decision-making, institutional adaptability, and performance impact to refine AI implementation. Finally, Future Scope in the exploration phase emphasizes long-term AI innovation, scalability, and global benchmarking for sustainable governance.
  Throughout all five phases, stakeholder participation is crucial to mitigate biases and reinforce human oversight, ensuring AI remains transparent, ethical, and aligned with higher education values.

**Table 7: Phased Unified Human Intelligence for Addressing AI Lifecycle Challenges**

| Phase | Phased Unified Human Intelligence | | | AI Lifecycle | | Literature Support |
|---|---|---|---|---|---|---|
| | **Phase Focus** | **Human Aspect** | **Technological Aspects** | **Phase Outcome** | **Validation Criteria** | |

<table>
<tr><td>Phase 1</td><td>Critical Evaluation</td><td rowspan="5">• Stakeholder's participation,<br>• Collaborative & Participative decision-making,<br>• Ethical & Inclusive AI decision-making,<br>• Real-time, Iterative & Continuous Feed-back,<br>• Training & Re-skilling,<br>• Capacity building,<br>• Stakeholder's confident, trust and satisfaction</td><td>SWOC Analysis, Institutional readiness, External Links, Risk Mitigation, Compliance & Funding Insights</td><td>Adoption</td><td>Feasibility Validation: Usefulness, Innovativeness, Resource Capabilities, Personalization, Performance and Effort Expectancy.</td><td>[6,23,24,178–180]</td></tr>
<tr><td>Phase 2</td><td>Resources Selection</td><td>AI Tool selection, Resource selection, Participatory design, Bias Minimization, Pilot testing, Processing</td><td>Design</td><td>Application & Benefits Validation: Teaching-Learning & Research, Predictive Analytics, Administrative Optimization, Admissions and Enrollment, Academic Integrity, Inclusivity and Accessibility, Performance Assessment.</td><td>[10,26,33,36,38]</td></tr>
<tr><td>Phase 3</td><td>Overcoming Challenges and Barriers</td><td>Barrier identification, Ethical compliance, Privacy protection, Institutional trust-building, Ethical AI acceptance, Transparency in AI Decision-making</td><td>Deployment</td><td>Ethical & Regulatory Validation: AI Governance Adherence, Data Security, Algorithmic Bias, Resources & Infrastructure Gaps, Human Balance, Equity & Inclusion.</td><td>[11,31,45,50,145, 181]</td></tr>
<tr><td>Phase 4</td><td>Evaluation of Outcomes and Benefits</td><td>Data-driven Decision Making, Accountability & Accuracy, Impact assessment, Outcome Performance matrix, Institutional adaptability, Performance & Institutional Trust Metrics</td><td>Evaluation</td><td>Outcome Validation: Lifelong Learning, Operational Efficiency, Inclusivity of Diversity, Improved Student Engagement, Strengthened Academic Integrity, Improved Academic Life.</td><td>[6,8,17,39,140]</td></tr>
<tr><td>Phase 5</td><td>Future Scope</td><td>SWOC Analysis, External Links, Sustainability & Scalability, Long-term AI Innovation, Global Collaboration & benchmarking, Global Ethical AI Partnership, Adaptive AI Regulations, Future AI Governance Standards, AI promotion for broader adoption.</td><td>Exploration</td><td>Future Orientation Validation: Scalability, Sustainability & Long-term Adaptability, Competitive Advantage, Benchmarking, Academic Expansion, Global Collaboration, Long-Term Benefits, AIHED Promotion.</td><td>[8,11,22,31,99]</td></tr>
</table>

### 5.3 Establishing and Executing Human-Driven AIHED

This study aims to develop a conceptual framework for the ethical, inclusive, and sustainable integration of AI in higher education (AIHED) while ensuring alignment with institutional goals, societal values, and ethical safeguards. To achieve this, a unified human synergy in decision-making and feedback processes is essential across all stages of AIHED, from adoption to deployment and long-term sustainability [6].

This unified synergy is designed to:

- **Facilitate Adaptability** – Feedback mechanisms enable AI systems to evolve in response to stakeholder needs and contextual changes [85].
- **Enhance Accountability** – Decision-making checkpoints ensure transparency and provide structured interventions when AI outcomes deviate from intended goals [8].

- **Promote Inclusivity** – Active stakeholder involvement incorporates diverse perspectives into strategic AI decision-making [175].
- **Support Ethical Compliance** – Regular audits and human oversight address data privacy, fairness, and responsible AI deployment [182].

Through a thematic and analytical approach, this study proposes a conceptual framework for the ethical, inclusive, and sustainable integration of AI in higher education (AIHED) while ensuring alignment with institutional goals, societal values, and ethical safeguards. The framework is structured into five distinct phases, representing Unified Human Intelligence corresponding to the AI lifecycle. These phases, spanning from AI adoption to exploration and long-term sustainability, are depicted as interconnected stages illustrating a sequential and systematic AI governance approach [26].

Human and external intelligence are represented as ovals, with five human connections assigned to each phase and a single external connection integrated into the system. Human decision-making points (DM), symbolized by green arrows, link strategic human roles to AI governance. Decision-making functions in two key modes [183]:

- Collaborative and participative – Ensuring stakeholder engagement in AI decision-making.
- Operative and directive – Providing structured implementation and strategic oversight.

Additionally, dotted red arrows form a dynamic feedback loop (FB) to enable continuous system reviews, refinements, and iterative enhancements.

To maintain efficiency, accountability, and adaptability, the framework integrates real-time feedback loops for continuous system improvements, while decision-making checkpoints ensure validation and governance oversight. These interconnected processes enhance precision, alignment, and AI optimization, reinforcing accountability, efficiency, and long-term adaptability [6].

The resulting model, termed Human-Driven AIHED (HD-AIHED), is illustrated in Figure 6. It represents a structured workflow where human intelligence, decision-making, and feedback mechanisms seamlessly integrate into AI-driven higher education, ensuring ethical AI governance, institutional sustainability, and adaptive transformation.

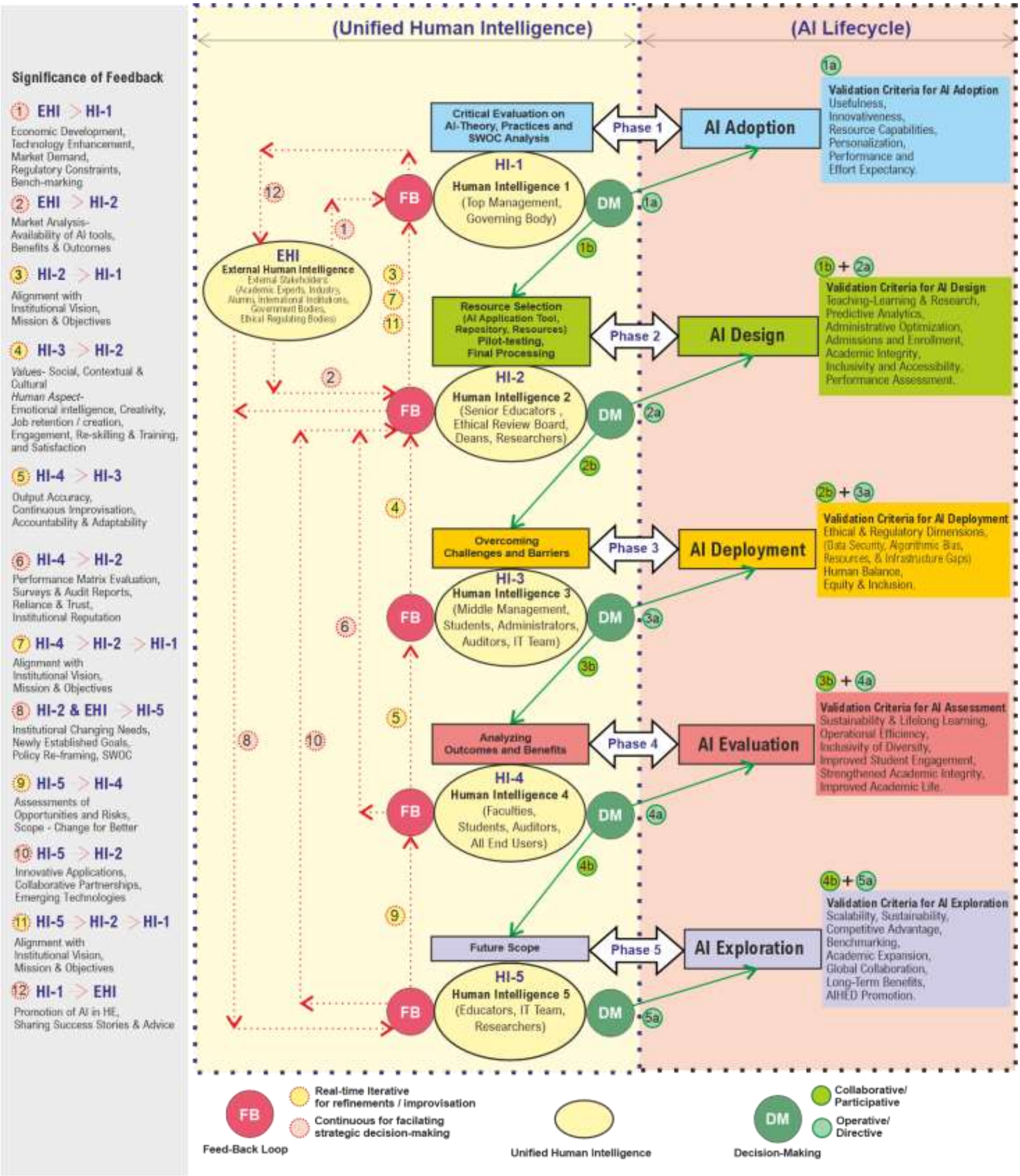

Figure 6: Human-Driven AIHED Model

### *5.3.1 Phase 1: Critical Evaluation (AI Adoption)*

Phase 1 of the AI adoption evaluation is spearheaded by Human Intelligence 1 (HI1), comprising the institution's top management and governing body. This phase is designed to critically assess the feasibility of AI integration while ensuring strategic alignment with institutional objectives and long-term sustainability [85].

To establish a robust foundation, HI1 engages in external intelligence gathering through External Human Intelligence (EHI), facilitated by Feedback Loop 1 (FB1). This continuous input is sourced from regulatory bodies, industry experts, and societal trends, providing valuable insights into emerging opportunities, potential challenges, and best practices in AI adoption [38].

A key focus of this phase is ethical compliance, ensuring that AI adoption adheres to internationally recognized frameworks, such as UNESCO's AI governance principles [31]. Additionally, socio-cultural adaptability is rigorously evaluated to assess AI's compatibility with institutional values, resource capabilities, and contextual applicability [6].

The evaluation framework incorporates an iterative feedback mechanism, wherein HI1 receives structured insights from Human Intelligence 2 (HI2)—a group comprising senior educators, researchers, and academic leaders. These insights facilitate an in-depth assessment of the institution's internal strengths and weaknesses concerning AI implementation.

In cases where unresolved issues persist, an additional iterative feedback loop (FB3) enables HI1 to reassess strategies, integrate newly acquired intelligence, and refine AI adoption approaches before advancing to the next phase. This dynamic and adaptive process ensures a high degree of transparency and responsiveness to emerging challenges.

A comprehensive SWOC (Strengths, Weaknesses, Opportunities, and Challenges) analysis is conducted to evaluate both internal institutional capabilities and external market dynamics. This structured assessment ensures that AI adoption decisions are data-driven, risk-aware, and aligned with institutional priorities [6].

Decision-making in this phase is twofold:

- Analytical decision-making (DM 1a) is employed to ensure a rigorous, data-driven evaluation of AI adoption, focusing on usefulness, innovativeness, resource efficiency, personalization, performance expectations, and socio-cultural adaptability.
- Directive decision-making (DM 1b) establishes a clear framework for operational execution, outlining the roles and responsibilities of HI2 to ensure a seamless transition to Phase 2.

The culmination of this phase is a validated AI adoption strategy that aligns with institutional objectives, ethical standards, and socio-cultural considerations. By leveraging continuous feedback loops, systematic evaluations, and strategic decision-making, Phase 1 ensures that AI adoption is methodically structured, ethically compliant, and institutionally sustainable before progressing to the next phase of implementation [85].

### *5.3.2 Phase 2: Selection of AI Tools and Resources, Pilot Testing and Processing (AI Design)*

Building on the critical evaluation in Phase 1, Phase 2 advances toward the selection, design, and integration of AI tools, ensuring their alignment with institutional objectives [66]. This phase is led by Human Intelligence 2 (HI2), comprising senior educators, deans, academicians, and researchers, and ethical review board who are responsible for assessing AI tools, managing their implementation, and designing AI frameworks tailored to institutional needs [11].

HI2 gathers external intelligence to evaluate AI tools and industry suppliers based on the AI applications recommended by HI1. This ensures that AI solutions are selected with a strong foundation of technological feasibility, ethical compliance, and institutional relevance.

The selection and design process is governed by strategic decision-making (DM2a), which is based on DM1a from Phase 1. This approach leverages the data-driven, analytical evaluation framework of DM1a, ensuring continuity in decision-making. DM2a is collaborative and participative, ensuring that AI solutions align with:

- Teaching-learning processes
- Predictive analytics and academic integrity
- Administrative optimization (admissions, enrollment, operations)
- Inclusivity and accessibility

AI design is a critical component of this phase, ensuring that AI solutions are contextually relevant, ethically responsible, and aligned with institutional goals. AI frameworks are developed with a focus on:

- Personalization to enhance user engagement
- Interoperability for seamless integration into institutional systems
- Scalability to support long-term adoption
- Compliance with regulatory and ethical standards

Upon the selection and design of AI tools, operative and directive decision-making (DM2b) ensures appropriate resource allocation, establishes clear roles and responsibilities within Human Intelligence 3 (HI3), and prepares for the transition to Phase 3.

Following AI selection and design, HI2 integrates AI tools into an institutional repository to facilitate pilot testing. This ensures that the AI system operates within a controlled environment, allowing for initial validation before full-scale deployment.

Pilot testing is continuously assessed through real-time iterative feedback (FB4) from HI3 (middle management, IT staff, and ethical review boards), confirming the effectiveness of AI implementation. Additionally, Feedback Loop 3 (FB3) from HI2 to HI1 ensures compliance with institutional AI adoption policies and strategic objectives.

Refinements to AI design are made dynamically based on feedback, ensuring the system's adaptability, accuracy, and efficiency before moving forward.

Once pilot testing is successfully completed, HI2 activates final processing, confirming that AI design and integration meet institutional standards. At this stage, HI2 facilitates operative, directive, and tactical decision-making to ensure a seamless transition of responsibilities to HI3 for full-scale implementation in Phase 3.

This decision-making framework promotes swift and effective AI selection, design, and implementation, while fostering a collaborative ecosystem involving both internal and external stakeholders.

Phase 2 transitions into tactical execution, addressing medium-term objectives such as repository integration, AI design validation, and pilot testing. HI2 oversees human role assignments, manages pilot programs, and monitors AI system functionality, ensuring that feedback continuously refines both selection and testing processes.

Upon the successful completion of pilot testing, Feedback Loop 3 (FB3) informs HI1 that the system aligns with AI adoption practices and institutional objectives, establishing a solid foundation for full-scale deployment in Phase 3.

### *5.2.3 Phase 3: Overcoming Challenges and Barriers (AI Deployment)*

Following the selection and pilot testing of AI tools in Phase 2, Phase 3 focuses on overcoming challenges and barriers to ensure a seamless AI deployment process [50]. This phase is led by Human Intelligence 3 (HI3), comprising middle management, IT teams, educators, students (end users), and auditors. Their role is to address ethical, regulatory, technical, and human-centered challenges through a structured decision-making framework.

HI3 adopts a collaborative and participatory approach to resolve barriers effectively. The decision-making process is structured into two key components:

- Analytical decision-making (DM3a) ensures a thorough evaluation of ethical, regulatory, and operational challenges in compliance with governing bodies.
- Operative and directive decision-making (DM3b) facilitates cross-functional collaboration, ensuring that issues are addressed efficiently, with a clear task allocation for stakeholders to implement the framework [184].

Tactical and operational decision-making plays a pivotal role in implementing solution-driven strategies, allowing HI3 to manage day-to-day operations while ensuring AI deployment remains aligned with institutional objectives.

HI3 is responsible for overcoming AI deployment challenges through DM3a, which includes:

- Ethical and regulatory compliance: Addressing data security, algorithmic bias, infrastructural limitations, and resource constraints to ensure AI operates within ethical and legal standards.
- Ensuring a human-centered approach: Maintaining a balance between technological advancement and human values, focusing on:

    - Emotional intelligence and creativity
    - Job retention and creation
    - Engagement, reskilling, and training
    - User satisfaction and adaptability

Real-time and iterative feedback (FB4) plays a crucial role in refining AI processes and tools while emphasizing ethical considerations. This continuous feedback mechanism ensures proactive issue resolution and iterative improvement in AI system deployment [7].

HI3 facilitates effective communication and engagement among stakeholders, ensuring timely problem resolution and continuous system refinement. Ethical compliance and inclusivity are key performance indicators guiding AI deployment adjustments.

Through directive and operative decision-making (DM3b), HI3 allocates tasks and responsibilities to HI4, ensuring a seamless transition to Phase 4, where AI deployment moves towards long-term sustainability, institutional integration, and strategic scaling.

### *5.3.4 Phase 4: Analysing Outcomes and Benefits (AI Evaluation)*

In Phase 4, Human Intelligence (HI4)—comprising senior faculty, administrators, auditors and students—plays a pivotal role in evaluating the effectiveness of the AIHED system. This assessment is based on critical metrics, including operational efficiency, ethical compliance, inclusivity, institutional trust, and performance benchmarks [10]. Analytical decision-making facilitates a data-driven evaluation of these outcomes, ensuring that AI adoption leads to meaningful educational and institutional benefits.

Strategic Decision-Making (DM4a) operates through a collaborative and participative approach, engaging HI4 to determine whether AI-driven outcomes justify enhanced operational efficiency, diversity and inclusivity, improved student engagement, academic integrity, and lifelong learning benefits.

A structured feedback mechanism supports continuous refinement:

- Continuous Feedback (FB6) collects audit reports, user performance data, and insights into institutional reputation for HI2, ensuring transparency and iterative system enhancements [45]. Further passed to HI1 through FB7.
- Iterative Feedback (FB5) helps identify gaps in performance metrics, enabling HI3 to refine strategies, resolve unresolved issues, and drive system improvements.

Operative and Directive Decision-Making (DM4b) ensures cross-stakeholder collaboration, integrating diverse perspectives for a comprehensive and balanced evaluation.

- Tactical decision-making focuses on implementing recommendations at the operational level.
- Directive decision-making assigns specific tasks and responsibilities to ensure effective execution.

By leveraging research and innovative methodologies, DM4b ensures that collective input leads to meaningful advancements. This multi-faceted and integrated framework optimizes the AIHED system and prepares it for a seamless transition to Phase 5

### *5.3.5 Phase 5: Future Scope (AI Exploration)*

Phase 5 of the HD-AIHED Model corresponds to future-proof strategies, emphasizing scenario planning and continuous research to ensure the long-term adaptability of AI systems to evolving societal and educational needs, as recommended by [99]. In the final phase, Human Intelligence (HI5), consisting of researchers and strategists, collaborates with HI2 and External Intelligence to evaluate emerging needs, technological advancements, and opportunities for enhancing global competitiveness through FB8 [26]. The SWOC analysis done by HI5 on future opportunity is communicated to HI2 then passed to HI1 through FB10 and FB11 respectively. After collaborative and participative discussion held at HI1 and HI2, FB8 provides feedback on AI exploration. Continuous feedback (FB8) from HI2 informs HI5 of areas requiring improvement, addressing evolving institutional needs, policy reforms, and newly established goals. Feedback loops (FB10) guide HI2 in exploring future partnerships and driving innovation to ensure adaptability in a dynamic landscape. Conceptual decision-making (DM5a) prioritizes scalability and long-term sustainability, ensuring AI remains an evolving enabler in higher education [80].

Feedback loops (FB9) provide iterative insights into potential opportunities and risks, enabling HI4 to refine strategies and prepare for the future. This dual feedback mechanism fosters transparent communication, adaptability, and alignment with long-term institutional objectives. It supports life-long learning opportunities and ensures that institutions remain agile and competitive. By integrating these iterative insights, institutions can proactively address future challenges, capitalize on advancements in AI, and align with global standards, ensuring sustained relevance and impact.

Lastly, after successful implementation and achieving satisfactory benefits and outcomes, Human Intelligence 1 (HI1) takes responsibility for communicating (FB12) success stories and with External Human Intelligence. This involves the publication of success stories and strategies to overcome obstacles associated with AIHED through various media channels to foster social appeal and demonstrate the institution's accomplishments. This communication reinforces transparency, enhances institutional reputation, and inspires broader engagement with the institution’s advancements in AI adoption.

The HD-AIHED operational flow (Refer Figure 5) ensures AI adoption, design, deployment, evaluation, and exploration remain ethically sound, institutionally integrated, and dynamically evolving. By embedding human oversight and external intelligence at every stage, this framework provides continuous adaptability through structured feedback loops, accountability and transparency via decision-making checkpoints, and scalability aligned with long-term educational needs. This integrated methodology strengthens institutional trust, fosters inclusive AI ecosystems, and ensures that AI remains an enabler of human intelligence rather than a replacement, preserving ethical, academic, and pedagogical integrity in higher education.

## 6. Key Insights from the HD-AIHED Model and Interpretation

### 6.1 Significance and Strengths of the HD-AIHED Model

The HD-AIHED framework envisions AI as an ethical force for innovation, inclusivity, and sustainability in higher education (HE), ensuring that AI enhances rather than replaces human intelligence [44,185]. It serves as a comprehensive and human-centric model that prioritizes responsible AI adoption, ensuring that AI-driven transformation in HE remains aligned with institutional goals, pedagogical values, and stakeholder needs [38,39].

By emphasizing strategic decision-making, dynamic feedback mechanisms, and phased implementation, the HD-AIHED Model provides a structured and ethical pathway for AI integration, addressing key challenges such as algorithmic bias, privacy concerns, and student agency in AI governance [9,20].

- ***Comprehensive Sequential Flow*:** The HD-AIHED Model encompasses the entire AI lifecycle, from critical evaluation during adoption to analyzing outcomes and planning for future scalability. This phased approach ensures that no essential aspect of AI implementation is overlooked, creating a holistic framework for integrating AI into higher education
- ***Human-Centric Approach:*** The model emphasizes human intelligence; oversight, integrating multiple feedback and strategic decision-making connections (e.g., HI1-HI5) throughout all phases. This human-centric approach ensures ethical compliance, alignment with institutional goals, and cultural sensitivity, fostering trust and inclusivity in AI systems through participatory design.
- ***Dynamic Feedback Mechanisms:*** Continuous and iterative feedback loops (FB1-12) embedded at every phase ensure iterative improvements, adaptability, and accountability in AI implementation. These mechanisms align with institutional objectives while addressing real-time inefficiencies and stakeholder concerns.
- ***Pilot Testing for Validation:*** The inclusion of pilot testing in Phase 2 validates AI tools against functional and institutional requirements before full deployment. This reduces risks associated with misalignment, errors, and inefficiencies, ensuring a smoother integration process.
- ***Addressing Ethical and Operational Barriers*:** Phase 3 focuses on overcoming barriers such as algorithmic bias, data security, and institutional resource gaps. By addressing these common challenges proactively, the framework ensures ethical and practical readiness for AI deployment through capacity building.
- ***Rigorous Outcome Evaluation and Scalability:*** Phases 4 and 5 are dedicated to evaluating AI outcomes, analyzing long-term impacts, and identifying opportunities and risks for growth and scalability. This future-oriented focus ensures the AI system remains relevant, effective, and aligned with technological advancements.
- ***Support Across the Student Lifecycle:*** The framework supports AI applications throughout the student lifecycle, from admissions and personalized learning to graduation. This ensures equitable opportunities, personalized engagement, and improved student outcomes, reflecting a commitment to inclusivity and student success.
- ***Benchmarking and Global Competitiveness:*** In Phase 5, the model incorporates benchmarking to assess global standards, ensuring competitiveness, scalability and sustainability. By aligning with emerging technologies and ethical principles, it prepares institutions for future advancements while maintaining compliance with regulatory norms.
- ***Knowledge Repository for Institutional Learning:*** The integration of lessons learned through feedback loops creates a valuable knowledge base, enabling institutions to refine strategies for future AI implementations. This facilitates organizational learning and innovation, building a robust foundation for continuous improvement.

- ***Ethical Oversight and Trust-Building:*** The model's emphasis on ethical oversight ensures compliance with principles such as beneficence, autonomy, and justice. Transparent processes foster trust among stakeholders by demonstrating accountability and fairness, promoting confidence in AI-driven systems.
- ***Global Integration:*** The Model underscores the significance of external connections to enhance its scalability, relevance, and impact with HI1 & HI2). By adhering to regulatory compliance, such as GDPR and global AI governance frameworks, it ensures ethical alignment and legal accountability. The model leverages technological advancements and market analyses to adapt AI-driven tools to emerging trends and innovations, while global benchmarking fosters competitiveness by aligning with international best practices. Additionally, it emphasizes the importance of cross-sector collaborations, partnerships with industries, and engagement with global stakeholders. A key strength lies in its commitment to communicating success stories and failures, enabling institutions to share insights, foster trust, and refine strategies. This transparent approach supports sustainable growth, innovation, and credibility within the AI-driven higher education ecosystem

### 6.2 HD-AIHED Model in Action: A Strategic Mapping and Solution for Overcoming Real-Time Challenges and Barriers

Table 8 serves as an analytical repository, mapping AI applications across global universities and assessing their adoption challenges, solutions, and institutional impact. It provides an evidence-based evaluation of how AI systems perform in diverse educational settings, identifying best practices, gaps, and opportunities for expansion.

By applying the HD-AIHED Model, this strategic mapping offers a validated framework to navigate AI adoption complexities, ensuring that institutions overcome governance barriers, mitigate ethical risks, and drive sustainable AI transformation in higher education. The table systematically analyzes AI applications across universities, aligning them with institutional goals and innovations while assessing their real-world effectiveness.

Each AI application is evaluated through five key phases—Adoption, Design, Deployment, Evaluation, and Exploration—providing a structured approach to overcoming institutional barriers and ethical challenges. This phased framework ensures AI integration remains strategic, adaptive, and aligned with human-driven governance, ultimately optimizing AI's role in shaping the future of higher education.

**Table 8: HD-AIHED Model: A Strategic Mapping and Solution for Overcoming Challenges and Barriers**

| Region/ University/ Institution | AI Application & AI Tool | Repository Input | Phase 1 - Adoption | Phase 2 - Design | Phase 3 - Deployment | Phase 4 - Evaluation | Phase 5 - Exploration |
|---|---|---|---|---|---|---|---|
| | | | **Critical Evaluation** | **Selection of Resources, Testing & Final Processing** | **Overcoming Challenges** | **Analysing Outcomes** | **Future Scope - Opportunities & Risks** |
| **Arizona State University (USA)** | Personalized Learning Pathways (AI platforms like Squirrel AI, ALEKS) | Student performance data, learning styles, and curriculum objectives | Usefulness, resource capabilities, and personalization validated through SWOC analysis | Pilot testing of personalized learning paths ensuring curriculum alignment | Address regulatory barriers and institutional compatibility | Analyze outcomes such as learning pace, engagement, and curriculum fit | Expand to support hybrid learning models and gamified experiences for broader implementation |
| **University of Toronto (Canada)** | AI-Powered Research Assistance (Tools like Research Rabbit, Iris.ai) | Institutional research outputs, journal databases, and researcher profiles | Evaluate the relevance and usefulness of AI tools for interdisciplinary research | Selection of tools based on literature gap identification and AI-driven recommendations | Address challenges like data accessibility, AI recommendation accuracy, and inclusivity | Assess research outcomes, interdisciplinary connections, and user satisfaction metrics | Scale tools for multilingual research synthesis and integrate with global research databases |
| **University of Melbourne (Australia)** | AI-Powered Chatbots for 24/7 Support (Chatbots like IBM Watson Assistant, Ada) | FAQs, institutional policies, student queries, and service logs | Determine chatbot usefulness for resolving routine queries and escalating complex cases | Pilot testing chatbot workflows within small departments before scaling up | Overcome challenges related to response accuracy, escalation processes, and language support | Analyze outcomes like improved service efficiency, query resolution rates, and user feedback | Expand chatbot capabilities to multilingual functionality and alumni/community engagement |
| **University of California (USA)** | Proctoring and Exam Integrity (Proctoring tools like ProctorU, Honorlock) | Exam question sets, student credentials, live webcam and keystroke monitoring data | Validate AI's effectiveness in proctoring against ethical and privacy considerations | Implementation of AI-based proctoring tools with faculty feedback and gradual integration | Address ethical concerns, false positives, and system accuracy challenges | Analyze outcomes such as proctoring effectiveness, fairness, and user satisfaction rates | Enhance AI-driven post-exam analytics and adaptive questioning systems |
| **Stanford University (USA)** | AI in Research & Patent Ethics (AI-driven plagiarism detection & research integrity tools) | Research papers, patent filings, and institutional policies | Evaluate AI's role in ensuring research integrity and detecting plagiarism | AI-driven analysis of patent applications and automated research verification | Address patent ownership disputes and legal ambiguity in AI-generated research | Assess AI's effectiveness in maintaining ethical research standards | Implement AI governance frameworks to manage AI-generated patents and research ethics |
| **Georgia Tech (USA)** | AI-Driven Recruitment Analytics (Recruitment tools like Eightfold.ai, Entelo) | Historical hiring data, skills databases, and institutional workforce needs | Evaluate tools for predictive hiring, skill gap analysis, and workforce planning | Selection of AI tools aligned with institutional workforce needs and recruitment priorities | Address challenges related to fairness, diversity, and predictive analytics outcomes | Analyze recruitment efficiency, diversity outcomes, and alignment with institutional goals | Scale AI to support long-term workforce planning and global talent acquisition |
| **National University of Singapore (Singapore)** | AI-Powered Accessibility for Neurodiverse Students (Tools like CogniFit, Glean) | Cognitive ability assessments, behavioral data, and neurodiverse student profiles | Assess inclusivity, usability, and alignment with institutional accessibility goals | Implementation of AI tools tailored to neurodiverse needs through pilot testing | Overcome accessibility barriers and validate inclusivity benchmarks | Measure learning outcomes, usability feedback, and cognitive adaptability | Expand with AR/VR tools for broader neurodiverse learning enhancements |
| **Harvard University (USA)** | AI in Accreditation (AI-based compliance tools) | Faculty performance metrics, accreditation frameworks | Validate AI for automating compliance tracking and performance evaluation | Implement AI-driven accreditation monitoring systems | Address bias in AI-based faculty performance assessments | Assess improvements in accreditation transparency and compliance rates | Develop adaptive AI credentialing frameworks for global accreditation |
| **MIT (USA)** | Virtual Reality (VR) Immersive Learning (VR tools like Oculus for Higher Education, zSpace) | Lesson plans, 3D models, and student progress data | Validate alignment of VR learning tools with curriculum goals and resource capabilities | Selection and pilot testing of VR tools for specific curriculum modules | Address barriers such as cost, resource limitations, and teacher training needs | Evaluate learning engagement, retention rates, and performance outcomes | Broaden VR integration across subjects and develop AR-based real-world learning systems |
| **IIT Bombay (India)** | AI in Campus Placements (AI-driven job recommendation systems) | Student resumes, employer requirements, industry hiring trends | Assess AI's potential in optimizing job matching and placement success | Implement AI-powered resume analysis and employer-student matchmaking | Address AI bias in resume shortlisting and candidate ranking | Evaluate job placement success rates and employer satisfaction | Develop human-AI hybrid hiring models for fairer recruitment processes |
| **Oxford University (UK)** | AI-Powered Emotional Intelligence Analysis (Sentiment analysis tools like Affectiva) | Video/audio data from classrooms, student survey responses, and interaction logs | Validate effectiveness in detecting emotions and engagement while ensuring privacy safeguards | Implementation of sentiment analysis tools and targeted emotional support programs | Overcome challenges related to emotional data interpretation and ethical concerns | Compare emotional well-being metrics pre- and post-intervention for outcome validation | Expand into AI-driven emotional development and teacher training programs |
| **Cambridge University (UK)** | AI-Powered Campus Energy Optimization (Smart campus tools like BuildingIQ, Enel X) | Real-time energy consumption data, weather patterns, and occupancy rates | Evaluate the cost-effectiveness and eco-friendly potential of energy optimization tools | Selection of tools for implementation with pilot energy audits and sustainability benchmarks | Overcome integration challenges with existing infrastructure and energy compliance standards | Assess energy savings, carbon footprint reduction, and user satisfaction metrics | Integrate renewable energy sources and improve sustainability across campuses |
| **Carnegie Mellon University (USA)** | AI in Industry Collaboration (AI tools for corporate-academic partnerships) | Research collaboration agreements, industry-funded projects | Assess AI's role in strengthening university-industry partnerships | Implement AI-driven collaboration platforms for project matchmaking | Address data ownership conflicts between universities and private firms | Evaluate research impact and industry engagement success rates | Establish contractual AI governance models to ensure institutional data control |
| **University of Cambridge (UK)** | AI-Driven Alumni Engagement (Engagement tools like Salesforce Einstein) | Alumni profiles, event participation data, and donation history | Evaluate tools for fostering meaningful alumni relationships and event participation | Pilot testing AI recommendations for mentorship, event planning, and fundraising efforts | Address challenges like alumni data quality and maintaining personalization at scale | Measure success rates in alumni engagement, participation, and fundraising | Scale AI-driven predictive models for long-term alumni collaboration and donor engagement |

# 7. Discussions

## 7.1 Implications and Recommendations

The AIED-HDMFB model's step-by-step approach integrates critical evaluation, tool selection, ethical compliance, and scalability into the adoption of AI in higher education. Its feedback loops ensure adaptability and inclusivity, aligning with the broader objectives of the Human-driven AIHED framework [36]. For instance, during the tool selection phase, human decision-makers validate AI tools for compliance with institutional missions and stakeholder needs, reflecting the principles of dynamic feedback and ethical alignment [10]. The framework proposed in this study is pivotal for ensuring the sustainable and ethical adoption of Artificial Intelligence (AI) in higher education (HE). It establishes a foundation for continuous improvement by aligning AI applications with institutional goals, addressing limitations, and balancing AI's capabilities with human roles. Beyond enhancing efficiency, the framework emphasizes AI as a tool for assistance rather than replacement, fostering innovation through collaboration, lifelong learning, and emerging technologies [155].

- ***Towards a new notion, Human-Driven AIHED: A Paradigm Shift***
  The model—HD-AIHED, conceptualized from this study, transitions the notion of "AIHED driving Humans" to "Human-driven AIHED", placing human agency at the core of AI integration. This approach highlights the collaborative role of both internal and external stakeholders—educators, students, policymakers, and technology providers in ensuring AI adoption aligns with ethical, cultural, and institutional values. Educators play a crucial role in co-designing AI systems to meet pedagogical goals, while institutions must establish frameworks where human oversight remains central to AI applications. Governments can facilitate this transition by developing policies that incentivize human-centric AI initiatives that uphold inclusivity and [10,36].
- ***Human-Driven AIHED: Empowering Not Replacing***

The HD-AIHED framework positions AI as an assistive tool that automates routine tasks, enabling educators to focus on mentorship, creativity, and strategic problem-solving. The integration of feedback loops and decision making ensures that AI evolves continuously to meet institutional and student needs through empowering human intelligence. To make this possible, professional development programs are essential for reskilling educators in AI-enabled methodologies, while students must be encouraged to engage actively with AI tools to supplement critical thinking and creativity. Ethical bodies need to establish interdisciplinary guidelines to ensure AI deployment is aligned with inclusivity and transparency, preserving education's human-centric essence [7,186]

- ***The Ethics of Equity: A Path to Inclusive HE***
  By integrating adaptive learning, multilingual support, and accessibility features, HD-AIHED aims to address cultural and physical disparities in education. Feedback mechanisms embedded in the framework help detect and mitigate biases, ensuring that AI-driven solutions provide equitable access for diverse learners. Governments can expand AI-driven initiatives to underserved populations through strategic funding and infrastructure investments, while institutions must embed cultural and contextual needs into their AI tools. Society at large should advocate for community-driven efforts that reduce digital divides and promote inclusivity in AI-powered education [6,10].
- ***Integrating Innovation and Institutional Values in a Digital World***
  The HD-AIHED framework incorporates innovative technologies such as blockchain, AR/VR, and advanced analytics to enhance transparency, security, and efficiency, while maintaining institutional values. These innovations, when supported by feedback mechanisms, allow institutions to adapt technologies responsibly and ensure alignment with evolving stakeholder needs. Industry partners are integral in co-developing AI systems tailored to educational requirements, while institutions must adopt structured AI life-cycle phases that balance innovation with core values. Regular audits by ethical bodies can ensure these technologies comply with equity and ethical standards, sustaining a mission-driven approach [10,148].
- ***A Shared Vision: Toward a Unified and Agile Future***
  Aligning with NEP 2020 of India, Higher Education 4.0, and Industry 4.0, HD-AIHED bridges learner-centric education with advanced technologies. The framework's scalability and adaptability prepare students for AI-augmented workplaces, while dynamic feedback loops ensure its responsiveness to workforce demands. Institutions should embrace global benchmarking practices and share success stories to enhance competitiveness, while policymakers must design regulations that promote scalable and responsible AI adoption. Students, in turn, can benefit from AI-powered career tools that offer personalized skill development and prepare them for global opportunities [8].
- ***Inspiring Global Transformation Through Collaborative Practices***
  The HD-AIHED framework emphasizes international collaboration to share best practices, address global challenges, and foster inclusivity. Benchmarking efforts and alignment with diverse cultural and educational contexts amplify its global impact, making it a transformative tool for inclusive HE. Institutions must actively engage in global forums to co-develop solutions and promote trust in AI's transformative potential. Public dialogue should be encouraged to address societal concerns about equity and transparency, while ethical bodies facilitate the development of harmonized global practices, creating an interconnected HE ecosystem [109,155].
- ***Catalysts of Transformation: Role and Responsibilities***
  Governments, institutions, educators, and industry stakeholders play critical roles in the implementation and promotion of HD-AIHED by addressing access gaps, fostering reskilling initiatives, and ensuring ethical AI integration. Governments can transform education by investing in infrastructure, fostering AI literacy, supporting curricula aligned with Industry 4.0, and constructing ethical regulations to ensure inclusivity and transparency [148]. Institutions must leverage AI to enhance operational efficiency and deliver personalized learning experiences that attract diverse learners while maintaining socio-cultural equity. By sharing success stories and adopting strategic enrollment approaches, institutions can build trust and forge global collaborations [10]

Educators, as ethical stewards, must embrace reskilling to adapt to AI-enhanced roles, focusing on mentorship, creativity, and innovation as AI automates routine tasks. Industry stakeholders can collaborate with academia to design scalable, ethical AI tools tailored to institutional needs, while creating jobs such as AI trainers, ethical auditors, and data analysts to support workforce transformation [7]. Students benefit from AI-driven personalization throughout their educational journey, from admissions to career preparation, gaining critical skills for AI-augmented industries.

By combining their efforts, these stakeholders ensure HD-AIHED is adopted inclusively, ethically, and innovatively, driving the creation of a sustainable and globally impactful AI-powered education ecosystem.

- ***HD-AIHED: A Universal Model for Borderless HE Future***

  The HD-AIHED framework serves as a universal model to address global challenges in higher education by harmonizing innovation with equity, transparency, and lifelong learning. It aligns with international initiatives such as SDG 4 and UNESCO's guidelines on AI ethics, promoting ethical AI adoption while ensuring inclusivity and accessibility for diverse populations. In developing nations, the framework bridges gaps in infrastructure and resources by leveraging affordable, culturally sensitive technologies that empower underserved communities with lifelong learning opportunities and skill development [8]. Advanced regions benefit from HD-AIHED's mechanisms for refining AI tools to enhance personalization, address algorithmic bias, and uphold data privacy while positioning themselves as innovation hubs for global research.

  For low-economic regions, the framework provides a cost-effective pathway for higher education reform, emphasizing ethical compliance and partnerships with governments and NGOs to tackle infrastructural challenges. Cross-border collaboration further strengthens its global impact, enabling institutions to share knowledge, co-develop scalable AI solutions, and benchmark best practices. These efforts align with the EU's AI Act and UNESCO's ethical AI guidelines, ensuring standardized, transparent, and equitable AI integration across diverse contexts [36].

  Through international partnerships and cultural sensitivity, the HD-AIHED framework transforms HE into an interconnected ecosystem, fostering innovation while preparing institutions and learners for a dynamic and equitable future. Institutions that actively adopt this framework and engage in global collaborations can position themselves as leaders in advancing ethical and inclusive AI in education.

- ***Long-term Adaptability: Awareness, Promotion and Training***

  Institutions should actively leverage social media platforms, research publications, and academic collaborations to drive awareness and engage stakeholders in AIHED adoption [187]. Social media can be a powerful tool for sharing success stories, challenges, and best practices, fostering a global knowledge-sharing ecosystem. Engaging with policymakers, faculty, students, and industry leaders through these platforms can help build trust and encourage cross-border collaborations [6,11].

  Regular publications in academic journals, magazines, and institutional reports can further reinforce the credibility of HD-AIHED, providing universities with frameworks, case studies, and real-world insights [30]. Hosting webinars, podcasts, and panel discussions will create an open dialogue about AI's evolving role in HE, fostering a community of practitioners, researchers, and decision-makers [38]. International conferences and institutional partnerships will enhance AI governance models, ensuring that HD-AIHED remains an evolving, globally adopted initiative [31].

  For responsible AI integration, faculty and students must be trained in AI ethics and governance [8]. Universities should implement AI ethics literacy programs that emphasize bias detection, data privacy, fairness, and responsible AI usage [20]. Educators must be equipped with knowledge to critically engage with AI-driven decision-making, ensuring that AI complements human expertise rather than replacing it [7]. Students should be trained in AI governance policies to enable their participation in institutional AI decision-making, ethical review boards, and AI auditing [26].

- ***Transformative Perspectives: AI as Critical Infrastructure***

The HD-AIHED framework reimagines AI not just as a tool but as critical infrastructure that fosters inclusive, ethical, and transformative progress in higher education. By emphasizing human agency, creativity, and ethical accountability, it positions AI as a collaborative partner that complements rather than replaces human intelligence. The framework incorporates dynamic feedback loops to ensure the continuous evolution of AI systems, enabling them to address the ever-changing challenges of higher education. Through the integration of SWOC analyses and the alignment of human and external intelligence across the phases of the AI life cycle, the framework bridges equity, inclusivity, and data ethics gaps.

It promotes practical strategies to enhance lifelong learning and sustainable innovation, making higher education more accessible and future-focused. Policymakers have a critical role in establishing metrics to evaluate AI's long-term societal and educational impacts, while ethical bodies advocate for iterative refinements to ensure AI aligns with institutional missions and broader societal values. This transformative approach ensures AI's adoption as a sustainable, equitable, and dynamic infrastructure that drives innovation and progress in HE globally [188,189].

- ***AI Bias Mitigation and Ethical AI Auditing***

  The HD-AIHED framework requires robust bias mitigation strategies to prevent algorithmic discrimination in grading, admissions, and learning analytics [68]. AI systems, if not carefully managed, can reinforce existing social inequalities and lead to unintended biases [60]. Ensuring fairness in AI decision-making is essential to maintaining trust, transparency, and inclusivity in higher education.

  To enhance fairness and accountability, following strategies should be adopted:

  *Bias Detection Frameworks*: Institutions should conduct periodic AI audits to assess fairness and eliminate biases in student performance predictions. This includes using algorithmic fairness metrics and AI transparency evaluations to ensure responsible AI deployment [39]. The AI Risk Management Framework developed by [63] offers an approach to assessing and mitigating AI risks, which could be applied in educational settings.

  *Explainable AI (XAI):* AI tools used in HE should incorporate explainable AI models that provide clear, interpretable decision-making processes [187]. Transparent AI models allow educators, administrators, and students to understand how AI-based decisions are made and intervene when necessary [26].

  *Interdisciplinary Ethics Committees:* Universities/institutions should establish AI Ethics Boards, composed of faculty, students, AI developers, and policymakers, to oversee AI deployment and governance in education. These committees would evaluate potential biases, risks, and ethical concerns related to AI systems in HE [8]. Several global frameworks, such as the UNESCO Recommendation on the Ethics of AI (2021) and the OECD Principles on Artificial Intelligence (2019), emphasize the importance of ethics-based AI governance in education [31,50].

  By integrating these mechanisms, HE institutions can ensure that AI enhances educational equity without reinforcing systemic biases. Adopting these strategies will not only improve AI fairness and transparency but also foster trust and inclusivity among students and educators in AI-driven learning environments.

## 7.2 Future Research Directions

### *7.2.1 Quantitative Survey for Validating Study's Framework*

To ensure the HD-AIHED framework is empirically validated, a quantitative survey combined with Structural Equation Modeling (SEM) is recommended. SEM has been widely recognized in educational research for assessing complex relationships between latent constructs, ensuring robust theoretical validation [190,191]. Given the increasing role of AI in higher education, it is essential to establish a data-driven approach that quantifies AI's impact across its adoption, design, deployment, performance, and sustainability phases [6]. Following Table 8 showcases future research directions. Table 8 presents a structured future research agenda aimed at empirically testing the HD-AIHED model through quantitative survey-based methodologies.

**Table 8: Future Research Outline: Testing HD-AIHED Model With Quantitative Survey**

| Steps | Description |
|---|---|
| **1. Define the Research Framework** | Develop hypotheses based on the five-phase AI lifecycle (adoption, design, deployment, analysis, and exploration). Identify key indicators such as ethical scalability, inclusivity, and human intelligence integration. Utilize technology adoption models like TAM, UTAUT, or DoI to examine AI's impact on educational institutions. |
| **2. Design a Quantitative Survey** | Create a structured questionnaire targeting higher education stakeholders, including faculty, administrators, and students. Incorporate Likert-scale questions to assess alignment with institutional ethics, operational effectiveness, stakeholder acceptance, and ethical compliance in AI integration. |
| **3. Conduct Data Collection** | Implement a large-scale survey across universities using random or stratified sampling methods. Collect data in either a cross-sectional or longitudinal format to ensure diverse representation and measure framework effectiveness over time. |
| **4. Perform Statistical Validation** | Use Structural Equation Modeling (SEM) to examine relationships between AI lifecycle phases and ethical outcomes. Apply factor analysis to identify key dimensions influencing AI implementation, and conduct regression analysis to predict AI's impact on learning outcomes and institutional efficiency. |
| **5. Assess Practical Implications** | Validate the scalability and adaptability of the HD-AIHED framework by comparing results across different geographical and institutional contexts. Provide actionable insights for policymakers and educators to enhance AI integration in higher education. |
| **6. Recommend Future Research Directions** | Integrate AI-driven data analytics to refine survey insights, combine quantitative and qualitative methods for deeper validation, and benchmark findings against global AI policy frameworks to assess ethical AI integration in higher education. |

A structured survey-based methodology should be employed, gathering responses from key stakeholders, including faculty, students, and administrators. Survey indicators should be designed based on well-established theoretical models such as the Technology Acceptance Model (TAM) [52] and the Unified Theory of Acceptance and Use of Technology (UTAUT) [54], ensuring reliability and construct validity. Likert-scale items measuring AI readiness, human-centered design, stakeholder-driven deployment, ethical considerations, and long-term sustainability should be incorporated [192].

To analyze these relationships, SEM should be applied, as it enables simultaneous testing of multiple hypotheses and constructs within a hierarchical framework [193]. This method will allow researchers to determine the extent to which AI adoption, ethical AI design, and stakeholder involvement contribute to institutional scalability and sustainability [194]. Furthermore, incorporating covariance relationships among first-order constructs will provide a comprehensive understanding of interdependencies in AI integration [195].

By validating the HD-AIHED framework through quantitative analysis and SEM, this study will offer generalizable insights into the operational, ethical, and pedagogical implications of AI in education. This approach will align AI deployment with institutional policies, ethical standards, and global governance frameworks, ensuring a scalable and human-centric AI transformation in higher education [8].

Table 9 exhibits key constructs and hypotheses aligned with the five phases of the HD-AIHED model, offering a structured approach for its empirical validation. By applying Structural Equation Modeling (SEM), this research proposes to test the relationships between constructs associated with various phases of the AI lifecycle. The table outlines relevant constructs, proposed hypotheses, and expected correlations, ensuring a data-driven assessment of the HD-AIHED model's effectiveness in AI-driven higher education.

**Table 9: Validating HD-AIHED Model through Structural Equation Model (SEM)**

| Higher-Order Construct | Phase | Sub-Constructs (Latent Variables) | Description | Hypothesis | Connection Description |
|---|---|---|---|---|---|
| **Human-Driven AI in Higher Education (HD-AIHED)** | **1. Adoption Phase** | **Institutional Readiness** | The level of infrastructure, policies, and technical support for AI adoption. | *H1:* The successful adoption of AI in higher education is positively associated with institutional readiness, faculty & students acceptance, and ethical policy frameworks. | Institutions must have the necessary infrastructure, policies, and stakeholder engagement for AI to be adopted effectively. |
| | | **Faculty and Student Acceptance** | The willingness of educators and learners to integrate AI into educational practices. | | |
| | | **Ethical and Regulatory Compliance** | The adherence to AI governance policies and ethical standards during adoption. | | |
| | **2. Design Phase** | **Human-Centered AI Design** | The extent to which AI systems integrate human intelligence, ethical considerations, and pedagogical principles. | *H2:* AI system design that integrates human intelligence, ethical safeguards, and institutional values enhances user trust and engagement among students and faculty. | AI should be designed with transparency, ethical considerations, and human intelligence integration to ensure responsible use. |
| | | **Transparency and Trust** | The clarity of AI decision-making processes and users' confidence in AI tools. | | |
| | | **Usability and Accessibility** | The ease of use and inclusivity of AI-driven learning systems. | | |
| | **3. Deployment Phase** | **Stakeholder Involvement** | The engagement of faculty, administrators, and students in AI implementation. | *H3:* The effectiveness of AI deployment in higher education is significantly influenced by stakeholder involvement, transparency in implementation, and compliance with ethical guidelines. | Effective AI implementation requires active involvement and ethical decision-making from faculty, students, and administrators. |
| | | **Implementation Transparency** | The degree of openness in AI integration, including clear communication about AI's role. | | |
| | | **Ethical AI Practices** | The extent to which AI deployment aligns with fairness, accountability, and privacy standards. | | |
| | **4. Analysis Phase** | **AI-Driven Decision-Making** | The effectiveness of AI in enhancing learning analytics, student performance tracking, and administrative decision-making. | *H4:* AI-driven analytics that incorporate real-time feedback and ethical oversight improve decision-making, learning outcomes, and institutional efficiency. | AI's success in HE depends on its ability to enhance learning outcomes, efficiency, and fairness while ensuring compliance with ethical standards. |
| | | **Feedback Integration** | The responsiveness of AI systems to human feedback for continuous improvement. | | |
| | | **Institutional Efficiency** | The impact of AI on streamlining administrative and academic operations. | | |
| | **5. Exploration Phase** | **Long-Term Sustainability** | The adaptability and continuous evolution of AI applications in education. | *H5:* The long-term sustainability and scalability of AI in higher education depend on continuous evaluation, policy evolution, and adaptive integration of emerging technologies. | The future of AI in education relies on continuous evaluation, policy evolution, and adaptability to emerging technologies. |
| | | **Policy Evolution and Scalability** | The ability of institutions to update policies and expand AI solutions. | | |
| | | **Emerging Technology Integration** | The readiness of institutions to adopt next-generation AI innovations. | | |

### *7.2.2 Other Emerging Research Priorities:*

While this study develops a conceptual framework for ethical AI adoption in HE, its empirical validation remains a critical next step. Future research should focus on longitudinal assessment, ethical compliance, comparative analysis, and iterative refinement to ensure AI's effectiveness, adaptability, and scalability across diverse institutional settings.

*Longitudinal Studies on AI's Institutional Impact:* Long-term impact studies should track AI adoption's influence on student outcomes, institutional governance, and educational equity over a multi-year period. By comparing pre-adoption and post-adoption performance metrics, these studies will provide insights into AI's role in academic innovation, faculty efficiency, and student support.

*Empirical Testing of the HD-AIHED Framework:* Pilot studies across diverse university settings should measure the effectiveness of the HD-AIHED framework in AI governance, faculty decision-making, and student engagement. Testing AI applications in different cultural, regulatory, and economic contexts will ensure their adaptability and scalability while refining ethical oversight mechanisms.

*Development of AI Bias Auditing Tools:* AI-powered decision-making systems in grading, admissions, and academic feedback must be regularly assessed for bias detection and mitigation. Research should focus on developing AI fairness auditing tools that ensure transparency, accountability, and ethical compliance with standards such as UNESCO AI Principles and GDPR regulations.

*Exploring AI's Role in Faculty Governance and Research Integrity:* AI's integration into peer review, faculty recruitment, and academic fraud detection requires further study to ensure human oversight and ethical accountability. Research should evaluate how AI can enhance academic quality assurance while preventing unintended biases or manipulation in research evaluation and faculty assessments.

*Advancing AI-Powered Student Support Systems:* Future studies should examine AI's role in student well-being, including AI-driven academic advising, mental health chatbots, and personalized learning platforms. Evaluating AI's impact on student retention, emotional intelligence, and inclusivity will ensure AI complements human educators rather than replacing critical human interactions.

*Cross-Institutional Comparative Analysis and AI Refinement:* AI adoption patterns across different academic, geographical, and socio-economic contexts should be analyzed to identify best practices and governance models for optimizing AI's role in HE worldwide. A real-time feedback-driven refinement model will enable AI governance frameworks to evolve iteratively, ensuring interdisciplinary collaboration, institutional adaptability, and strategic AI expansion.

Through empirical, data-driven research, these future directions will strengthen AI integration in HE, ensuring alignment with institutional goals, ethical governance, and the evolving academic landscape while fostering a human-centered, transparent, and sustainable AI ecosystem.

### 7.3 AIHED Research Milestones And Accomplishments

This study successfully meets its research objectives by conducting a comprehensive review of literature, case studies, and global reports, critically analyzing existing research to identify key gaps in AIED frameworks, particularly in the domains of ethics, inclusivity, and human agency. This foundational analysis supports the development of a comprehensive and human-centered framework, ensuring AI's responsible integration into higher education.

- This directly addresses Research Question 1 (RQ1) by identifying deficiencies in existing AIHED integration frameworks and proposing evidence-based strategies to address these gaps. Through a comprehensive review of literature, case studies, and global reports, the study evaluates key shortcomings in ethics, inclusivity, and human agency, laying the foundation for a human-centered AI framework that ensures responsible AI integration into higher education.

- Proposing Solutions for Ethical and Inclusive AI: The research presents strategies for responsible AI adoption, emphasizing transparency, fairness, and inclusivity in AI-driven educational environments.
- Laying the Foundation for a Human-Centered AI Framework: The findings support the development of a comprehensive AIHED model that prioritizes ethical governance, human collaboration, and institutional alignment.
- Addressing AI's Dual Role in Higher Education: The research examines AI as both a transformative driver and a collaborative tool, highlighting global opportunities and challenges associated with AI deployment. This analysis directly addresses Research Question 2 (RQ2) by balancing technological advancements with human-centric values, ensuring that AI enhances, rather than replaces, human agency in educational settings.
- Development of the Human-Driven AIHED (HD-AIHED) Framework: Integrating core principles of accountability, adaptability, and empathy, this framework ensures AI adoption remains ethical and human-centric. This aligns with Research Question 3 (RQ3) by fostering a balanced, inclusive, and responsible AI implementation strategy in higher education.
- Evaluating Framework Operability in Real-Time Global Contexts: The study assesses AI deployment across diverse educational settings, directly responding to Research Question 4 (RQ4). By incorporating feedback mechanisms and adaptive design, the framework remains scalable, relevant, and accessible, even in resource-constrained regions, promoting equitable AI integration.
- Promoting Interdisciplinary Collaboration and Participatory Co-Design: The research provides actionable insights for ethical AI adoption, directly contributing to Research Question 5 (RQ5). These insights foster trust, inclusivity, and alignment with institutional and educational values, ensuring AI applications are developed and implemented with a strong ethical foundation.
- Strategic Recommendations for Key Stakeholders: Finally, the study meets its objective of offering strategic recommendations for educators, policymakers, and technology developers, addressing Research Question 6 (RQ6). These recommendations ensure effective AI integration while mitigating challenges related to equity, accountability, and cultural sensitivity, fostering a more sustainable and ethical AI-driven educational ecosystem.

## 8. Conclusions

This study provides a critical analysis of Artificial Intelligence in Higher Education (AIHED), highlighting its dual role as both a transformative driver and a collaborative tool. The proposed Human-Driven AIHED (HD-AIHED) framework establishes a structured and ethical roadmap for AI adoption, ensuring that AI integration in higher education remains human-centered, ethically sound, and strategically aligned with institutional objectives. By positioning AI as a facilitator rather than a replacement for human intelligence, the framework preserves the core values of accountability, adaptability, and inclusivity, reinforcing ethical integrity, transparency, and accessibility. The model underscores that the future of higher education does not rest in AI's capabilities alone but in human intelligence guiding AI towards its highest potential.

The HD-AIHED framework's applicability spans both the AI life cycle and the student life cycle, ensuring a dual-layered synergy that integrates Human Intelligence at each phase of AI development and deployment. Through its phased approach—adoption, design, deployment, and exploration—the framework ensures AI remains context-sensitive, adaptable, and responsive to evolving higher education landscapes. By mapping global real-time AI challenges, the HD-AIHED model provides actionable insights on how institutions can leverage AI-driven models without compromising human-centric values, thus ensuring scalability, sustainability, and cross-border collaboration in AI governance.

The HD-AIHED framework serves as a strategic enabler, leveraging AI's potential to enhance personalized learning, optimize administrative processes, and strengthen institutional decision-making while embedding a strong human-centered ethical foundation. In developed nations, the framework supports institutional competitiveness, workforce readiness, and advanced research capabilities by integrating governance models, ethical safeguards, and continuous feedback mechanisms into AI implementation. In developing regions, the framework plays a democratizing role, fostering scalable, cost-effective AI solutions that bridge digital and infrastructural gaps, ensuring equitable access to AI-driven education. Recognizing the risks associated with AI adoption, the framework proactively addresses bias, data privacy, algorithmic transparency, and student agency, reinforcing human oversight and ethical compliance as fundamental pillars of responsible AI integration.

To strengthen ethical AI governance, institutions must establish dedicated AI governance boards to oversee bias mitigation, privacy protection, and regulatory compliance. Aligning AI policies with UNESCO AI ethics guidelines, GDPR, and Sustainable Development Goal 4 (SDG 4) ensures that AI adoption remains globally relevant, ethically responsible, and institutionally aligned. The HD-AIHED framework emphasizes that AI applications must be customized to institutional goals, ensuring AI enhances adaptive learning, predictive analytics, and administrative efficiency, while placing students at the center of AI integration.

Stakeholder engagement plays a crucial role in AI governance, requiring participation from internal and external stakeholders, including faculty, students, policymakers, industry partners, and regulatory bodies. A participatory decision-making approach combined with dynamic feedback mechanisms fosters AI systems that are responsive, inclusive, and aligned with institutional values.

To ensure strategic alignment, ethical compliance, and institutional readiness, the implementation of SWOC (Strengths, Weaknesses, Opportunities, and Challenges) analysis at both the adoption phase and the AI exploration phase is critical. This structured evaluation allows institutions to identify risks, leverage opportunities, and establish a sustainable AI adoption strategy that is scalable, adaptable, and future-ready. Furthermore, AI tools must be culturally and linguistically adaptable, ensuring inclusivity and diversity in educational environments.

In addition, the promotion of AIHED is imperative for institutions to learn from both success stories and failures, enabling them to refine AI strategies, mitigate risks, and enhance best practices. By fostering knowledge-sharing and institutional collaboration, AI integration can be continuously improved, ensuring its long-term impact and ethical alignment in higher education.

Bridging the AI readiness gap in higher education necessitates significant investment in AI literacy programs, equipping faculty, administrators, and students with the skills necessary to navigate AI's evolving role. Institutions in low-resource settings should develop collaborative partnerships to build AI infrastructure and implement capacity-building initiatives that foster sustainable AI adoption. Additionally, real-time feedback mechanisms must be embedded in AI models, ensuring continuous assessment, refinement, and alignment with institutional and policy shifts. The iterative testing and improvement of AI applications will help institutions remain responsive to emerging technological advancements and evolving educational needs.

Future research should prioritize the empirical validation of the HD-AIHED framework through pilot implementations, longitudinal assessments, and cross-institutional comparative studies. Testing the framework in real-world educational settings will provide insights into its scalability, effectiveness, and governance impact. Multi-institutional pilot programs should be initiated to evaluate AI's influence on student engagement, institutional decision-making, and ethical AI adoption.

The development of a standardized AIHED governance roadmap is essential to define best practices, compliance mechanisms, and long-term monitoring strategies for responsible AI integration in higher education. Research must also focus on designing AI bias auditing tools to detect and mitigate algorithmic bias, ensuring fairness, accountability, and inclusivity in AI-driven decision-making.

Additionally, exploring regional AIHED adaptation strategies is crucial for understanding how policy frameworks, cultural dynamics, and infrastructure disparities shape AI adoption across diverse educational landscapes. Identifying context-sensitive governance models will help institutions develop adaptive AI policies that align with global ethical standards while addressing local challenges.

By adopting the HD-AIHED model proposed in this study, universities and higher education institutions can harness AI's transformative power not just as a 'tool', but as a core 'infrastructure' that upholds ethical integrity and human-centered progress. This Human-Driven AIHED Framework ensures that AI serves as a catalyst for empowerment, rather than exclusion, fostering a future where technology and human expertise collaborate to advance equitable and sustainable education.

Ultimately, the HD-AIHED framework envisions AI as an inclusive, ethical, and transformative force in higher education rather than a disruptive replacement for human intelligence. By fostering collaboration between governments, institutions, industry stakeholders, and students, the framework ensures that AI integration remains human-driven, ethically aligned, and strategically positioned to enhance global education. As AI continues to redefine higher education, its success will depend on our collective ability to balance technological progress with ethical imperatives, ensuring that AI in HE remains a beacon of knowledge, accessibility, and integrity for future generations.

## References

1. Tegmark M. Life 3.0: Being human in the age of artificial intelligence. Vintage; 2018.
2. Tegmark M. Being human in the age of artificial intelligence. 2019.
3. Brynjolfsson BYE, McAfee A. Artificial intelligence for real. Harv Bus Rev. 2017;July: 1–31. Available: https://starlab-alliance.com/wp-content/uploads/2017/09/AI-Article.pdf
4. Luckin R, Holmes W. Intelligence Unleashed: An argument for AI in Education. UCL Knowledge Lab: London, UK. . 2016. Available: https://www.pearson.com/content/dam/corporate/global/pearson-dot-com/files/innovation/Intelligence-Unleashed-Publication.pdf
5. Holmes W, Bialik M, Fadel C. Artificial intelligence in education promises and implications for teaching and learning. Center for Curriculum Redesign; 2019.
6. Zawacki-Richter O, Marín VI, Bond M, Gouverneur F. Systematic review of research on artificial intelligence applications in higher education–where are the educators? Int J Educ Technol High Educ. 2019;16: 1–27.
7. Pedro F, Subosa M, Rivas A, Valverde P. Artificial intelligence in education: Challenges and opportunities for sustainable development. 2019.
8. Floridi L, Cowls J. A unified framework of five principles for AI in society. Mach Learn city Appl Archit urban Des. 2022; 535–545.
9. Wu C, Zhang H, Carroll JM. AI Governance in Higher Education: Case Studies of Guidance at Big Ten Universities. Futur Internet. 2024;16. doi:10.3390/fi16100354
10. Bond M, Khosravi H, De Laat M, Bergdahl N, Negrea V, Oxley E, et al. A meta systematic review of artificial intelligence in higher education: a call for increased ethics, collaboration, and rigour. Int J Educ Technol High Educ. 2024;21. doi:10.1186/s41239-023-00436-z
11. Crompton H, Burke D. Artificial intelligence in higher education: the state of the field. Int J Educ Technol High Educ. 2023;20. doi:10.1186/s41239-023-00392-8
12. Lin CC, Huang AYQ, Lu OHT. Artificial intelligence in intelligent tutoring systems toward sustainable education: a systematic review. Smart Learn Environ. 2023;10. doi:10.1186/s40561-023-00260-y
13. Oyebola Olusola Ayeni, Nancy Mohd Al Hamad, Onyebuchi Nneamaka Chisom, Blessing Osawaru, Ololade Elizabeth Adewusi. AI in education: A review of personalized learning and educational technology. GSC Adv Res Rev. 2024;18: 261–271. doi:10.30574/gscarr.2024.18.2.0062
14. Baker RS. Challenges for the future of educational data mining: The Baker learning analytics prizes. J Educ data Min. 2019;11: 1–17.
15. Wang T, Lund BD, Marengo A, Pagano A, Mannuru NR, Teel ZA, et al. Exploring the Potential Impact of Artificial Intelligence (AI) on International Students in Higher Education: Generative AI, Chatbots, Analytics, and International Student Success. Appl Sci. 2023;13. doi:10.3390/app13116716
16. PwC. Sizing the prize: What's the real value of AI for your business and how can you capitalise? PwC; 2019. Available: https://www.pwc.com/gx/en/issues/analytics/assets/pwc-ai-analysis-sizing-the-prize-report.pdf
17. Alotaibi NS. The Impact of AI and LMS Integration on the Future of Higher Education: Opportunities, Challenges, and Strategies for Transformation. Sustain. 2024;16. doi:10.3390/su162310357
18. Lytras MD, Serban AC, Alkhaldi A, Aldosemani T, Malik S. What's Next in Higher Education: The AI Revolution 2030. In: Lytras MD, Serban AC, Alkhaldi A, Malik S, Aldosemani T, editors. Digital Transformation in Higher Education, Part A. Emerald Publishing Limited; 2024. pp. 155–172. doi:10.1108/978-1-83549-480-620241008
19. Digital Education Council. Digital Education Council Global AI Student Survey 2024 AI or Not AI : What Students Want. 2024. Available: https://www.digitaleducationcouncil.com/post/digital-education-council-global-ai-student-survey-2024
20. Memarian B, Doleck T. Fairness, Accountability, Transparency, and Ethics (FATE) in Artificial Intelligence (AI) and higher education: A systematic review. Comput Educ Artif Intell. 2023;5: 100152. doi:10.1016/j.caeai.2023.100152
21. Khatri BB, Karki PD. Artificial Intelligence (AI) in Higher Education: Growing Academic Integrity and Ethical Concerns. Nepal J Dev Rural Stud. 2023;20: 1–7. doi:10.3126/njdrs.v20i01.64134
22. Batista J, Mesquita A, Carnaz G. Generative AI and Higher Education: Trends, Challenges, and Future Directions from a Systematic Literature Review. Inf. 2024;15. doi:10.3390/info15110676
23. Airaj M. Ethical artificial intelligence for teaching-learning in higher education. Educ Inf Technol. 2024;29: 17145–17167. doi:10.1007/s10639-024-12545-x
24. Al-Zahrani AM, Alasmari TM. Exploring the impact of artificial intelligence on higher education: The dynamics of ethical, social, and educational implications. Humanit Soc Sci Commun. 2024;11: 1–12. doi:10.1057/s41599-024-03432-4

25. Alfredo R, Echeverria V, Jin Y, Yan L, Swiecki Z, Gašević D, et al. Human-centred learning analytics and AI in education: A systematic literature review. Comput Educ Artif Intell. 2024;6. doi:10.1016/j.caeai.2024.100215

26. Holmes W, Miao F. Guidance for generative AI in education and research. UNESCO Publishing; 2023. Available: https://unesdoc.unesco.org/ark:/48223/pf0000386693.locale=en

27. Maslej N, Fattorini L, Brynjolfsson E, Etchemendy J, Ligett K, Lyons T, et al. Artificial intelligence index report 2023. arXiv Prepr arXiv231003715. 2023.

28. Hamid OH. Impact of Artificial Intelligence (AI) in Addressing Students at-Risk Challenges in Higher Education (HE). The Evolution of Artificial Intelligence in Higher Education: Challenges, Risks, and Ethical Considerations. Emerald Publishing Limited; 2024. pp. 183–194.

29. Stanford University. AI Index Report 2023: Measuring trends in Artificial Intelligence. AI Index Rep. 2023. Available: https://aiindex.stanford.edu/report/

30. Marengo A, Pagano A, Pange J, Soomro KA. The educational value of artificial intelligence in higher education: a 10-year systematic literature review. Interact Technol Smart Educ. 2024.

31. UNESCO. Recommendation on the Ethics of Artificial Intelligence. Paris: UNESCO; 2021.

32. Sambasivan N, Kapania S, Highfill H, Akrong D, Paritosh P, Aroyo LM. “Everyone wants to do the model work, not the data work”: Data Cascades in High-Stakes AI. Proceedings of the 2021 CHI Conference on Human Factors in Computing Systems. New York, NY, USA: Association for Computing Machinery; 2021. doi:10.1145/3411764.3445518

33. Ozmen Garibay O, Winslow B, Andolina S, Antona M, Bodenschatz A, Coursaris C, et al. Six Human-Centered Artificial Intelligence Grand Challenges. Int J Hum Comput Interact. 2023;39: 391–437. doi:10.1080/10447318.2022.2153320

34. Rakowski R, Polak P, Kowalikova P. Ethical Aspects of the Impact of AI: the Status of Humans in the Era of Artificial Intelligence. Society. 2021;58: 196–203. doi:10.1007/s12115-021-00586-8

35. Postman N. Technopoly: The surrender of culture to technology. Vintage; 2011.

36. Castañeda L, Selwyn N. More than tools? Making sense of the ongoing digitizations of higher education. Int J Educ Technol High Educ. 2018;15. doi:10.1186/s41239-018-0109-y

37. Mustafa MY, Tlili A, Lampropoulos G, Huang R, Jandrić P, Zhao J, et al. A systematic review of literature reviews on artificial intelligence in education (AIED): a roadmap to a future research agenda. Smart Learning Environments. Springer Nature Singapore; 2024. doi:10.1186/s40561-024-00350-5

38. Alqahtani N, Wafula Z. Artificial Intelligence Integration: Pedagogical Strategies and Policies at Leading Universities. Innov High Educ. 2024. doi:10.1007/s10755-024-09749-x

39. Crawford J, Allen KA, Pani B, Cowling M. When artificial intelligence substitutes humans in higher education: the cost of loneliness, student success, and retention. Stud High Educ. 2024;49: 883–897. doi:10.1080/03075079.2024.2326956

40. Holmes W, Porayska-Pomsta K, Holstein K, Sutherland E, Baker T, Shum SB, et al. Ethics of AI in Education: Towards a Community-Wide Framework. Int J Artif Intell Educ. 2022;32: 504–526. doi:10.1007/s40593-021-00239-1

41. Alexander B, Ashford-Rowe K, Barajas-Murphy N, Dobbin G, Knott J, McCormack M, et al. Educause Horizon report: 2019 Higher Education edition. 2019.

42. Heidegger M. The question concerning technology. 1977.

43. Heidegger M. The question concerning technology. Readings Philos Technol. 2009; 9–24.

44. Kamali J, Alpat MF, Bozkurt A. AI ethics as a complex and multifaceted challenge: decoding educators’ AI ethics alignment through the lens of activity theory. Int J Educ Technol High Educ. 2024;21: 1–20.

45. Alshahrani BT, Pileggi SF, Karimi F. A Social Perspective on AI in the Higher Education System: A Semisystematic Literature Review. Electron. 2024;13: 1–27. doi:10.3390/electronics13081572

46. Walsh D, Downe S. Meta-synthesis method for qualitative research: a literature review. J Adv Nurs. 2005;50: 204–211.

47. Finfgeld-Connett D. Introduction To Theorygenerating Meta- Synthesis Research. A Guid to Qual Meta-Synthesis. 2018; 1–12. doi:10.4324/9781351212793-1

48. Heaton J. Secondary analysis of qualitative data: An overview. Hist Soc Res Sozialforsch. 2008; 33–45.

49. Zimmer L. Qualitative meta-synthesis: a question of dialoguing with texts. J Adv Nurs. 2006;53: 311–318.

50. OECD. OECD principles on artificial intelligence. OECD Paris, France; 2019.

51. IEEE. Ethics of Autonomous and Intelligent Systems. New York: IEEE; 2020.

52. Davis FD, Bagozzi RP, Warshaw PR. Technology acceptance model. J Manag Sci. 1989;35: 982–1003.

53. Rogers EM. Diffusion of innovations, 5th edn London. UK Free Press. 2003.

54. Venkatesh V, Bala H. Technology acceptance model 3 and a research agenda on interventions. Decis Sci. 2008;39: 273–315.

55. Trow M. Reflections on the transition from elite to mass to universal access: Forms and phases of higher education in modern societies since WWII. International handbook of higher education. Springer; 2007. pp. 243–280.

56. Bourdieu P. The forms of capital.(1986). Cult theory An Anthol. 2011;1: 949.

57. Sandelowski M, Barroso J. Handbook for synthesizing qualitative research. springer publishing company; 2006.

58. Noblit GW. Meta-ethnography: Synthesizing qualitative studies. Stage Publ. 1988.

59. Hannes K, Lockwood C. Synthesizing qualitative research: choosing the right approach. John Wiley & Sons; 2011.

60. Kordzadeh N, Ghasemaghaei M. Algorithmic bias: review, synthesis, and future research directions. Eur J Inf Syst. 2022;31: 388–409.

61. Artyukhov A, Wołowiec T, Artyukhova N, Bogacki S, Vasylieva T. SDG 4, Academic Integrity and Artificial Intelligence: Clash or Win-Win Cooperation? Sustainability. 2024;16: 8483.

62. Heintz F. ARTIFICIAL INTELLIGENCE DUALITY. Comput Think Educ K-12 Artif Intell Lit Phys Comput. 2022;143.

63. NIST. AI Risk Management Framework. Gaithersburg: National Institute of Standards and Technology; 2023.

64. Thomas J, Harden A. Methods for the thematic synthesis of qualitative research in systematic reviews. BMC Med Res Methodol. 2008;8: 1–10.

65. Leshem S, Trafford V. Overlooking the conceptual framework. Innov Educ Teach Int. 2007;44: 93–105.

66. Ahmad SF, Alam MM, Rahmat MK, Mubarik MS, Hyder SI. Academic and Administrative Role of Artificial Intelligence in Education. Sustain. 2022;14: 1–11. doi:10.3390/su14031101

67. Hu S, Ke F, Vyortkina D, Hu P, Luby S, O'Shea J. Artificial Intelligence in Higher Education: Applications, Challenges, and Policy Development and Further Considerations BT - Higher Education: Handbook of Theory and Research: Volume 40. In: Perna LW, editor. Cham: Springer Nature Switzerland; 2025. pp. 1–52. doi:10.1007/978-3-031-51930-7_13-1

68. Baker RS, Hawn A. Algorithmic bias in education. Int J Artif Intell Educ. 2022; 1–41.

69. Singh N. Keeping artificial intelligence real: The importance of AI governance. McKinsey & Company; 2024. Available: https://www.mckinsey.com/capabilities/mckinsey-digital/our-insights/keeping-artificial-intelligence-real

70. Nguyen A, Hong Y, Dang B, Huang X. Human-AI collaboration patterns in AI-assisted academic writing. Stud High Educ. 2024;49: 847–864. doi:10.1080/03075079.2024.2323593

71. Dobrev D. A Definition of Artificial Intelligence. 2012; 1–7. Available: http://arxiv.org/abs/1210.1568

72. Russell SJ, Norvig P. Artificial intelligence: a modern approach. Pearson; 2016.

73. Kaplan A, Haenlein M. Siri, Siri, in my hand: Who's the fairest in the land? On the interpretations, illustrations, and implications of artificial intelligence. Bus Horiz. 2019;62: 15–25. doi:10.1016/j.bushor.2018.08.004

74. Roll I, Wylie R. Evolution and Revolution in Artificial Intelligence in Education. Int J Artif Intell Educ. 2016;26: 582–599. doi:10.1007/s40593-016-0110-3

75. Chen X, Xie H, Zou D, Hwang GJ. Application and theory gaps during the rise of Artificial Intelligence in Education. Comput Educ Artif Intell. 2020;1: 100002. doi:10.1016/j.caeai.2020.100002

76. Mohammed AT, Velander J, Milrad M. A Retrospective Analysis of Artificial Intelligence in Education ( AIEd ) Studies : Perspectives , Learning Theories , Challenges , and Emerging Opportunities. Springer Nature Singapore; 2024. doi:10.1007/978-981-97-8638-1

77. Claudio S, Sirois J, Owen K, Marya V, Krishnan C, Brasca C. How technology is shaping learning in higher education. McKinsey Co. 2022; 1–9. Available: https://www.mckinsey.com/industries/education/our-insights/how-technology-is-shaping-learning-in-higher-education

78. Agarwal P, Swami S, Malhotra SK. Artificial Intelligence Adoption in the Post COVID-19 New-Normal and Role of Smart Technologies in Transforming Business: a Review. J Sci Technol Policy Manag. 2024;15: 506–529. doi:10.1108/JSTPM-08-2021-0122

79. Luckin R, Holmes W. Intelligence unleashed: An argument for AI in education. 2016.

80. Perrault R, Clark J. Artificial Intelligence Index Report 2024. 2024.

81. Grand View Research. Artificial Intelligence (AI) in Education Market Report. 2025. Available: https://www.grandviewresearch.com/industry-analysis/artificial-intelligence-ai-education-market-report

82. Anon. Pearson 2024 End of Year AI Report for Higher Education. Pearson; 2024.

83. Marques-Cobeta N. Artificial Intelligence in Education: Unveiling Opportunities and Challenges BT - Innovation and Technologies for the Digital Transformation of Education: European and Latin American Perspectives. In: García-Peñalvo FJ, Sein-Echaluce ML, Fidalgo-Blanco Á, editors. Singapore: Springer Nature Singapore; 2024. pp. 33–42. doi:10.1007/978-981-97-2468-0_4

84. Luckin R, Holmes W, Griffiths M, Forcier LB. Artificial intelligence in education: Opportunities and challenges for a new era. Educ Technol. 2021;61: 16–22.

85. Holstein K, Aleven V, Rummel N. A conceptual framework for human–AI hybrid adaptivity in education. Lecture Notes in Computer Science (including subseries Lecture Notes in Artificial Intelligence and Lecture Notes in Bioinformatics). Springer International Publishing; 2020. doi:10.1007/978-3-030-52237-7_20

86. George B, Wooden O. Managing the Strategic Transformation of Higher Education through Artificial Intelligence. Adm Sci. 2023;13. doi:10.3390/admsci13090196

87. Spivakovsky O V., Omelchuk SA, Kobets V V., Valko N V., Malchykova DS. Institutional Policies on Artificial Intelligence in University Learning, Teaching and Research. Inf Technol Learn Tools. 2023;97: 181–202. doi:10.33407/itlt.v97i5.5395

88. Chan CKY. A comprehensive AI policy education framework for university teaching and learning. Int J Educ Technol High Educ. 2023;20: 38.

89. Barnett R. Being a university. Routledge; 2010.

90. Zhang D, Mishra S, Brynjolfsson E, Etchemendy J, Ganguli D, Grosz B, et al. 2021 AI Index Report. AI Index Steer Committee, Stanford Univ Human-Centered Artif Intell Institute, Stanford Univ. 2021; 1–222. Available: https://aiindex.stanford.edu/report/

91. Rogers EM, Singhal A, Quinlan MM. Diffusion of innovations. An integrated approach to communication theory and research. Routledge; 2014. pp. 432–448.

92. Chapman M. Constructive evolution: Origins and development of Piaget’s thought. Cambridge University Press; 1988.

93. Bruner J. Humon Development Celebrating Divergence: Piaget and Vygotsky1. Hum Dev. 1997;40: 63–73.

94. Zhang D, Mishra S, Brynjolfsson E, Etchemendy J, Ganguli D, Grosz B, et al. The AI index 2021 annual report. arXiv Prepr arXiv210306312. 2021.

95. Urmeneta A, Romero M. Creative Applications of Artificial Intelligence in Education. Springer Nature; 2024.

96. Arsen’eva N V, Putyatina LM, Tarasova N V, Tikhonov G V. Advantages and Disadvantages of Using Artificial Intelligence in Higher Education. Russ Eng Res. 2024;44: 1687–1690. doi:10.3103/S1068798X24702794

97. Osorio C, Fuster N, Chen W, Men Y, Juan AA. Enhancing accessibility to analytics courses in higher education through AI, simulation, and e-collaborative tools. Information. 2024;15: 430.

98. Simeonov S, Feradov F, Marinov A, Abu-Alam T. Integration of AI Training in the Field of Higher Education in the Republic of Bulgaria: An Overview. Educ Sci. 2024;14. doi:10.3390/educsci14101063

99. Boustani NM, Sidani D, Boustany Z. Leveraging ICT and Generative AI in Higher Education for Sustainable Development: The Case of a Lebanese Private University. Adm Sci. 2024;14. doi:10.3390/admsci14100251

100. Pacheco-Mendoza S, Guevara C, Mayorga-Albán A, Fernández-Escobar J. Artificial Intelligence in Higher Education: A Predictive Model for Academic Performance. Educ Sci. 2023;13. doi:10.3390/educsci13100990

101. Sinha RK, Deb Roy A, Kumar N, Mondal H. Applicability of ChatGPT in Assisting to Solve Higher Order Problems in Pathology. Cureus. 2023;15. doi:10.7759/cureus.35237

102. Ade-Ibijola A, Okonkwo C. Artificial Intelligence in Africa: Emerging Challenges. Responsible AI in Africa: Challenges and Opportunities. Springer; 2023. pp. 101–117. Available: https://link.springer.com/chapter/10.1007/978-3-031-08215-3_5#citeas

103. Wang S, Wang F, Zhu Z, Wang J, Tran T, Du Z. Artificial intelligence in education: A systematic literature review. Expert Syst Appl. 2024;252: 124167. doi:https://doi.org/10.1016/j.eswa.2024.124167

104. Kerimbayev N, Adamova K, Shadiev R, Altinay Z. Intelligent educational technologies in individual learning: a systematic literature review. Smart Learn Environ. 2025;12: 1–30. doi:10.1186/s40561-024-00360-3

105. Sova R, Tudor C, Tartavulea CV, Dieaconescu RI. Artificial Intelligence Tool Adoption in Higher Education: A Structural Equation Modeling Approach to Understanding Impact Factors among Economics Students. Electronics. 2024;13: 3632.

106. Mahmood NZ, Ahmed SR, Al-Hayaly AF, Algburi S, Rasheed J. The Evolution of Administrative Information Systems: Assessing the Revolutionary Impact of Artificial Intelligence. 2023 7th International Symposium on Multidisciplinary Studies and Innovative Technologies (ISMSIT). 2023. pp. 1–7. doi:10.1109/ISMSIT58785.2023.10304973

107. Automation Anywhere. University of Melbourne saves 10,000 hours annually with Automation Anywhere. 2024. Available: https://www.automationanywhere.com/resources/customer-stories/university-of-melbourne-aau

108. Zong Z, Guan Y. AI-Driven Intelligent Data Analytics and Predictive Analysis in Industry 4.0: Transforming Knowledge, Innovation, and Efficiency. J Knowl Econ. 2024; 1–40.

109. Mariani MM, Machado I, Magrelli V, Dwivedi YK. Artificial intelligence in innovation research: A systematic review, conceptual framework, and future research directions. Technovation. 2023;122: 102623. doi:10.1016/j.technovation.2022.102623

110. Ferk Savec V, Jedrinović S. The Role of AI Implementation in Higher Education in Achieving the Sustainable Development Goals: A Case Study from Slovenia. Sustain. 2025;17: 1–22. doi:10.3390/su17010183

111. Funda V, Piderit R. A review of the application of artificial intelligence in South African Higher Education. 2024 Conf Inf Commun Technol Soc ICTAS 2024 - Proc. 2024; 44–50. doi:10.1109/ICTAS59620.2024.10507113

112. Roscoe RD, Salehi S, Nixon N, Worsley M, Piech C, Luckin R. Inclusion and equity as a paradigm shift for artificial intelligence in education. Artificial Intelligence in STEM Education. CRC Press; 2022. pp. 359–374.

113. Gemiharto I, CMS S. Inclusivity and Accessibility in Digital Communication Tools: Case Study of AI-Enhanced Platforms in INDONESIA. J Pewarta Indones. 2024;6: 78–88. doi:10.25008/jpi.v6i1.154

114. Schwab K. The fourth industrial revolution. Crown Currency; 2017.

115. Sajja R, Sermet Y, Cikmaz M, Cwiertny D, Demir I. Artificial Intelligence-Enabled Intelligent Assistant for Personalized and Adaptive Learning in Higher Education. 2023; 1–23. doi:10.3390/info15100596

116. Mohamed Hashim MA, Tlemsani I, Matthews R. Higher education strategy in digital transformation. Educ Inf Technol. 2022;27: 3171–3195. doi:10.1007/s10639-021-10739-1

117. Knox J. Artificial intelligence and education in China. Learn Media Technol. 2020;45: 298–311. doi:10.1080/17439884.2020.1754236

118. Kalyanakrishnan S, Panicker RA, Natarajan S, Rao S. Opportunities and Challenges for Artificial Intelligence in India. Proceedings of the 2018 AAAI/ACM Conference on AI, Ethics, and Society. New York, NY, USA: Association for Computing Machinery; 2018. pp. 164–170. doi:10.1145/3278721.3278738

119. Sharma S, Singh G, Sharma CS, Kapoor S. Artificial intelligence in Indian higher education institutions: a quantitative study on adoption and perceptions. Int J Syst Assur Eng Manag. 2024. doi:10.1007/s13198-023-02193-8

120. World Economic Forum. State of the education report for India, 2022: artificial intelligence in education; here, there and everywhere. World Economic Forum; 2022. Available: https://unesdoc.unesco.org/ark:/48223/pf0000382661

121. Mumtaz S, Carmichael J, Weiss M, Nimon-Peters A. Ethical use of artificial intelligence based tools in higher education: are future business leaders ready? Educ Inf Technol. 2024. doi:10.1007/s10639-024-13099-8

122. O'Dea X. Generative AI: is it a paradigm shift for higher education? Stud High Educ. 2024;49: 811–816. doi:10.1080/03075079.2024.2332944

123. Salas-Pilco SZ, Yang Y. Artificial intelligence applications in Latin American higher education: a systematic review. Int J Educ Technol High Educ. 2022;19: 21. doi:10.1186/s41239-022-00326-w

124. Rios-Campos C, Zambrano EOG, Cantos MFM, Anchundia-Gómez O, León MEC, Moya GEM, et al. Universities and Artificial Intelligence. South Florida J Dev. 2024;5: e4016. doi:10.46932/sfjdv5n6-010

125. Thagard P. The Ethics of Artificial Intelligence. Bots Beasts. 2021; 225–248. doi:10.7551/mitpress/14102.003.0010

126. UNESCO. Recommendation on the Ethics of AI. Unesco. 2021; 1–21. Available: https://unesdoc.unesco.org/ark:/48223/pf0000380455

127. NASSCOM. Responsible AI: Principles and Practices. New Delhi: National Association of Software and Service Companies; 2021.

128. Aayog N. National Strategy for Artificial Intelligence. New Delhi: NITI Aayog; 2018.

129. U.S. Department of Education. Guidelines for AI in Education. Washington, D.C.: Department of Education; 2021.

130. Katsamakas E, Pavlov O V, Saklad R. Artificial intelligence and the transformation of higher education institutions: A systems approach. Sustainability. 2024;16: 6118.

131. European Commission. Ethics Guidelines for Trustworthy AI. Brussels: European Commission; 2019.

132. European Commission. EU Artificial Intelligence Act. Brussels: European Commission; 2021.

133. GPAI. Ethical AI in Practice. Montreal: Global Partnership on AI; 2020.

134. Silberg J, Manyika J. Notes from the AI frontier : Tackling bias in AI ( and in humans ). Mckinsey Glob Inst. 2019; 1–8. Available: https://www.mckinsey.com/~/media/mckinsey/featured insights/artificial intelligence/tackling bias in artificial intelligence and in humans/mgi-tackling-bias-in-ai-june-2019.pdf

135. Prem E. From ethical AI frameworks to tools: a review of approaches. AI Ethics. 2023;3: 699–716. doi:10.1007/s43681-023-00258-9

136. Pedro F, Subosa M, Rivas A, Valverde P. Artificial intelligence in education: Challenges and opportunities for sustainable development. 2019.

137. Kaplan A, Haenlein M. Siri, Siri, in my hand: Who's the fairest in the land? On the interpretations, illustrations, and implications of artificial intelligence. Bus Horiz. 2019;62: 15–25. doi:10.1016/j.bushor.2018.08.004

138. Essien A, Bukoye OT, O'Dea X, Kremantzis M. The influence of AI text generators on critical thinking skills in UK business schools. Stud High Educ. 2024;49: 865–882. doi:10.1080/03075079.2024.2316881

139. Pinto M, Leite C. Digital technologies in support of students learning in higher education: Literature review. Digit Educ Rev. 2020; 343–360. doi:10.1344/DER.2020.37.343-360

140. Hunt P, Leijen Ä, van der Schaaf M. Automated feedback is nice and human presence makes it better: Teachers' perceptions of feedback by means of an e-portfolio enhanced with learning analytics. Educ Sci. 2021;11. doi:10.3390/educsci11060278

141. UNESCO. Foundation models such as ChatGPT through the prism of the UNESCO Recommendation on the Ethics of Artificial Intelligence. 2023; 13.

142. Escotet MÁ. The optimistic future of Artificial Intelligence in higher education. Prospects. 2023. doi:10.1007/s11125-023-09642-z

143. Gligorea I, Cioca M, Oancea R, Gorski A-T, Gorski H, Tudorache P. Adaptive learning using artificial intelligence in e-learning: a literature review. Educ Sci. 2023;13: 1216.

144. Stanford University. Introduction to the AI Index Report 2024. 2024.

145. Alordiah CO. Proliferation of Artificial Intelligence Tools: Adaptation Strategies in the Higher Education Sector. Propellers J Educ. 2023;2: 53–65. Available: https://ijvocter.com/pjed/article/view/68

146. Correa-Peralta M, Vinueza-Martínez J, Castillo-Heredia L. Evolution, topics and relevant research methodologies in Business Intelligence and Data Analysis in the Academic Management of Higher Education Institutions. A literature review. Results Eng. 2024; 103782. doi:https://doi.org/10.1016/j.rineng.2024.103782

147. Sharma S, Singh G. Adoption of artificial intelligence in higher education: an empirical study of the UTAUT model in Indian universities. Int J Syst Assur Eng Manag. 2024. doi:10.1007/s13198-024-02558-7

148. Tang PM, Koopman J, Mai KM, De Cremer D, Zhang JH, Reynders P, et al. No person is an island: Unpacking the work and after-work consequences of interacting with artificial intelligence. J Appl Psychol. 2023.

149. MA W, MA W, HU Y, BI X. The who, why, and how of ai-based chatbots for learning and teaching in higher education: A systematic review. Educ Inf Technol. 2024. doi:10.1007/s10639-024-13128-6

150. Köbis L, Mehner C. Ethical Questions Raised by AI-Supported Mentoring in Higher Education. Front Artif Intell. 2021;4: 1–9. doi:10.3389/frai.2021.624050

151. George AS. Artificial Intelligence and the Future of Work: Job Shifting Not Job Loss. Partners Univers Innov Res Publ. 2024;2: 17–37.

152. Lopez M, Gonzalez I. Artificial Intelligence Is Not Human: The Legal Determination of Inventorship and Co-Inventorship, the Intellectual Property of AI Inventions, and the Development of Risk Management Guidelines. J Pat Trademark Off Soc'y. 2024;104: 135.

153. Cremer D De, Koopman J. Using AI at Work Makes Us Lonelier and Less Healthy. In: Havard Business Review [Internet]. 2024 [cited 26 Dec 2024]. Available: https://hbr.org/2024/06/research-using-ai-at-work-makes-us-lonelier-and-less-healthy

154. Danaher J, Nyholm S. Automation, work and the achievement gap. AI Ethics. 2021;1: 227–237. doi:10.1007/s43681-020-00028-x

155. Nemorin S, Vlachidis A, Ayerakwa HM, Andriotis P. AI hyped? A horizon scan of discourse on artificial intelligence in education (AIED) and development. Learn Media Technol. 2023;48: 38–51. doi:10.1080/17439884.2022.2095568

156. Palomares I, Martínez-Cámara E, Montes R, García-Moral P, Chiachio M, Chiachio J, et al. A panoramic view and swot analysis of artificial intelligence for achieving the sustainable development goals by 2030: progress and prospects. Applied Intelligence. 2021. doi:10.1007/s10489-021-02264-y

157. UNESCO. Challenges and Opportunities for Sustainable Development Education Sector. UNESCO Work Pap Educ Policy, No 7 Fr Pedró (Ed) . 2019; 1–48. Available: https://en.unesco.org/themes/education-policy-

158. van de Poel I. Embedding Values in Artificial Intelligence (AI) Systems. Minds Mach. 2020;30: 385–409.

doi:10.1007/s11023-020-09537-4

159. Crawford LM. Conceptual and theoretical frameworks in research. Research Design and Methods: An Applied Guide for the Scholar-Practitioner. 2019. pp. 35–48. Available: https://uk.sagepub.com/sites/default/files/upm-binaries/105274_ch03_14.pdf

160. Yang SJH, Ogata H, Matsui T, Chen NS. Human-centered artificial intelligence in education: Seeing the invisible through the visible. Comput Educ Artif Intell. 2021;2: 100008. doi:10.1016/j.caeai.2021.100008

161. Bond RR, Mulvenna M, Wang H. Human centered artificial intelligence: Weaving UX into algorithmic decision making. RoCHI 2019 Int Conf Human-Computer Interact. 2019; 2–9. Available: https://pure.ulster.ac.uk/en/publications/human-centered-artificial-intelligence-weaving-ux-into-algorithmi

162. Pedro F, Subosa M, Rivas A, Valverde P. Artificial Intelligence in Education: Challenges and Opportunities for Sustainable Development Education Sector United Nations Educational, Scientific and Cultural Organization. Minist Educ. 2019; 1–46. Available: https://en.unesco.org/themes/education-policy-

163. Cowls J, Floridi L. Prolegomena to a white paper on an ethical framework for a good AI society. 2018.

164. Guilbault M. Students as customers in higher education: The (controversial) debate needs to end. J Retail Consum Serv. 2016; 0–1. doi:10.1016/j.jretconser.2017.03.006

165. Yelkur R. Customer Satisfaction and the Services Marketing Mix. J Prof Serv Mark. 2000;21: 105–115. doi:10.1300/J090v21n01

166. Boddington P. Towards a Code of Ethics for Artificial Intelligence. Springer; 2017.

167. Cope B, Kalantzis M, Searsmith D. Artificial intelligence for education: Knowledge and its assessment in AI-enabled learning ecologies. Educ Philos Theory. 2021;53: 1229–1245. doi:10.1080/00131857.2020.1728732

168. Ojha M, kumar Mishra A, Kandpal V, Singh A. The Ethical Dimensions of AI Development in the Future of Higher Education: Balancing Innovation with Responsibility. Ethical Dimensions of AI Development. IGI Global; 2025. pp. 401–436.

169. Seldon A, Abidoye O, Metcalf T. The Fourth Education Revolution Reconsidered: Will Artificial Intelligence Enrich Or Diminish Humanity? Legend Press Ltd; 2020.

170. Bucea-Manea-ţoniş R, Kuleto V, Gudei SCD, Lianu C, Lianu C, Ilić MP, et al. Artificial Intelligence Potential in Higher Education Institutions Enhanced Learning Environment in Romania and Serbia. Sustain. 2022;14: 1–18. doi:10.3390/su14105842

171. Slimi Z, Villarejo-Carballido B. Unveiling the Potential: Experts' Perspectives on Artificial Intelligence Integration in Higher Education. Eur J Educ Res. 2024;13.

172. Murdan AP, Halkhoree R. Integration of Artificial Intelligence for educational excellence and innovation in higher education institutions. 2024 1st International Conference on Smart Energy Systems and Artificial Intelligence (SESAI). IEEE; 2024. pp. 1–6.

173. Leoste J, Jõgi L, Õun T, Pastor L, San Martín López J, Grauberg I. Perceptions about the future of integrating emerging technologies into higher education—the case of robotics with artificial intelligence. Computers. 2021;10. doi:10.3390/computers10090110

174. Lukichyov PM, Chekmarev OP. Artificial intelligence in higher education. Russ J Innov Econ. 2023;13: 485–502.

175. Luckin R. Machine Learning and Human Intelligence. The future of education for the 21st century. UCL institute of education press; 2018.

176. Holstein K, McLaren BM, Aleven V. Designing for complementarity: Teacher and student needs for orchestration support in AI-enhanced classrooms. Artificial Intelligence in Education: 20th International Conference, AIED 2019, Chicago, IL, USA, June 25-29, 2019, Proceedings, Part I 20. Springer; 2019. pp. 157–171.

177. Ivanov S. The dark side of artificial intelligence in higher education. Serv Ind J. 2023;43: 1055–1082.

178. Helmiatin, Hidayat A, Kahar MR. Investigating the adoption of AI in higher education: a study of public universities in Indonesia. Cogent Educ. 2024;11: 2380175.

179. Luckin R. Machine Learning and Human Intelligence: The Future of Education for the 21st Century. UCL IOE Press. 2018. Available: https://eric.ed.gov/?id=ED584840

180. Dwivedi YK, Hughes L, Ismagilova E, Aarts G, Coombs C, Crick T, et al. Artificial Intelligence (AI): Multidisciplinary perspectives on emerging challenges, opportunities, and agenda for research, practice and policy. Int J Inf Manage. 2021;57. doi:10.1016/j.ijinfomgt.2019.08.002

181. Bhatnagar H. Artificial Intelligence-New Horizon in Indian Higher Education. J Learn Teach Digit Age. 2020;2020: 30–34.

182. Xu W, Member IEEE S, Gao Z, Dainoff M. An HCAI Methodological Framework: Putting It Into Action to Enable Human-Centered AI. arXiv Prepr arXiv231116027. 2023. Available: https://arxiv.org/abs/2311.16027v3

183. Dwivedi YK, Hughes L, Ismagilova E, Aarts G, Coombs C, Crick T, et al. Artificial Intelligence (AI): Multidisciplinary perspectives on emerging challenges, opportunities, and agenda for research, practice and policy. Int J Inf Manage. 2021;57: 101994.

184. Floridi L, Cowls J. A unified framework of five principles for AI in society. Mach Learn City Appl Archit Urban Des. 2022; 535–545. doi:10.1002/9781119815075.ch45

185. Kurtz G, Amzalag M, Shaked N, Zaguri Y, Kohen-Vacs D, Gal E, et al. Strategies for Integrating Generative AI into Higher Education: Navigating Challenges and Leveraging Opportunities. Educ Sci. 2024;14. doi:10.3390/educsci14050503

186. Xu W, Ouyang F. A systematic review of AI role in the educational system based on a proposed conceptual framework. Educ Inf Technol. 2022;27: 4195–4223. doi:10.1007/s10639-021-10774-y

187. Rai A. Explainable AI: from black box to glass box. J Acad Mark Sci. 2020;48: 137–141. doi:10.1007/s11747-019-00710-5

188. Laplante P, Amaba B. Artificial Intelligence in Critical Infrastructure Systems. Computer (Long Beach Calif). 2021;54: 14–24. doi:10.1109/MC.2021.3055892

189. Chaudhry J, Pathan ASK, Rehmani MH, Bashir AK. Threats to critical infrastructure from AI and human intelligence. J Supercomput. 2018;74: 4865–4866. doi:10.1007/s11227-018-2614-0

190. Hair JF. Multivariate data analysis. 2009.

191. Green T. A methodological review of structural equation modelling in higher education research. Stud High Educ. 2016;41: 2125–2155. doi:10.1080/03075079.2015.1021670

192. Teo T. Handbook of quantitative methods for educational research. Springer Science & Business Media; 2014.

193. Byrne BM. Structural Equation Modeling With AMOS. Structural Equation Modeling With AMOS. 2013. doi:10.4324/9781410600219

194. Kline RB. Principles and practice of structural equation modelling (4th ed.). Methodol Soc Sci. 2015; 1–554.

195. Hu L, Bentler PM. Cutoff criteria for fit indexes in covariance structure analysis: Conventional criteria versus new alternatives. Struct Equ Model a Multidiscip J. 1999;6: 1–55.